\documentclass[useAMS,usenatbib]{mn2e}
\usepackage{aas_macros}
 \usepackage{epsfig}
 \usepackage{figs}
 \usepackage{stfloats}
 \usepackage{fixltx2e}
 \usepackage{relsize}
\usepackage[usenames,dvipsnames]{color}

\newcommand{\jwb}[1]{\textcolor{black}{#1}}
\newcommand{\jwg}[1]{\textcolor{black}{#1}}
\newcommand{\jwr}[1]{\textcolor{black}{#1}}

\def\eq#1{Eq.~(\ref{#1})}
\def\Eq#1{Eq.~(\ref{#1})}
\def\sec#1{\S\ref{#1}}

\def\twosec#1#2{\S\ref{#1} and \ref{#2}}
\def\Fig#1{Fig.~\ref{#1}}
\def\fig#1{Fig.~\ref{#1}}

\def\etal{et al.}
\def\ie{i.e.}
\def\eg{e.g.}
\def\be{\begin{equation}}
\def\ee{\end{equation}}
\def\prop{\propto}
\def\ifm#1{\relax\ifmmode#1\else$\mathsurround=0pt #1$\fi}
\def\kms{\ifmmode\,{\rm km}\,{\rm s}^{-1}\else km$\,$s$^{-1}$\fi}

\def\msun{{\rm  M}_{\odot} }
\def\Msun{{\rm  M}_{\odot} }

\def\lteq{\; \buildrel < \over = \;}
\newcommand {\gtsima} {\ {\raise-.5ex\hbox{$\buildrel>\over{\mathsmaller \sim}$}}\ }
\newcommand {\ltsima} {\ {\raise-.5ex\hbox{$\buildrel<\over{\mathsmaller \sim}$}}\ } 

\def\Ms{M_{\rm *}}

\def\Mh{M_{\rm h}}
\def\Mcrit{M_{\rm crit}}

\def\Rvir{R_{\rm vir}}

\def\Ha{{\rm H}\alpha}
\def\Hb{{\rm H}\beta}
\def\OII{\ion{O}{2}}

\def\Dthree{\log(1+\delta_3)}
\def\Dfive{\log(1+\delta_5)}
\def\dproj{d_{\rm proj}}
\def\Dist{D}

\title[Quenching by Halo Environment]
{\jwb{Dependence of Galaxy Quenching on Halo Mass and Distance from its Centre}}

\author[Woo \etal] 
    {\parbox{\textwidth}{Joanna Woo$^{1}$\thanks{joaw@phys.huji.ac.il}, Avishai
    Dekel$^{1}$, S. M. Faber$^{2}$, Kai Noeske$^{3}$, David
    C. Koo$^{2}$, 
    Brian F. Gerke$^{4}$, Michael C. Cooper$^{5,6}$,
    Samir Salim$^{7}$, Aaron A. Dutton$^8$, 
    Jeffrey Newman$^{9}$, Benjamin
    J. Weiner$^{10}$, Kevin Bundy$^{11}$, Christopher
    N. A. Willmer$^{10}$, 
    Marc Davis$^{12}$, Renbin Yan$^{13}$}
\vspace{0.4cm}\\
\parbox{\textwidth}{ 
$^{1}$Racah Institute of Physics, The Hebrew University, Jerusalem
91904 Israel;  \\
$^{2}$University of California Observatoires/Lick Observatory,
 University of California, Santa Cruz, CA 95064 \\
$^{3}$Harvard–Smithsonian Center for Astrophysics, 60 Garden Street, Cambridge, MA 02138, USA\\
$^{4}$Kavli Institute for Particle Astrophysics and Cosmology, Department of Physics, Stanford University, 
Stanford, CA 94305, USA\\
$^{5}$Center for Galaxy Evolution, Department of Physics
 and Astronomy, University of California, Irvine, 4129 Frederick
 Reines Hall, Irvine, CA 92697, USA; m.cooper@uci.edu\\
$^{6}$Hubble Fellow\\
$^{7}$Department of Astronomy, Indiana University, Bloomington, IN 47404, USA \\
$^{8}$Department of Physics, University of California, Santa Barbara, CA
 93106, USA\\
$^{9}$Department of Physics and Astronomy, University of Pittsburgh, 3941 O’Hara Street, Pittsburgh, PA 15260, USA \\
$^{10}$Steward Observatory, University of Arizona, 933 N. Cherry Avenue, Tucson, AZ 85721, USA \\
$^{11}$Kavli Institute for the Physics and Mathematics of the
 Universe, University of Tokyo, Kashiwa, 277-8583, Japan \\
$^{12}$Department of Astronomy, University of California, Berkeley, CA 94705, USA\\
$^{13}$Center for Cosmology and Particle Physics, Department of Physics, New York University, 4 Washington Place, New York, NY 10003, USA \\
}
}

\pagerange{\pageref{firstpage}--\pageref{lastpage}} \pubyear{2012}

\begin{document}

\label{firstpage}

\maketitle

\begin{abstract}
  We study the dependence of star-formation quenching on galaxy mass
  and environment, in the SDSS ($z \sim 0.1$) and the AEGIS ($z \sim
  1$).  \jwg{It is crucial that we define quenching by low
    star-formation rate rather than by red colour, given that one
    third of the red galaxies are star forming.  We address stellar
    mass $\Ms$, halo mass $\Mh$, density over the nearest $N$
    neighbours $\delta_N$, and distance to the halo centre $\Dist$.}
  \jwb{The fraction of quenched galaxies appears more strongly
    correlated with $\Mh$ at fixed $\Ms$ than with $\Ms$ at fixed
    $\Mh$, while for satellites quenching also depends on $\Dist$.}
  \jwg{We present the $\Ms$-$\Mh$ relation for centrals at $z \sim 1$.
    At $z \sim 1$, the dependence of quenching on $\Ms$ at fixed $\Mh$
    is somewhat more pronouced than at $z \sim 0$, but the quenched
    fraction is low (10\%) and the haloes are less massive.}  \jwb{For
    satellites, $\Ms$-dependent quenching is noticeable at high
    $\Dist$, suggesting a quenching dependence on sub-halo mass for
    recently captured satellites.  At small $\Dist$, where satellites
    likely fell in more than a few Gyr} ago, quenching strongly
  depends on $\Mh$, and not on $\Ms$.  The $\Mh$-dependence of
  quenching is consistent with theoretical wisdom where virial shock
  heating in massive haloes shuts down accretion and triggers
  ram-pressure stripping, causing quenching.  The interpretation of
  $\delta_N$ \jwg{is complicated by the fact that it} depends on the
  number of observed group members compared to $N$, motivating the use
  of $\Dist$ as a better measure of local environment.
\end{abstract}


\section{Introduction}
\label{intro}

Galaxies can be crudely divided into two classes: blue star-forming
disc galaxies and red ellipticals dominated by old stars
\citep{str01,kau03a,bla03,bald04}.  The blue galaxies lie on a sequence in
the star formation rate (SFR)-stellar mass ($\Ms$) plane in that the
more massive galaxies form stars at a higher rate
\citep[\eg,][]{bri04,sal07,noe07,gilbank10}.  \cite{noe07} find that this
sequence has persisted since $z \sim 1$, but has decreased in
normalisation (zero point) to the present day.  This sequence is often
called the ``main sequence'', but we will refer to it as the
``star-forming sequence'' (SF sequence) following \cite{sal07}.  In
a plot of SFR vs. stellar mass, the intrinsically red galaxies tend to
lie below the SF sequence.

The galaxy bimodality is manifest not only in their star formation
properties, but also in their masses and the number density of
surrounding galaxies.  In the local universe, less dense environments
usually feature less massive blue star-forming disc galaxies while
denser surroundings like the cluster environment tend to host massive
red and dead S0 and elliptical galaxies \citep[first measured by
][]{oem74,dav76,dre80}.  Pure ellipticals tend to be the most massive
of intrinsically red galaxies and reside in the centres of large
clusters \citep{kormendy12}.

Large surveys of hundreds of thousands of galaxies have made it
possible to study density correlations with unprecedented statistical
power.  In the Sloan Digital Sky Survey \citep[SDSS - see][]{york00},
\cite{hog03} find that for the blue cloud in the colour-magnitude
diagram, the environment density correlates with colour (redder
galaxies lie in denser environments) \citep[see also
][]{bal04,blanton05b,baldry06}.  Likewise, \cite{kau04} found that
density correlates with several galaxy properties, and most strongly
with mean SFR/$\Ms$ (or specific star formation rate - SSFR), in that
the average SFR/$\Ms$ of galaxies is lower in denser environments.
Yet the SFR-density relation seems not to apply to the SF sequence
itself but is manifested in the {\it ratio} of quenched galaxies to
star forming galaxies
\citep{carter01,bal04,tanaka04,rines05,wolf09,vonderLinden10,peng10}.

There is evidence that environment-related quenching operates
differently between the most massive galaxies of a group or cluster,
usually assumed to be residing at the centres of the group potential
well (and hence are called centrals), and those that are in orbit
around them (satellites).  \cite{hog03} find that for the red sequence
in the colour-magnitude diagram of galaxies, the most and least
luminous galaxies, \ie, the centrals and satellites respectively - see
\cite{berlind05}, lie in higher density environments while
intermediately luminous galaxies on the red sequence lie in a density
``saddle''.  Some \citep{vandenBosch08,peng10,peng11} even suggest
that density-related quenching affects only satellites and not
centrals.

The picture at $z\sim 1$ is similar.  In the DEEP2 galaxy survey
\citep{davis03}, just as the colour bimodality persists out to $z \sim
1$ \citep{lin99,im01,bel04,wil06,faber07}, the mean colour-density
relation \citep{coo06} and the SSFR-density \citep{coo08} relation are
also present at $z \sim 1$, while, at these redshifts, \cite{bundy06}
find that red-sequence galaxies are still more abundant in dense
environments than their star-forming counterparts.  A recent study
\citep{quadri12} shows that this star-formation-density relation
persists out to $z\sim1.8$.

Some have suggested that these relations are driven by stellar mass
\citep{vandenBosch08,marc08,grutzbauch11}, while others
(\citealp{cooper10} at $z\sim 1$ and \citealp{peng10,peng11} at $z\sim
0$) show that the SFR-density relation is separable from the
colour-mass relation.  It has also been a point of contention whether
or not the ``environment-related'' quenching, which is usually
measured on group scales, is fundamentally driven by the presence of
near neighbours, or the halo mass or both \citep[\eg,
][]{weinmann06,vandenBosch08,skibba09a,skibba09b,wetzel12,peng11}.
Thus in order to understand the drivers of quenching, it is necessary
to understand first how environment density, stellar mass and halo
mass relate to each other, properly treating centrals and satellites
separately in a coherent picture, as we attempt to do in this study.

\subsection{Density, stellar mass, halo mass and satellite distance}
\label{introrel}

To first order, haloes are self-similar in that they all have the same
density profile $\rho(r/\Rvir)$ in units of the virial radius
regardless of their mass.  Thus, if the density of a galaxy's
environment (\ie, the number density of other galaxies around it -
let's call it $\delta$) follows the halo mass density $\rho$, the
galaxy's environment density $\delta$ is not expected to be correlated
with its halo mass.  However, on closer inspection, the $\delta$-halo
mass relation is more complex and depends strongly on the $\delta$
measure, the number of observed galaxies within a group, and the
limiting mass/magnitude of the galaxies counted in a group.  For
example, environment density is often measured by the distance to a
galaxy's $N$th nearest neighbour (for helpful reviews of density
measures see \citealp{coo05,muldrew12}).  On reflection it is clear
that such a density, $\delta_N$, has two modes depending on whether or
not the galaxy is isolated or is embedded in a massive halo containing
many satellites.  For isolated galaxies (and for groups with fewer
than $N$ satellites), $\delta_N$ is a measure of the proximity to the
next big haloes, and is therefore a measure of field density over a
large volume (this has been pointed out by for example,
\citealp{weinmann06}).  We call this the {\it cross-halo mode} of
$\delta_N$.  For a galaxy belonging to a massive halo, $\delta_N$
measures the density of galaxies within its own halo.  We call this
the {\it single-halo mode} of $\delta_N$.  In the single-halo mode the
number density of galaxies is expected to correlate with the halo mass
under the assumption that the distribution of satellites follows the
dark matter distribution which is described by something similar to
the NFW profile \citep{NFW97}.  For more massive haloes, the dark
matter profile falls off less steeply with radius than for less
massive haloes, so that if you are a central galaxy in a more massive
halo, the probability of finding the $N$th nearest satellite close to
you is higher.  Similarly, if you are a satellite in a large halo, the
probability that you lie closer to the centre (and thus to other
satellites) is higher in more massive haloes.

Because of these dual modes of density, the $\delta_N$-$\Ms$ and
$\delta_N$-$\Mh$ relations, and hence quenching in these spaces, will
depend on the group membership of the galaxy.  It turns out that most
centrals reside in groups with fewer members than the typical $N$
(3-5), so the interpretation of $\delta_N$ for centrals is relatively
straightforward.  However, a significant number of satellites will
reside in groups of more than $N$ members and thus the interpretation
of $\delta_N$ will be not be simple.  This motivates the search for a
better measure of ``environment'' for satellites.  We propose in this
study the use of the relative projected distance of the satellite from
its group centre ($\Dist$).

These complex relations have not been clearly understood to date
although some have made major steps toward this end (see for example
\citealp{wilman10} who studied how halo mass relates to environment
measures on different scales, and also \citealp{haas12} who studied
the relation between several environment measures and halo mass in
dark matter simulations).  Recent group catalogues by \cite{gerke05}
(for the All-Wavelength Extended Groth Strip International Survey -
AEGIS) and \cite{yang07} (for SDSS) make it possible to disentangle
these relations and inform our intuition.

Also important to quenching studies is the understanding of the
relation between a galaxy's stellar mass and halo mass.  \cite{yang08}
used their group catalogue of the SDSS to show that stellar mass for
centrals correlates strongly with halo mass up to a certain $\Ms$
after which centrals of the similar massive $\Ms$ can reside in a wide
range of halo masses.  This spread in halo masses at high $\Ms$
presents an opportunity to discriminate between halo-dependent quenching
and $\Ms$-dependent quenching.

With a proper understanding of these relations between density, mass,
and distance for centrals and satellites, we can distinguish between
different mechanisms of quenching, which we will describe next.

\subsection{Quenching}
\label{introquenching}

Several broad categories of quenching mechanisms have been proposed,
with the mechanisms acting on central galaxies being different from the
mechanisms acting on satellites.

Quenching mechanisms for central galaxies fall into two classes, those
that remove existing gas from central galaxies, which we term
``internal'', and those that prevent new gas from falling in and
cooling, which we term ``external''.  Internal mechanisms include
active galactic nucleus (AGN) feedback \citep{granato04,springel05},
quasar-driven winds generated by collisions of galaxies
\citep{springel05}, and stellar (radiation, winds and/or supernova)
feedback \citep{dek86,fall10}.

Among the external mechanisms is what we call ``halo quenching''.  In
this picture, haloes more massive than the critical mass for shock
heating, $M_{\rm crit} \sim {\rm few} \times 10^{12}\msun$ are capable
of sustaining a virial shock \citep[\eg, ][]{bir03,ker05,dekbir06}.
The heating by such an expanding virial shock \citep{birnboim07}, by
the energy of gravitational infall \citep{dek08}, or by AGN feedback
that naturally couples to the hot dilute gas when it is present, shuts
down the cold gas supply and suppresses star formation in galaxies
that reside in that massive halo \citep{dekbir06}.  For a population of
haloes, this should not be interpreted as a sharp threshold but rather
as a mass range extending over one or two decades where there is
gradual growth of the fraction of quenched galaxies as a function of
halo mass.  

Broadly speaking, the above proposed external mechanism for quenching
centrals should depend on halo properties rather than properties of
the galaxies themselves.  The internal mechanisms, on the other hand,
depend on conditions or properties of the galaxy itself, such as
black-hole mass, merger state, or star-formation rate.  This
distinction is useful for testing the importance of quenching
mechanisms as we attempt to do in this study.  However we recognise
that quenching could be a combination of external and internal
mechanisms.  For example, \cite{dekbir06} suggested that halo quenching
is possibly connected to the ``radio mode'' of AGN feedback from
central black holes which inhibits central cooling in hot haloes
\citep[see][]{croton06}.

As for satellite galaxies, the above mechanisms may have operated on
them while they were still centrals, and may continue to operate on
them to some degree after they become satellites by falling into a
more massive halo.  However, there are additional quenching mechanisms
that are unique to satellites only.  One such mechanism is falling
into a more massive and hotter halo than the satellite's original
halo, impeding gas cooling onto that satellite, usually referred to as
strangulation \citep{larson80,balogh00,dekbir06,choi09}.  Another
mechanism is ram pressure stripping \citep{gunn72,abadi99} which
occurs only when the satellite orbital velocity within the larger halo
greatly exceeds its own internal dynamical velocities.  Ram pressure
stripping is efficient where gas densities are higher, \ie, in the
central regions of host haloes.  Tidal stripping \citep{read06} and
harassment \citep{moore96,moore98} are two more quenching mechanisms
unique to satellites.  Their efficiencies depend on the mass ratio of
the interaction, orbital eccentricity, and high environment densities
\citep{villalobos12}.

Among the first to explore the importance of external vs. internal
quenching was a study by \cite{weinmann06} who found that halo mass
was a better predictor of quenching for centrals and satellites than
galaxy luminosity, pointing to external mechanisms of quenching.
However, since the range of luminosity is large for a given $\Ms$, any
sharp trend of quenching with $\Ms$ may have been diluted, leaving the
question of internal vs. external quenching still uncertain.

\cite{kimm09} also studied quenching in relation to $\Ms$ and $\Mh$
but could not decide which was more important for centrals.  For
satellites, they found that they were equally important, but did not
address the importance of environment density or central proximity.  

Two recent studies of quenching \citep{peng10,peng11} propose an
empirical model of quenching in order to explain the observed
correlations between quenching, environment density and stellar mass.
\jwb{They propose that internal quenching mechanisms for centrals should
correlate with SFR. Without choosing a specific mechanism, they
suggested that such a mechanism would be consistent with any kind of
stellar feedback or AGN feedback.  Through the SFR-$\Ms$ relation of
star forming galaxies (\ie, the SFR sequence), a correlation between
quenching and $\Ms$ is observed.  In addition to this internal
SFR-dependent quenching, the observed correlation between quenching
and density is assumed to be satellite-specific quenching of the sort
discussed above.  This idea of two kinds of mechanisms for quenching,
one mass-dependent and one satellite-specific, has been implemented in
semi-analytical models (see for example, \citealp{croton06}).}

We build on this picture by putting everything in the context of the
halo model, which likewise predicts different behaviour of quenching
for satellites and centrals.  However, a difference is that the halo
\jwr{quenching} model predicts stronger dependence on $\Mh$ than with $\Ms$,
for both centrals and satellites, and this different quenching
dependence proves to be observable.  In the process, we also choose a
better indicator of environmental density, $\Dist$ in preference to
$\delta_N$, which provides additional insight for satellites.  \jwb{In
  addition, we aim to study these quenching relations at $z\sim 1$
  where the number density of haloes above $M_{\rm h,crit}$ is lower,
  so halo quenching is not expected to be as important since fewer
  galaxies reside in large enough halos for virial shock quenching.}
Finally, we improve on previous studies by using dust-corrected SFR
instead of colour to separate quenched galaxies from star forming
galaxies.  It turns out that the choice of colour versus SFR to
classify a quenched galaxy yields different quenching behaviour.

\subsection{The goals of this paper}

In this paper, our goals are twofold.  First, we aim to form an
intuitive and consistent understanding of the complex relations between
$\delta_N$, $\Ms$ and $\Mh$ for central galaxies with the addition of
$\Dist$ for satellites using the group information available
for the SDSS and DEEP2 surveys.  We demonstrate that keys to
understanding these relations include separating the relations for
centrals and satellites and the fact that $\delta_N$ behaves
differently in cross-halo and single-halo modes.

With this understanding, our second goal is to test the physically
motivated hypothesis that quenching scales with $\Mh$ as predicted by
halo quenching \citep{bir03,ker05,dekbir06}, but differently for
satellites and centrals when viewed in the diagrams of $\Ms$, $\Mh$,
$\delta_N$ and $\Dist$.  Along the way, we compare quenching as
measured by dust-corrected SFR with quenching as measured by colour to
see how much of an effect dust has on quenching behaviour.

Our detailed mappings of quenched fractions as a function of $\Ms$,
$\Mh$, $\delta_N$ and $\Dist$ provide a rich repertoire of data for
future testing of models.  Matching all of these relationships
quantitatively is a severe challenge.  \jwb{Several current semi-analytical
models have problems reproducing some aspects of the environmental
dependencies of quenching \citep{weinmann06b,kimm09} and therefore
quenching trends need to be observed more closely.  This is
possible because star formation rates are becoming quantitatively
accurate under a wide range of conditions.}

The paper is organised as follows: In \sec{data}, we describe the data
sets, particularly the $\Ms$, environment density measures, and the
group catalogues used to estimate $\Mh$, and SFR measures for the SDSS
and the Extended Groth Strip (EGS) field of DEEP2 (taken from the
All-Wavelength EGS International Survey - AEGIS).  In \sec{SFvscolour}
we show that using dust-corrected SFR gives a better definition of
star forming and quenched galaxies than rest-frame $U-B$ colour.  In
\sec{tutorial}, we present and discuss the relations between $\Mh$,
$\Ms$, $\delta_N$ and $\Dist$ for the SDSS.  Our main results are
presented in \sec{quenching} where we study the quenched fraction as a
function of these quantities for central and satellite galaxies in the
SDSS, and compare our results to previous work.  In
\sec{tutorialaegis} we do the same exercise for the DEEP2 survey
examining quenching in a context where halo quenching is not expected
to be important.  In \sec{discuss}, we summarise our results.

All SFR and $\Ms$ values are calibrated using the \cite{kro01} initial
mass function (IMF), and concordance cosmology ($H_o = 70~{\rm km~s^{-1}
  Mpc^{-1}}, \Omega_M = 0.3, \Omega_\Lambda=0.7$).  All magnitudes are
given in the AB system.  Halo masses are converted to this 
cosmology with $\sigma_8=0.9, \Omega_b =0.04$.

\section{The Data}
\label{data}

\subsection{SDSS}
\label{sdssdata}

\subsubsection{The sample}
\label{sdsssample}

The SDSS sample used throughout this analysis is limited to the DR6
area due to the coverage of the environment density catalogue (to be
described below).  We limit the redshift range to $0.005 < z < 0.2$
(abbreviated as $0 < z < 0.2$ or often loosely referred to as $z\sim
0$).  After matching all catalogues described below, the resulting
sample size is 368,781.  Matching all catalogues except the group
catalogue, which is limited to $M_r - 5\log h \lteq -19.5$, the sample
size is 459,174.  This second sample is used in \sec{SFvscolour}, but
the first sample is used in the main analysis
(\twosec{tutorial}{quenching}).

We obtained $ugriz$ photometry (``petro'' values) and redshifts
from the NYU-VAGC (DR7) catalogue
\citep{blanton05,adelman08,padmanabhan08} and used these data as input
to the K-correction utilities of \cite{blanton07} (v4\_2) to calculate
$V_{\rm max}$ from the $r$-band limit of the spectroscopic survey of
17.77 mag and \cite{bessell90} $U$ and $B$ rest-frame absolute
magnitudes (adjusted to $h=0.7$).  In the analyses to follow, we
weight each galaxy by its $1/V_{\rm max}$ multiplied by the inverse of
its spectroscopic completeness (also obtained from the NYU-VAGC).  All
quoted galaxy fractions and densities are weighted.

\subsubsection{SFR} 
\label{sdsssfr}

SFR estimates are an updated version of those derived in \cite{bri04}.
These new estimates are taken from the DR7 catalogue and are provided
online by J. Brinchmann et
al. (http://www.mpa-garching.mpg.de/SDSS/DR7/).  As in \cite{bri04},
these SFR estimates, calibrated to the \cite{kro01} IMF, measure SFR
within the fiber directly from $\Ha$ and $\Hb$ lines for star forming
galaxies.  For galaxies with weak lines or showing evidence for AGN, they
use the D4000 break calibrated to SFR using the star forming galaxies.
Outside the fiber, the method for estimating SFR is improved over
\cite{bri04}.  Following the method of \cite{sal07}, but without using
UV flux points, they fit stochastic models of stellar populations to
the observed photometry outside the fiber and constructed a
probability distribution function (PDF) of SFR estimates.  The means
of the PDFs were added to the SFRs in the fiber for a total estimate
of SFR.  J. Brinchmann (private communication) estimate the typical
error of these SFRs to be about 0.4 dex for star forming galaxies, 0.7
dex for more quiescent galaxies and growing to 1 dex or more for dead
galaxies (but for dead galaxies, these SFRs are more upper limits than
measurements).  Our own analysis comparing these SFR estimates with
those of \cite{sal07}, which are derived from SED fitting of UV and
optical light from GALEX, indicate that the error is closer to 0.2 dex
for star forming galaxies.

Our results are not changed significantly if we use the SFR estimates
from \cite{sal07} since our analysis depends only on a division
between star forming and passive galaxies (\sec{SFvscolour}) rather
than on an absolute SFR value.  However since the \cite{sal07} SFR
estimates cover only part of the DR4 (or DR7) area, we opted to use
the updated \cite{bri04} SFR estimates which increase our sample size
10-fold.

\subsubsection{The group catalogue and halo masses}
\label{groupcat}

\cite{yang07} constructed a group catalogue and calculated group halo
masses for the SDSS DR4 sample.  Yang et. al. also applied the same
algorithm to the DR7 sample, as described in \cite{yang12}, and we use
this catalogue throughout this study.  We briefly describe their group
finder and calculation of halo masses below.

The group finder consists of an iterative procedure that starts with
an initial guess of group centres and membership (the result of a
friends-of-friends algorithm).  For each of these tentative groups
they calculated the group size, mass and velocity dispersion and used
these to calculate a profile for the number density contrast of dark
matter particles based on a spherical NFW profile \citep{NFW97}.
Group membership was then recalculated based on the number density
contrast expected at the distance of the potential group member to the
group centre.  Then the centres, sizes, masses and velocity
dispersions were recalculated using the new membership, applying
completeness corrections to account for missing members.  
Then the process was repeated until there were no further changes to group
memberships.  They tested their group finder on an SDSS mock catalogue
found that their group finder successfully selected more than $90\%$
of true haloes more massive than $10^{12}\Msun$.

Once the group catalogue was constructed, \cite{yang12} assigned halo
masses using an abundance matching method.  They rank-ordered the
groups by group stellar mass and assigned halo masses according to the
halo mass function of \cite{warren06} with the transfer function of
\cite{eisenstein98}.  Comparing these halo masses with groups found in
a mock catalogue, they find a rms scatter of about 0.3 dex.  They also
calculate halo masses by rank-ordering group luminosities, but we
choose to use the rank-ordered group masses because stellar mass is a
better indicator of halo mass than luminosity \citep[see for
  example][]{more10}.

Using this group catalogue we defined the most massive member in each
group to be the central galaxy of the group and all other members to
be satellites.  Galaxies that were deemed by the group finder to be
isolated are also defined as centrals.

\subsubsection{Stellar masses}
\label{mssdss}
Stellar mass $\Ms$ is taken from the $\Ms$ catalogue provided online
by J. Brinchmann et al. (http://www.mpa-garching.mpg.de/SDSS/DR7/).
They derive $\Ms$, calibrated to the \cite{kro01} IMF, through SED
fitting similar to the method that \cite{sal07} used for estimating
SFR's, but without using UV constraints.  The method is also related
to that used by \cite{kau03a}, who fitted spectral features instead of
photometry.  These stellar mass estimates differ from \cite{kau03a} by
less than 0.1 dex for $\Ms \gtsima 10^{9}\Msun$ for more than
two-thirds of the galaxies.  The formal $1$-$\sigma$ errors derived
from the 95$\%$ confidence intervals of the probability distribution
are typically about 0.05 dex or less.

\subsubsection{Environment density}
\label{sdssenv}

The environment density measure $\log(1+\delta_N)$ is defined in
\cite{coo05}, but we briefly describe it here.  The number density
(per unit area) of a galaxy is calculated from the projected distance
\jwb{to the $N$th nearest neighbour (within a velocity window of
$\pm1500\kms$ to exclude foreground and background objects) and then
divided by the median surface density of the whole sample at the
galaxy's redshift to give an overdensity relative to the median
density $1+\delta_N$.  This median density is calculated in bins of
$\Delta z=0.02$.  Edge effects are minimised by removing all galaxies
within $1 h^{-1}$ comoving Mpc from the survey's edges.  For the SDSS
(DR6) \cite{cooper09} calculated $1+\delta_N$ for $N=5$ and these are
the density estimates used here.  Typical $1$-$\sigma$ uncertainties
for this density measure are about 0.5 dex.  This environment density
measure was computed for the galaxies of the SDSS DR6 photometric
catalog which is limited to objects brigher than $r=22.2$ in asinh
magnitudes \citep{lupton99}}.

\jwb{In this work we also define the relative distance $\Dist$ of
satellites to the centres of their haloes as a measure of environment
density.  Using the group catalogue in \sec{groupcat} to define
centrals and satellites, we calculate the projected distance $\dproj$
of each satellite to the central galaxy of its group and divide that
by the virial radius $\Rvir = 120 (\Mh/10^{11}\Msun)^{1/3}
{\rm kpc}$ (\eg, \citealp{dekbir06}).  Thus $\Dist \equiv \dproj/\Rvir$.}

\subsection{AEGIS}
\label{aegisdata}

\subsubsection{The sample}
\label{aegissample}

AEGIS is a collaboration between many survey teams using a variety of
space- and ground-based facilities to observe the Extended Groth Strip
(EGS) field over varied areas \citep[][Newman \etal, in
  preparation]{davis07}.  Our sample is limited to areas covered
simultaneously by {\it Spitzer} MIPS 24$\mu$m photometry
\citep{davis07}, CFHT $BRI$ photometry \citep{coil04}, DEEP2 and DEEP3
spectroscopy \citep{davis07,cooper11,cooper12}, Palomar $K$-band
photometry \citep{bundy06}, and GALEX UV photometry \citep{sal09}.
Furthermore, only galaxies with redshift quality 3 or 4 (which means
that at least two certain spectral features were identified by eye -
see \citealp{davis07}) were included this analysis.  Objects also had
to have a probability of being a galaxy of greater than 0.4
(unresolved galaxies are assigned a probability of 0-1 of being a
galaxy based on size, colour and magnitude - see \citealp{newman12}).
The sample size for these overlapping data sets for our redshift range
of interest ($0.75 < z < 1$) is 1846.

Using the $K$-correction utilities of \cite{blanton07} (v4\_2) we
calculated $V_{\rm max}$ from the $R$-band limit of the spectroscopic
survey of 24.1 mag.  We weight each AEGIS galaxy in this analysis by
$1/V_{\rm max}$ multiplied by the ``optimal'' weights calculated by
\cite{wil06} which account for colour-dependent redshift success rate
(see \citealp{wil06} for details).

\cite{wil06} also computed $U$ and $B$ absolute rest-frame magnitudes
using their own $K$-corrections, and these are the magnitudes used in
this analysis, adjusted for $h=0.7$.

\subsubsection{SFR}
\label{aegissfr}

Several star-formation tracers are available from the multi-wavelength
AEGIS field \citep{davis07}:

\begin{itemize}
\item IR light from reprocessed UV photons from young, massive stars
  that are dust-enshrouded.  The rest-frame 15$\mu$m luminosity of
  galaxies is tightly correlated with total IR luminosity \citep[][and
    references therein]{lef05,bel05} which can be used to estimate SFR
  \citep{ken98}.  The IR data are from the Spitzer MIPS 24$\mu$m
  photometry (which corresponds to rest-frame 12-15$\mu$m), which are
  used to derive the total IR luminosity using SED templates of
  \cite{cha01}.  \cite{lef05} find that the dispersion
  between IR luminosities estimated from several different SED
  libraries can reach up to 0.2 dex.
\item Non-extinction-corrected rest-frame UV luminosity from young,
  massive stars, derived from the observed $B$ flux.  Following
  \cite{bel05}, the rest-frame 2800\AA~ luminosity is used to estimate
  the total integrated 1216-3000\AA~ UV luminosity, which comes from
  $\sim 100$ Myr old populations.  The UV luminosity is in turn used
  to estimate SFR following \cite{ken98}.
\item Extinction-corrected UV luminosity derived from GALEX which
  observes in the near UV (NUV).  Combined with the CFHT $u^*$ band,
  one may calculate the UV slope $\beta$ between the rest-frame far UV
  (FUV) and NUV.  Each galaxy is corrected for extinction using the
  correlation between $\beta$ and FUV attenuation \citep{sei05}.
  (However, note that the UV slope may also depend on age and not only
  on dust - see \citealp{kong04,sal09}.  We do not attempt to correct
  for this.)  The FUV luminosity is then used to calculate SFR
  following \cite{ken98}.  This SFR derivation is described in more
  detail in \citep{schiminovich07}.
\item $\Ha$, $\Hb$, [\OII] emission lines from ionising photons from
  very massive, young stars \citep{wei07}.  For DEEP2 spectroscopy,
  the $\Ha$ line is visible out to $z \sim 0.37$ (outside our redshift
  range of interest, $0.75< z < 1$), the $\Hb$ line is visible at $0.35
  \ltsima z \ltsima 0.86$, and the [\OII] line is visible at $0.77
  \ltsima z \ltsima 1.42$.  Only those galaxies with S/N $>$ 2 are
  considered a detected emission line SFR.  (Note that line SFR's
  carry some limitations including the uncertainties of 
  estimating dust extinction for the Balmer lines and the dependence
  of the oxygen lines on metallicity; see \citealp{mostek12} for details.)
  We used the mass-dependent flux ratios $F_{\Hb}/F_{\Ha}$ and
  $F_{OII}/F_{\Ha}$ (for galaxies for which these lines observable)
  to estimate $\Ha$ from $\Hb$ and [\OII] at higher redshifts, and
  also to estimate the reddening correction \citep{wei07}.  Then the
  prescription of \cite{ken98} is applied to calculate SFR from $\Ha$.
\end{itemize}

\cite{sal09} and Noeske et al. (in preparation) find that the combined
SFR estimates from IR and non-extinction-corrected UV (IR+UV) yield
good agreement on average with extinction-corrected SFR estimates
from GALEX, and, where available, estimates from $\Ha$.  Emission
line indicators yield SFR estimates that also agree with the other
indicators except that they are systematically higher (0.2-0.4 dex).

With this wealth of SFR indicators, we created a fiducial sample
combining several indicators in the following way that we believe
takes advantage of the strengths of each indicator (see also
\citealp{noeske12}).  Since IR+non-extinction-corrected UV (hereafter
IR+UV for brevity) is growing in popularity as a reliable indicator of
star formation \citep[\eg, ][]{elb07,noe07,sal09}, we use IR+UV as our
first estimate of SFR, and where this is not available, we use the
extinction-corrected estimates from GALEX.  Where neither IR+UV nor
GALEX UV are available, we use the SFR estimates from emission lines.
In the redshift region where the ranges for detecting $\Hb$ and [\OII]
overlap, we select $\Hb$ as the preferred SFR indicator.  Since these
indicators produce SFR estimates that are offset from each other, we
normalise the other indicators to IR+UV so that their average offsets
are zero.  To calculate these offsets, we used a wide range in log SFR
of -1.2 to 3.  Overall, 27$\%$ of this mix is derived from IR+UV,
43$\%$ from GALEX UV, and 30$\%$ from emission lines.  We checked
whether restricting the SFR sample to any single method used in the
fiducial combination makes any difference.  Results were not affected,
and we do not discuss these other approaches any further.

\subsubsection{The group catalogue and halo masses}

\cite{gerke12} constructed a group catalogue for the AEGIS field using
a Voronoi-Delaunay method (VDM) group finder \citep[first implemented
  by ][]{marinoni03}.  This method makes use of a Voronoi partition of
the galaxy catalogue into unique polyhedrons each containing one
galaxy, and it also makes use of the Delaunay mesh which is a system
of line segments connecting neighbouring galaxies.  Starting with the
smallest Voronoi volumes, the group finder finds all the neighbouring
galaxies to a test galaxy via the Delaunay mesh within a certain
cylinder.  The size of the cylinder is determined iteratively as a
function of the number of group members, and its free parameters are
optimised to produce the best success in a mock AEGIS catalogue
\jwr{(described in detail in \cite{gerke12})}.  Success is measured by
completeness and purity defined as follows.  One can check the mock
catalogue to see if a group matches the halo that contains a plurality
of the group's members and if a plurality of the halo's members are
part of the group.  Such a situation is considered a two-way match.
Completeness is the fraction of haloes that have a two-way match to a
group found by the VDM group finder, and purity is the fraction of VDM
groups that have a two-way match to a halo in the mock catalogue.  The
AEGIS group catalogue has a completeness of $\sim 74\%$ and a purity
of $\sim 67\%$.


Using this group catalogue, we defined the most massive member in each
group to be the central galaxy of the group and all other members to
be satellites.  We also defined as centrals those galaxies that were
deemed by the group finder to be isolated.  To test the effectiveness
of massiveness as an identifier of central and satellite galaxies, we
identified the 
most massive members of the groups found by the group finder on 40
AEGIS mock catalogues.  When compared to
the mocks' underlying dark matter distribution (from the N-body
simulation), we find that
massiveness correctly identifies a central or satellite on average
87$\pm 1\%$ of the time, the $\pm 1\%$ being the standard deviation
around this mean over the 40 mocks.  \jwb{The fraction of centrals
identified in this way that are actually satellites is $12\pm 1\%$ and the
fraction of satellites that are actually centrals is $29\pm 1\%$. }

In order to estimate host halo masses, we assumed that each group
roughly represents a distinct halo and estimated the halo masses of
each group using an ``abundance matching'' method similar to that used
by \cite{yang07} (described in \sec{sdssdata}).  For each group, we
calculated the 
total
stellar mass $M_{\rm *,tot}$ for all group members above $\Ms =
10^{9.5}\Msun$ in each group.  Each group member was weighted by the
``optimal'' weights computed by \cite{wil06} to account for the
spectroscopic completeness of the AEGIS survey.  Then we
self-calibrated the groups to account for missing members following
\cite{yang07}.  Specifically, we selected groups found in AEGIS
between $0.2 < z < 0.4$ and calculated the fraction of the group
mass that is contributed by members above the stellar mass limits of
(10,10.1,10.2) corresponding to the limits at the redshifts
(0.7,0.9,1.1) \citep{bundy06}.  We then fit an exponential function to
these fractions as a function of group mass for each redshift bin, and
these exponentials became the redshift- and mass-dependent fractions
by which we divided each group mass (see \citealp{yang07} for further
details).

We then assigned a halo mass to each group by rank ordering 
$M_{\rm *,tot}$ and matching them with the predicted
distribution of halo masses in redshift bins of 0.1.  The halo mass
function is that of \cite{tinker08} using the mass transfer function
of \cite{eisenstein98}, computed using a code kindly provided by
M. Cacciato.  This method assumes that the halo mass function and the
luminosity function of groups match one-to-one.  This is a rather
simplistic assumption, but we are interested mainly in average trends
with halo mass, and we expect that the relative masses between groups
of very different numbers of members to be quite robust.  

Using the
same algorithm on the groups found by the group finder on the AEGIS
mock catalogues, we find that the halo masses agree well with those of
the mocks with a mean rms scatter of 
0.37 dex for the rank-ordered stellar mass.

\jwb{After matching all catalogs listed in \sec{aegissample}, our sample
includes 274 groups with more than one member, and 1361 isolated galaxies.}

\subsubsection{Stellar masses}

The stellar masses $\Ms$ are derived in \cite{bundy06}.  These authors
use $BRIK$ colours, as well as the $J$-band from Palomar when
available in the EGS, and spectroscopic redshifts to fit observed SEDs to
synthetic SEDs from the stellar population models of \cite{bru03}.
These model SEDs span a large range of star formation histories, ages,
metallicities and dust content.  From the fits, they construct a PDF
for stellar mass-to-light ratio $\Ms/L_K$ for each galaxy and use the
median of the PDF as the best estimate mass-to-light ratio to
calculate $\Ms$.  Uncertainties in the model fitting combined with the
uncertainties of the observed $K$-band luminosity yield typical final
$1$-$\sigma$ uncertainties of 0.2-0.3 dex.

In order to increase the statistical power of the AEGIS sample, we
supplemented these stellar masses with those calculated using the
K-correction utilities of \cite{blanton07} (v4\_2) whenever $K$-band
stellar masses were not available.  Since the K-correction stellar
masses were on average smaller than the $K$-band masses by 0.05 dex we
added this value to all the K-correction masses used in the sample.
Supplementing the stellar masses in this manner adds 425 galaxies to
the $0.75 < z < 1$ sample (for a total of 1846).

\subsubsection{Environment density}
\label{aegisenv}

\cite{coo08} calculated $\log(1+\delta_N)$ for the EGS field using
$N=3$ and a \jwb{velocity window of $\pm 1000\kms$}, and these are the
density estimates used here.  Log$(1+\delta_N)$ is described above in
\sec{sdssdata}.  Using the mock catalogues of \cite{yan04},
\cite{coo05} compared the effectiveness and accuracy of several
measures of environment density and found that the projected distance
to the third nearest neighbour provided the most accurate measure of
environment density for the DEEP2 survey.  Edge effects are minimised
by removing all galaxies within $1 h^{-1}$ comoving Mpc from the
survey's edges.  Although the $N$ we use is different for the SDSS and
for AEGIS, the number densities used to calculate $\log(1+\delta_N)$
are higher in the SDSS than in AEGIS, and so the distances probed are
comparable.  The typical distance to the third nearest neighbour in
AEGIS is about 1.5 $h^{-1}$ comoving Mpc while the typical distance to
the fifth nearest neighbour in SDSS is about 1.7 $h^{-1}$ comoving
Mpc.  Typical $1$-$\sigma$ uncertainties for this density measure
are about 0.5 dex.

\begin{figure}
\epsscale{1}
\plotone{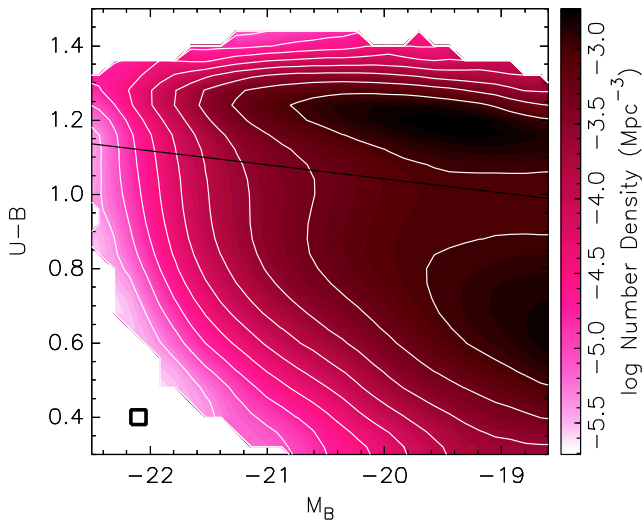}
\caption{\small The colour-magnitude diagram of the SDSS galaxies ($0
  < z < 0.2$).  The solid line divides the colour bimodality and is
  defined as $U-B = -0.032*(M_B+21.62)+1.175$ \citep{coo08}.  The size
  of a pixel used to smooth the diagram is indicated at bottom left.
}
\label{cmdsdss}
\end{figure}

\begin{figure*}
\epsscale{2}
\plotone{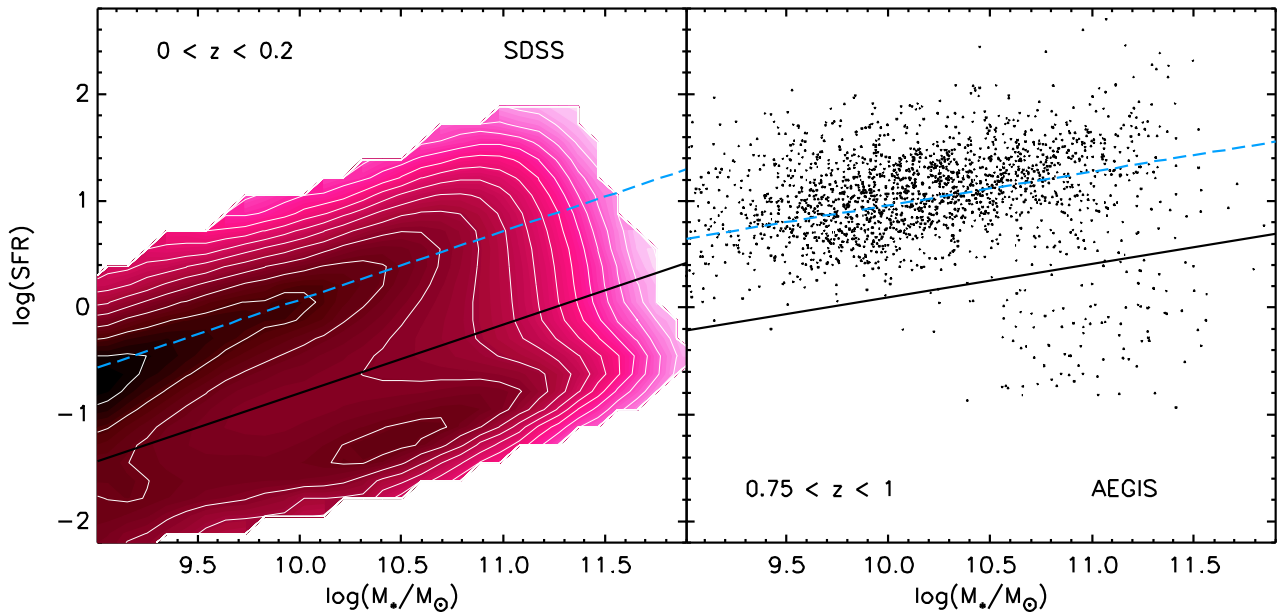}
\caption{\small Log SFR vs. log $\Ms$ for galaxies in two redshift
  bins.  The dashed blue line is a fit to the SF sequence.  The solid black
  line shows the division between the ``SF sequence'' of galaxies
  (above the line) and the ``low-SFR'' or passive galaxies.  The text describes
  how these lines are determined.  The weighted ratio of low-SFR galaxies
  to the total number of galaxies is the ``quenched fraction''.}
\label{fourmstwo}
\end{figure*}

\begin{figure}
\epsscale{1}
\plotone{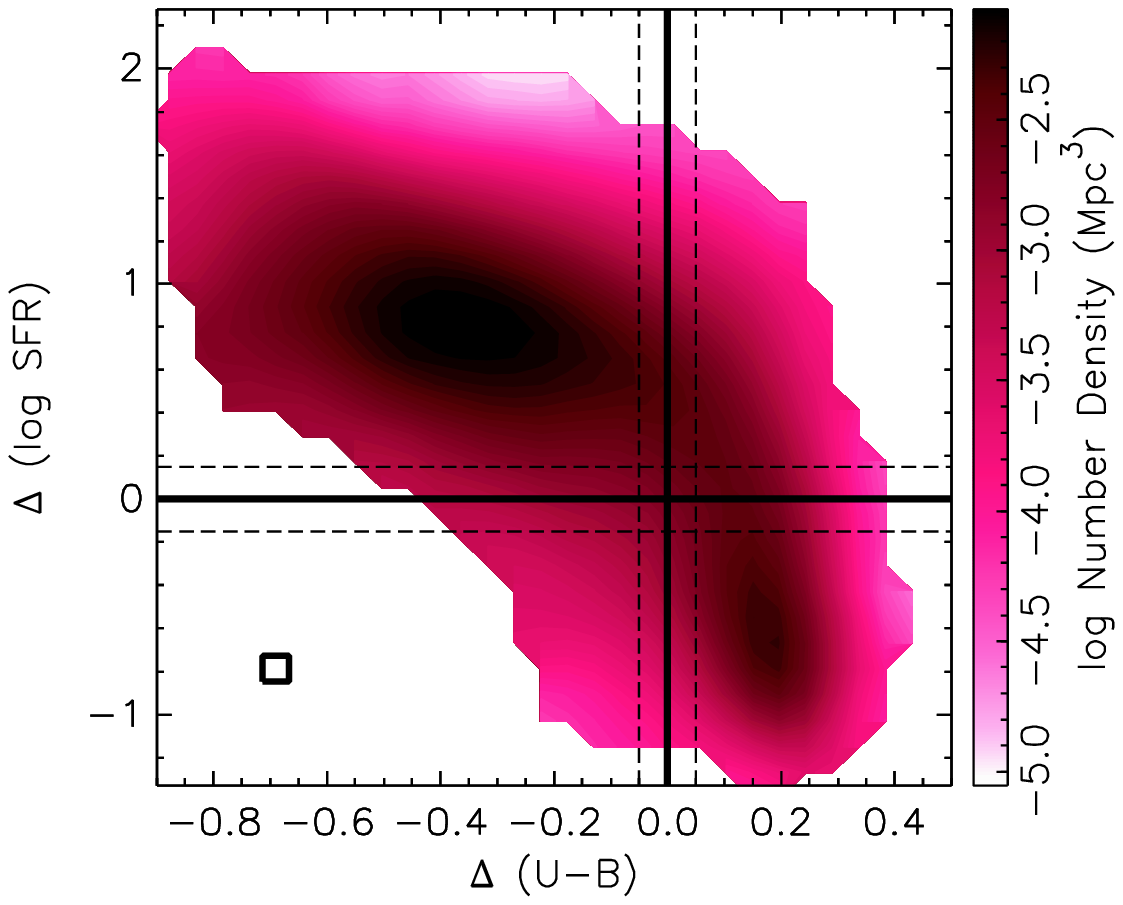}
\caption{\small \jwb{The residuals of the SFR division (solid line in
  \fig{fourmstwo}$a$) plotted against the residuals of the rest-frame
  $U-B$ colour division (solid line in \fig{cmdsdss}) for the SDSS ($0
  < z < 0.2$).  The colour bar on the right shows the number density
  in a pixel smoothed over adjacent pixels.  The size of a pixel is
  indicated in the bottom left.  The galaxies with intermediate SFR
  lie roughly between the two horizontal dashed lines, with the
  central horizontal solid line marking the SFR division.  The vertical
  solid line marks the colour division, while the dashed lines roughly
  delineate the green valley.}  The SFR bimodality mostly corresponds
  to the colour bimodality but 30$\%$ of red sequence galaxies lie on
  the SF sequence while only 4$\%$ of blue cloud galaxies lie below
  the SF sequence.  Also, being in the green valley is not synonymous
  with transitioning from star formation to quiescence.}
\label{colourvssfrresid}
\end{figure}

\section{Bimodality in SFR versus Color}
\label{SFvscolour}

The galaxy bimodality is usually discussed in terms of the bimodality
in the colour-magnitude diagram.  \jwb{\fig{cmdsdss} shows the
colour-magnitude diagram of the SDSS sample with 
the line dividing the colour bimodality used in \cite{coo08}:
\be 
U-B = -0.032*(M_B+21.62)+1.175.
\label{gvline}
\ee 
The red sequence, which lies above this line,} is assumed to
consist mostly of galaxies that have ceased to form stars while the
blue cloud (below the line) consists of currently star-forming
galaxies.  Such a division is convenient because photometry of large
samples is easy to do with relatively good accuracy.  However,
techniques have improved for determining star formation rates for
large samples, and we can now divide the galaxies by star formation
directly rather than by colour as a proxy.

We show the ``physical'' version (\ie, reflecting the physical
processes of star formation) of the colour-magnitude diagram, namely
the SFR-$\Ms$ diagram for the SDSS and the AEGIS in \fig{fourmstwo}.
The star-forming galaxies lie on a sequence as described by
\cite{noe07}, which we call the ``SF sequence,'' while there is a
population of galaxies lying under it, which we interchangeably call
``low-SFR'', ``passive'' or ``quenched'' galaxies.  \jwb{We defined the
division between these populations in the following way.  We assigned
an initial line to divide the population by eye.  Then we performed a
weighted linear least-squares fit to the points above that line and
calculated the distance of each point from the line.  After a Gaussian
fit to these distances, we then exclude from the SF sequence those
points that lie more than $2.5\sigma$ below the line, and re-perform
the fit.  The process is repeated until no more points are excluded.
Quenched galaxies are galaxies that do not lie in the SF sequence as
defined in this way.  We show the resulting divisions in
\fig{fourmstwo} with the equation for the dividing line (solid black)
being: \be \log~{\rm SFR} = a\log \Ms - b
\label{sfrdiv}
\ee 
where $a$ = [0.64,0.31] and $b$ = [-7.22,-3.04] for the redshift bins
[$0<z<0.2$, $0.75<z<1$].}

When we refer to ratios of the low-SFR galaxies to the total number of
galaxies, we call this the ``quenched fraction'' rather than the ``red
fraction'' to remind the reader that we divide the galaxies by SFR and
not by colour.

In order to compare the SFR bimodality with the colour bimodality, we
\jwb{show in \fig{colourvssfrresid} the residuals from the SFR division
$\Delta$ log SFR = log SFR - \Eq{sfrdiv} versus the residuals from the
colour division $\Delta(U-B) = (U-B) -$ \eq{gvline} for the SDSS.  The
horizontal solid line in \fig{colourvssfrresid} corresponds to the
solid black line in \fig{fourmstwo}$a$.  As explained above, we call
everything above this line a star forming galaxy, or SF sequence
galaxy, while everything below it is quenched.  Similarly, the
vertical solid line in \fig{colourvssfrresid} corresponds to the line
in \fig{cmdsdss} that divides the red sequence (right of the vertical
solid line in \fig{colourvssfrresid}) from the blue cloud (left of the
vertical solid line).  The two horizontal dashed lines are $\pm0.15$
around the SFR division and are meant to guide the eye to the area of
intermediate SFR.  The vertical dashed lines are $\pm0.05$ around the
colour division and mark the area of the green valley.}

\jwb{From \fig{colourvssfrresid}, we observe that the SFR bimodality mostly
corresponds to the colour bimodality, but there are some notable
differences.  In particular, green valley galaxies are
not necessarily intermediate in SFR.  Rather about $80\%$ of green
valley galaxies are still star forming (lying above the horizonal
solid line).}  Furthermore, about $30\%$ of red sequence galaxies lie
in the SF sequence (these are star-forming and are probably reddened
by dust - see \citealp{brammer09} and also \citealp{lotz08} who find a
similar result for $z\sim 1$) whereas only $4\%$ of the blue cloud lie
below the SF sequence (these may be post-starburst galaxies, those
that exhibit no signs of ongoing star formation but experienced recent
star formation within the last $\sim$Gyr - see for example,
\citealp{dressler83,quintero04}).  Therefore dividing the galaxies by
$U-B$ colour does not accurately divide the population by star
formation mainly because dusty star formers appear red.  We recognise
that dust may not explain fully explain the large population of star
forming red galaxies since $U-B$ and SFR contain large scatter and are
constructed differently.  However, these findings are consistent with
the results of \cite{maller09}, who find that about a third of red
sequence galaxies in a volume-limited sample of the SDSS at $0.02 < z
< 0.22$ actually lie in the blue cloud after applying an inclination
correction to their $M_g-M_r$ colours, implying a dusty origin to the
observed red colours of these galaxies.

These dusty star formers are also the more massive members of the SFR
sequence - the dusty star formers are on average 0.4 dex more massive
than the entire SF sequence population.  This is understandable as the
more massive members of the SF sequence are more metal-rich
\citep{tre04} and therefore have proportionately more dust.  This is
consistent with the results of \cite{maller09} who find that the
inclination-dependent colour correction also depends on absolute $K$
magnitude in that brighter galaxies have higher attenuation than
dimmer galaxies of the same inclination angle.  The fact that massive
star-forming galaxies are dusty will cause {\it more massive galaxies
  to appear quenched if one defines quenching only by colour}.  This
will be important in comparing our results with those of others who
define quenching by colour only.  We discuss this further in
\sec{comppeng}.

\section{Relations between Masses and Environment}
\label{tutorial}

In this section, we describe the complex relations between stellar
mass $\Ms$, halo mass $\Mh$, environment density $\delta_N$ and
satellite group-centric distance $\Dist$ for central and satellite
galaxies in order to prepare the reader for a discussion of quenching
as a function of these quantities in the following section.
Summarising this section in brief, $\Mh$ correlates with $\Ms$ for
centrals at low masses, but shows a large spread at high $\Ms$,
offering a chance to separate quenching effects between $\Ms$ and
$\Mh$ for centrals.  $\delta_N$ for centrals is a simple two-mode
function of $\Mh$ while for satellites $\delta_N$ is given by a
complicated interplay between $\Mh$ and the location within a halo.
The dividing line of interpretation of $\delta_N$ is where the number
of observed group members is less than $N$ (the cross-halo mode) or
greater than $N$ (the single-halo mode).  Due to the complex nature of
$\delta_N$, the better measure of local ``environment'' for satellites
may be radial distance from the group centre which is highly
anti-correlated with $\delta_N$, but only in the single-halo mode.
Readers interested mainly in the results of quenching as a function of
$\Ms$, $\Mh$ and $\delta_N$/radial distance may skip to
\sec{quenching}.

Note that in all the figures that follow for the SDSS analysis, one
quantity is plotted as contours as a function of the $x$- and
$y$-axes.  These contours are computed in the following way: values of
this third quantity are computed in pixels of the indicated size in
the plots as long as the pixel contains more than $m$ galaxies
(``good'' pixels).  $m$ is equal to $2\%$ of the average number of
galaxies in each pixel which contains at least one galaxy.  The values
are then smoothed over twice by computing the mean over adjacent good
pixels.  The pixel sizes are the same in all plots except where noted:
0.2 dex for both the mass quantites ($\Ms$ and $\Mh$), 0.3 dex for
$\Dfive$ and 0.15 dex for $\log(d_{\rm prog}/R_{\rm Vir})$.  Our pixel
sizes and smoothing scheme were chosen to produce acceptably smooth
contour lines in order to bring out the main trends.  Secondary
features should not be taken too seriously.

\subsection{$\Ms$ and $\Mh$ for centrals}

We begin with a discussion of the relation between $\Ms$ and $\Mh$.
\jwb{For central galaxies, it has been established with high confidence
(\eg,
\citealp{cacciato09,more09,mandelbaum09,leauthaud11,leauthaud12}) that
the relation with the halo mass is somewhat steeper at low mass than
at high mass.  This is a manifestation of the fact that the halo mass
function and the galaxy mass function are considerably different at
the low and high mass ends.  The result is that $\Ms$ should correlate
closely with $\Mh$ up to around $M_{\rm crit}$; above this $\Ms$
begins to level off while $\Mh$ increases \citep[\eg,][their Figure
  2]{cattaneo07}.  So $\Ms$ is a measure of parent $\Mh$ for $\Ms$
below a few $\times 10^{10}\Msun$, and above that $\Ms$ serves as a
lower limit indicator of $\Mh$.}

\begin{figure*}
\epsscale{2.2}
\plotthree{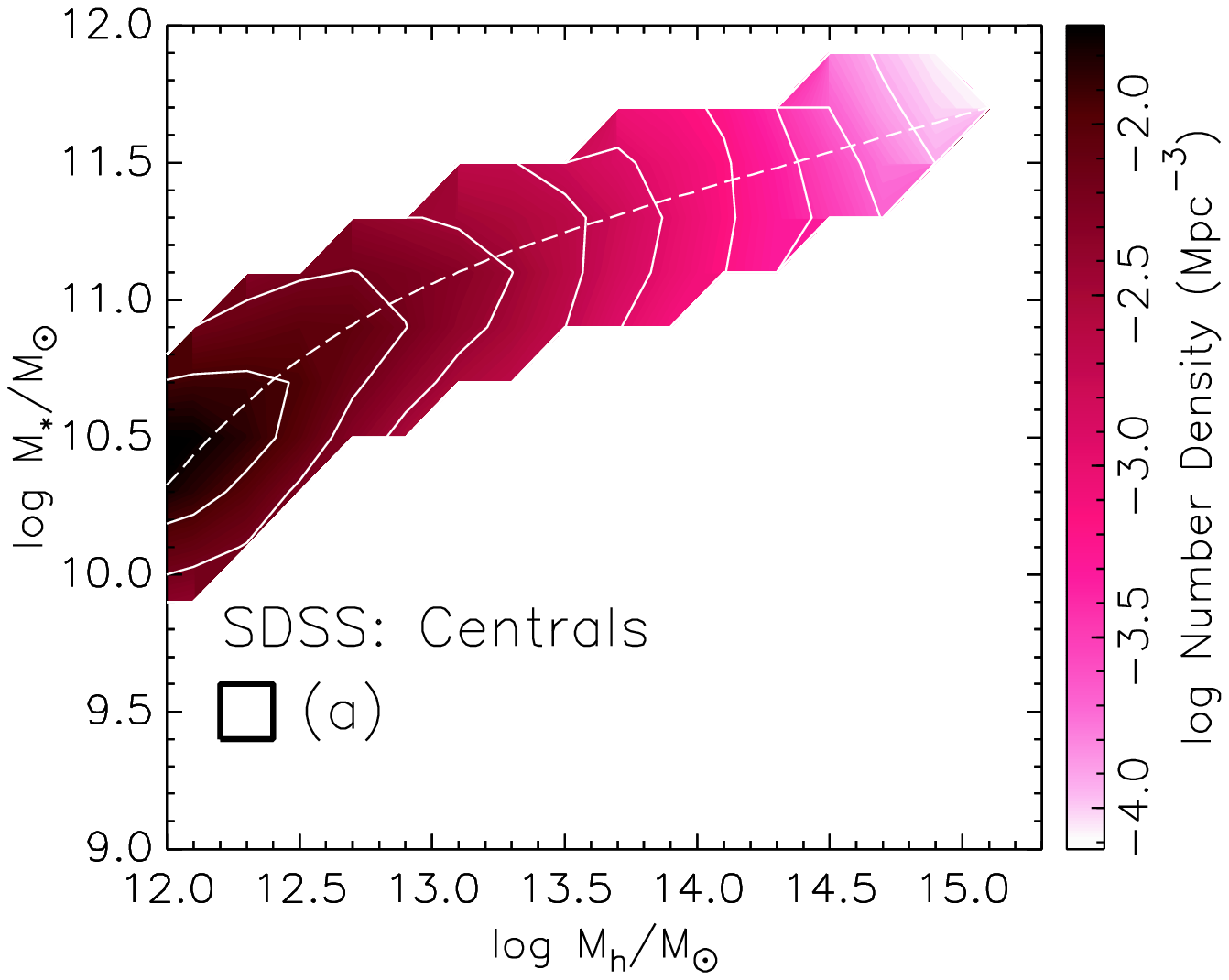}{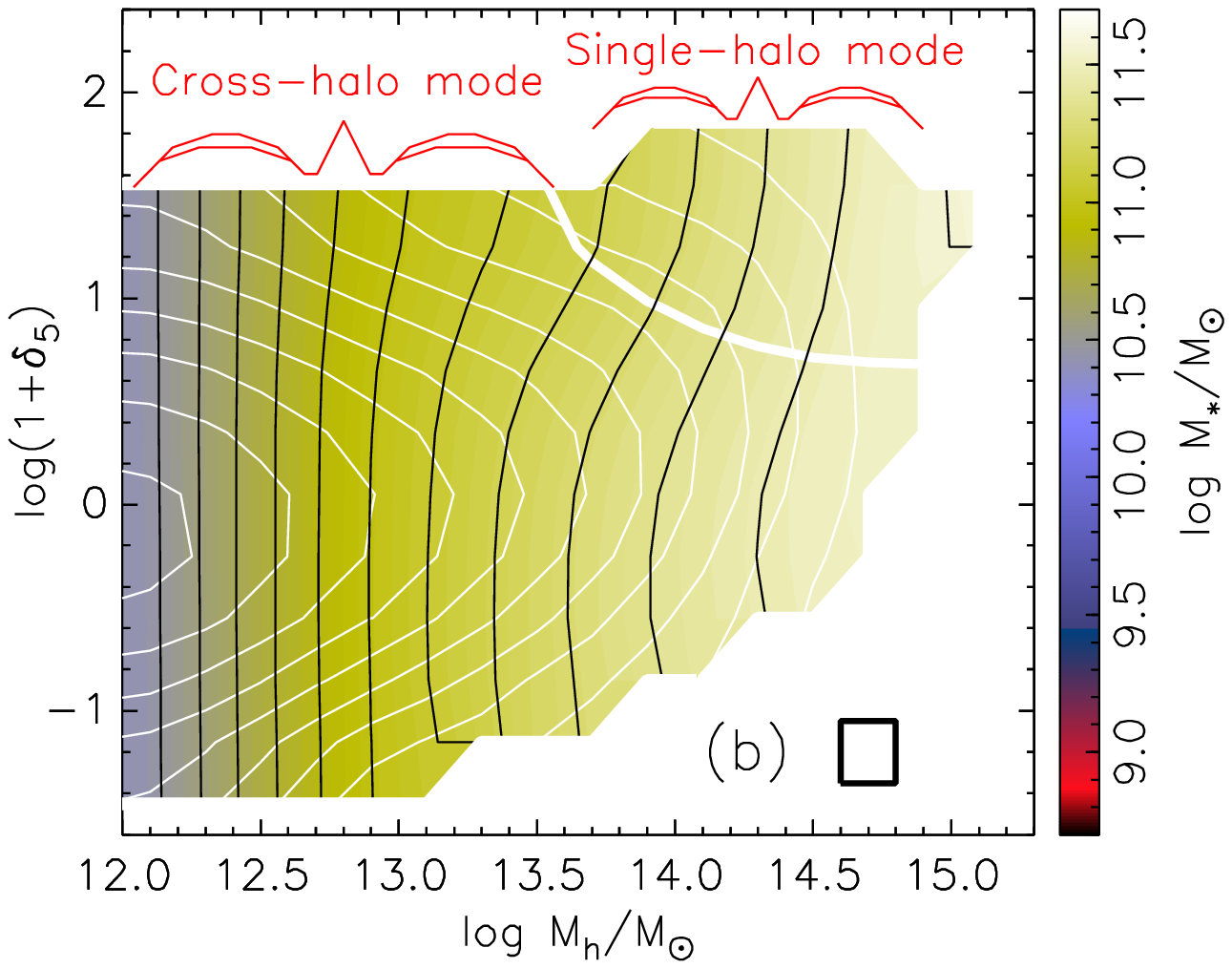}{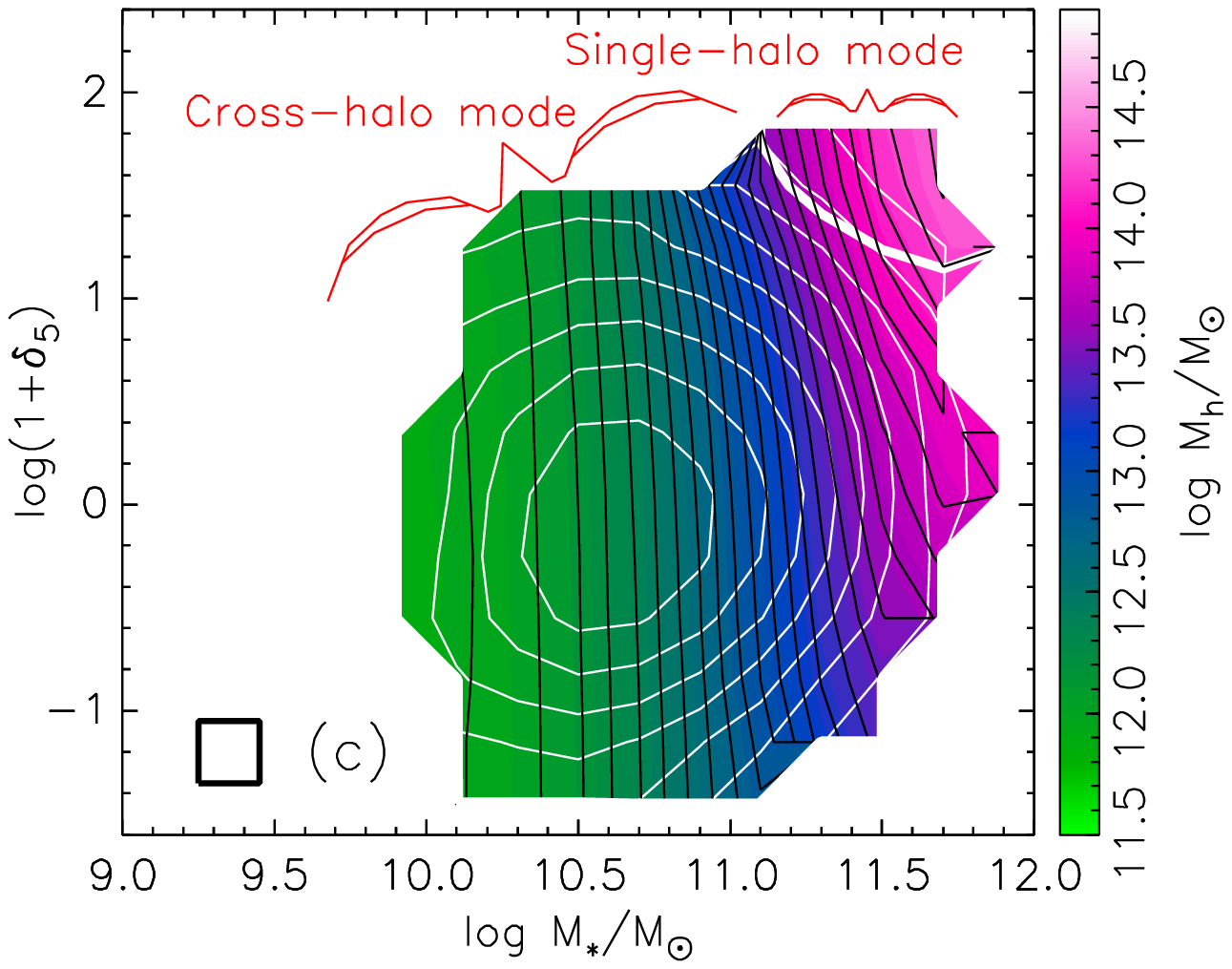}
\caption{\small (a). Stellar mass $\Ms$ versus halo mass $\Mh$ for
  SDSS central galaxies ($0 < z < 0.2$).  The dashed curve is a fit to
  Eq. 7 of Yang et al. (2008).  (b) Mean stellar mass $\Ms$ as a
  function of density $\Dfive$ and halo mass $\Mh$.  (c). Mean halo
  mass $\Mh$ as a function of density $\Dfive$ and stellar mass $\Ms$.
  The thin white contours in each panel mark the weighted number
  density of points and are separated by 0.25 dex in density.  The
  thick white lines in (b) and (c) mark where the median membership of
  a galaxy's group is six.  Membership is greater above this line.
  The black contours in (b) and (c) follow the coloured shading and
  are separated by 0.1 dex.  The number densities in (a) and the means
  in (b) and (c) were computed in pixels of the displayed size and
  smoothed over adjacent pixels.  These panels show that: (1) $\Ms$
  correlates with $\Mh$ for low masses, but massive centrals occupy a
  large range of halo masses; (2) the single-halo mode is
  characterised by an increased $\delta_5$ {\it and} $\Mh$ while there
  is no correlation between $\delta_5$ and $\Mh$ in the cross-halo
  mode; (3) for low $\Ms$, $\Mh$ increases with $\Ms$, and for high
  $\Ms$, $\Mh$ also increases with $\delta_5$.}
\label{tutcen}
\end{figure*}

In \fig{tutcen}$a$ we show the relation between $\Ms$ and
$\Mh$ for SDSS central galaxies.  \cite{yang08} find that galaxies
above $\Mh \sim 10^{12.5}$ follow a very weak scaling of $\Ms \propto
\Mh^{0.22}$ while below this mass, $\Ms$ scales more steeply with
$\Mh$, scaling as $\Ms \propto \Mh^{1.83}$ for galaxies well below
$\Mh \sim 10^{12.5}\Msun$ \citep{yang08}.  We reproduce this result in
\fig{tutcen}$a$ (dashed curve) by fitting their Eq. 7 and obtaining
very similar parameters.  This flattening of the $\Ms$-$\Mh$ relation
at high masses reflects the fact that massive centrals occupy groups
with a large range of group membership and group masses (since $\Mh$
corresponds to group $\Ms$), and presents an excellent opportunity to
separate the effects of quenching between $\Ms$ and $\Mh$ as we will
see in \sec{quenching}.

\begin{figure*}
\epsscale{2.2}
\plotthree{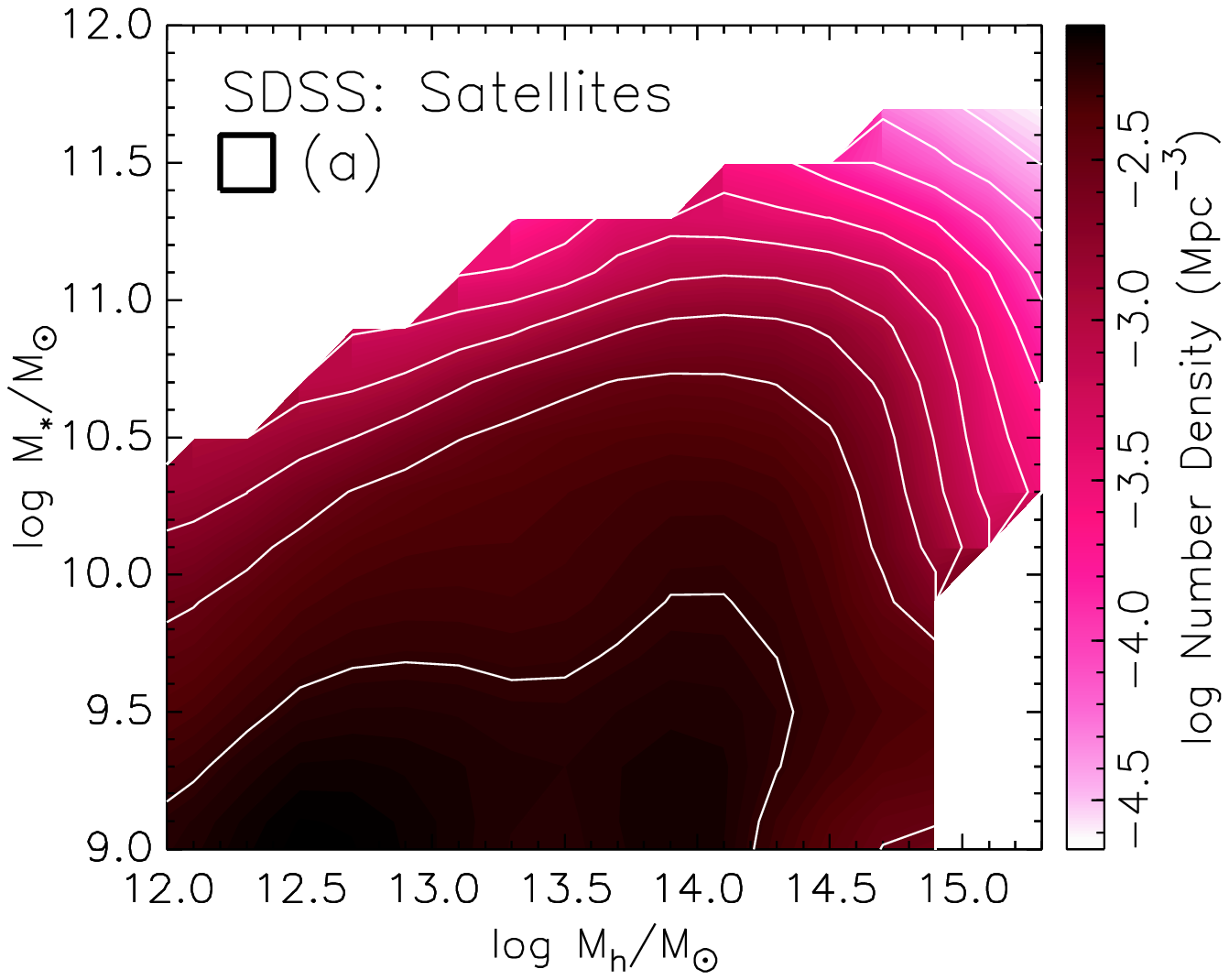}{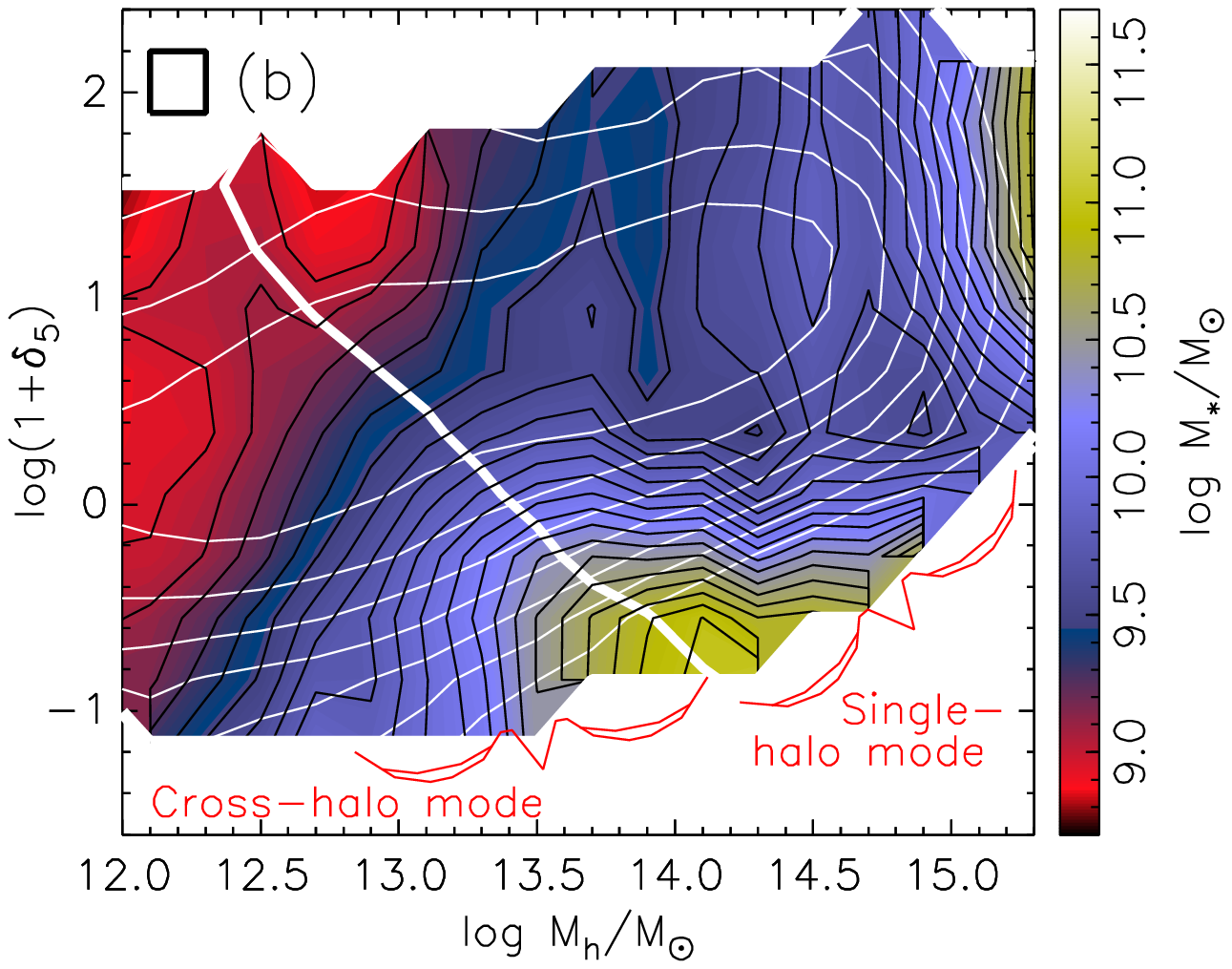}{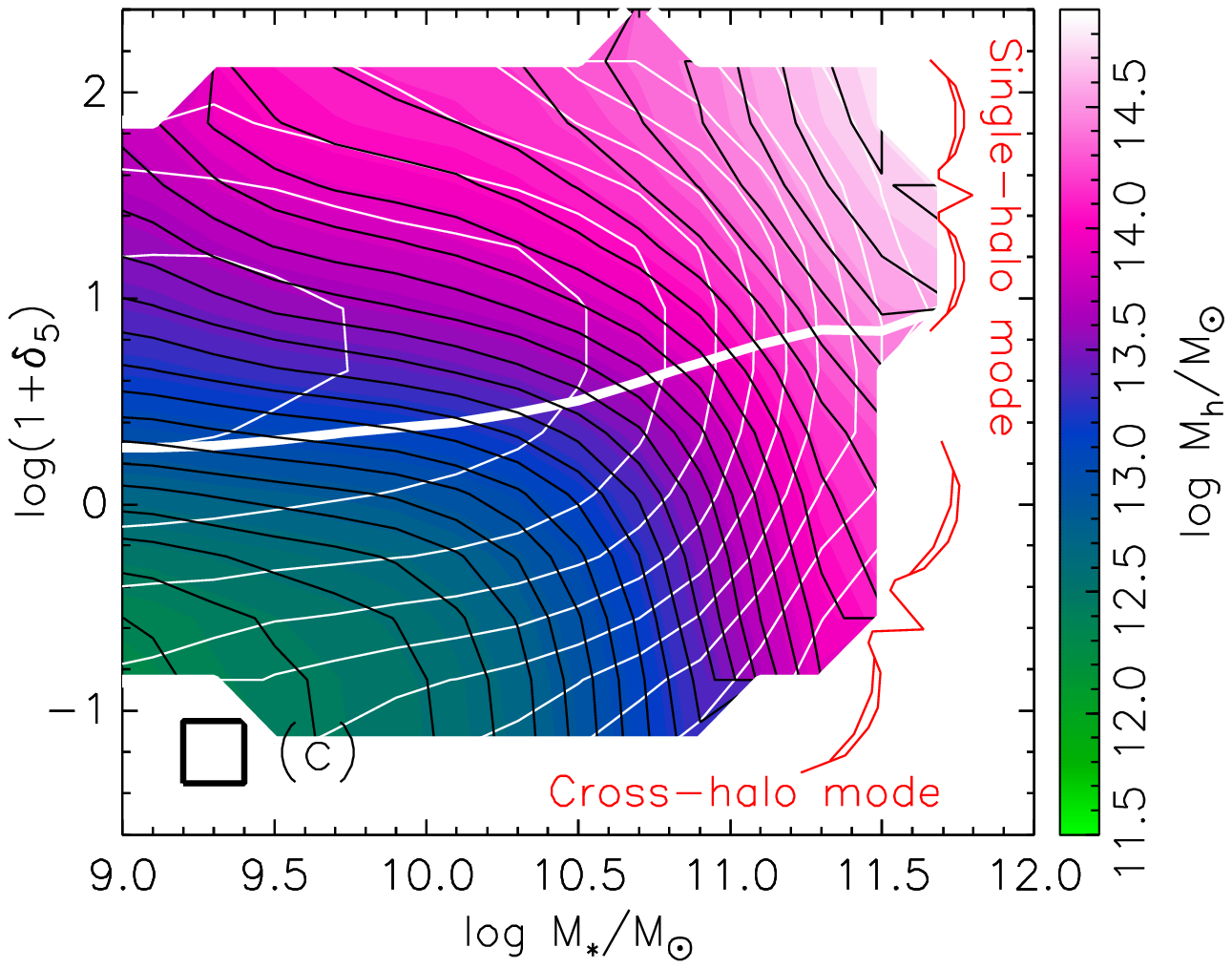}
\caption{\small (a). Stellar mass $\Ms$ versus halo mass $\Mh$ for
  SDSS satellites ($0 < z < 0.2$).  (b) Mean stellar mass $\Ms$ as a
  function of density $\Dfive$ and halo mass $\Mh$.  (c). Mean halo
  mass $\Mh$ as a function of density $\Dfive$ and stellar mass $\Ms$.
  The thin white contours in each panel mark the weighted number
  density of points and are separated by 0.25 dex in density.  The
  thick white line in (b) and (c) marks where the median membership of
  a galaxy's group is six.  Membership is greater above this line.
  The black contours in (b) and (c) follow the coloured shading and
  are separated by 0.1 dex.  The number densities in (a) and the means
  in (b) and (c) were computed in pixels of the displayed size and
  smoothed over adjacent pixels.  These panels show that: (1) there is
  very little correlation between $\Ms$ and $\Mh$ for satellites
  overall; (2) however $\Ms$ correlates with $\Mh$ in the cross-halo
  mode; and (3) in the single-halo mode $\delta_5$ correlates with
  $\Mh$ as expected in a virialised halo with a halo-occupation
  function that scales with $\Mh$. }
\label{tutsat}
\end{figure*}

\subsection{$\delta_N$ and mass for centrals}
\label{tutorialcen}

As described in \sec{introrel}, environment density $\delta_N$ for
centrals has two modes: \jwr{the single halo mode measuring field
  density and the cross-halo mode measuring
halo mass}.
Figs. \ref{tutcen}$b$ and $c$ show the distribution of SDSS centrals
in the $\delta_5$-$\Mh$ and $\delta_5$-$\Ms$ planes.  The thick white
curve in the top right corner of panels $b$ and $c$ marks where the
median number of group members is six.  (More \etal, in preparation,
give a careful treatment of why lines of constant $N_{\rm members}$
should run more or less parallel to the thick white lines in
Figs. \ref{tutcen}$b$ and \ref{tutsat}$b$.)  Above this curve, the
centrals are mostly in the single-halo mode of $\delta_5$ while below
the line, $\delta_5$ measures distances to the next halo.  The main
distribution of centrals in \fig{tutcen}$b$ (marked by the thin white
contours) shows very little correlation between $\delta_5$ and $\Mh$
except in the upper right corner where single-halo mode centrals are
characterized by the highest $\delta_5$ and $\Mh$.  This region of
high $\Mh$, high $\delta_5$ corresponds nicely to the increase of
$\Mh$ with $\delta_5$ in the upper right corner of \fig{tutcen}$c$
(coloured shading).  \jwb{We also see a deviation from the strong
$\Ms$-$\Mh$ relation in these region of high $\delta_5$.  This simply
reflects the fact that higher densities in a halo roughly means more
satellites, and the summed mass of the satellites will begin to
matter, compared to the mass of the central, in the determination of
the halo mass.}

\subsection{$\Ms$, $\Mh$ and $\delta_N$ for satellites}
\label{tutorialsat}

\fig{tutsat}$a$ shows the $\Ms$-$\Mh$ relation for satellites.  Unlike
for centrals, there is very little correlation between $\Ms$ and
$\Mh$.  For small haloes, $\Ms$ correlates somewhat with $\Mh$ in that
low-mass haloes do not host massive satellites.  But the least massive
satellites occupy host haloes of all masses.

Environment density is more complex for satellites than for centrals.
More than half (58\%) of SDSS satellites reside in groups of six or
more members.  \fig{tutsat}$b$ and $c$ show the distribution of
satellites in the $\delta_5$-$\Mh$ and $\delta_5$-$\Ms$ planes, again
with a thick white line marking where the median number of group
members is six.  Above the line, the satellites reside in groups of
more than six members (the single-halo mode).

In the cross-halo mode of $\delta_N$, satellites mostly reside in
small haloes ($\Mh \ltsima 10^{13}\Msun$ - see \fig{tutsat}$b$) with
only a few other galaxies.  In the cross-halo regions of
\fig{tutsat}$b$ and $c$ (under the thick white line), $\Mh$ roughly
correlates with $\Ms$ (see the coloured shading in these panels).
There are at least two reasons why $\Ms$ for satellites is expected to
correlate with $\Mh$ in this regime.  1) We expect a massive satellite
to live in a massive halo, or else it would be the central.  Less
massive satellites on the other hand are more likely to be merging
into a low mass halo, since there are many more of these than massive
haloes.  2) The timescale for dynamical friction is inversely
proportional to the ratio of the satellite mass to the halo mass
\citep{bin87}.  So very large satellites relative to the halo will
have disappeared into the middle of the halo in a very short time.
Therefore, the largest remaining satellite must be some fraction of
the halo mass, but smaller than the ratio of the central mass to the
halo mass.

In the single-halo mode, $\delta_N$ correlates with $\Mh$ for
satellites (see the white density contours of \fig{tutsat}$b$ and the
coloured shading of \fig{tutsat}$c$ above the thick white line) just
as for centrals again because of the halo profile.  Since the halo
profile falls off more slowly in a more massive halo, a
satellite in a more massive halo will most likely be found near
the centre of its halo than if it were in a smaller halo, and its
$N$th nearest neighbour would more likely be nearby.  For a simpler
way to understand the $\delta_N$-$\Mh$ relation for satellites in the
single halo mode, assume a uniform distribution of satellites
throughout the halo (which is wrong).  Recalling that the number of
satellites in a halo is $N_{\rm sat} \prop \Mh^{\alpha}$ where
$\alpha$ is close to 1 (see \citealp{yang08}), and recalling that in
the spherical collapse model a virialised halo follows $\Mh \prop
R^3$, then $1+\delta_N \prop N/R^2 \prop \Mh^{\alpha-2/3} \sim
\Mh^{1/3}$.  Therefore environment density is a measure of the halo
mass following something like $\Dfive \prop \frac{1}{3}\log\Mh$ which
is close to the scaling of the SDSS satellites (\fig{tutsat}$b$).

\begin{figure}
\epsscale{1}
\plotone{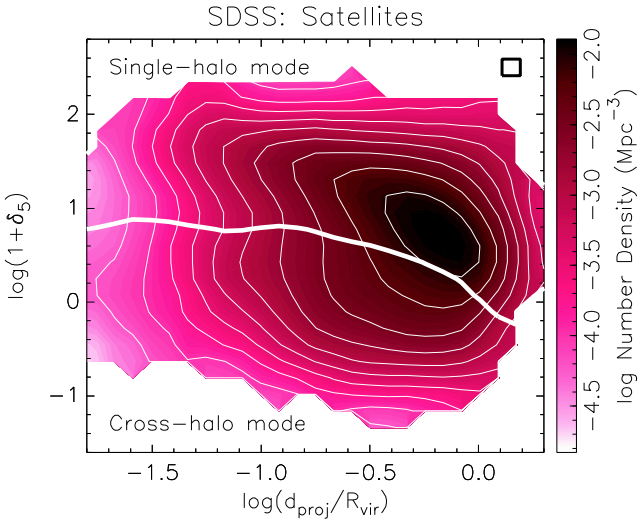}
\caption{\small The environment density of SDSS satellites ($0 < z <
  0.2$) versus their projected distance to their group centres
  relative to their virial radii ($d_{\rm proj}/R_{\rm vir}$) .  The
  thick white line marks the dividing line between groups with six or
  more members (above the line) and fewer than six members (below the
  line).  Single-halo mode is above this line, cross-halo mode is
  below.  Note how the valley in the contour lines tends to follow
  this line.  In the single-halo mode, proximity to the group centre
  is clearly associated with an increase in $\delta_5$, while in
  cross-halo mode, there is no relationship.  This is expected based
  on the statistics of these relationships.}
\label{satdist}
\end{figure}

In the single-halo mode for satellites, we also expect $\delta_N$ to
be a measure of radial position within a halo.  \fig{satdist} shows
$\delta_5$ versus the projected distance $d_{\rm proj}$ of SDSS
satellites to their group centres divided by their group virial radii
[$R_{\rm vir} = 1.2\times10^5(\Mh/10^{11}\Msun)^{1/3} {\rm pc}$
  \citep{dekbir06}].  As mentioned in \sec{sdssenv}, we define $\Dist
\equiv \dproj/\Rvir$.  In the single-halo mode (above the thick white
line which marks where the median group membership is six), higher
density indeed means closer proximity to the group centre.  In the
cross-halo mode, the two quantities do not seem to be related, which
is what we expect given that $\delta_5$ measures distances to external
haloes.  (Note that in \fig{satdist} the pixel size is smaller than
the usual 0.3 dex for $\delta_5$ and 0.15 dex for $\log\Dist$ in order
to highlight the valley in the contours that separates the single-halo
mode from the cross-halo mode.)

A further complication to this already complex nature of $\delta_N$ is
that whether or not a galaxy is in the single-halo or cross-halo mode
of $\delta_N$ depends not only on the number of members in its group
but also on the limiting magnitude of the survey, which is
$z$-dependent.  A group of the same mass at $z=0.005$, for example,
which appears to be in the single-halo mode will appear to be in the
cross-halo mode at $z=0.2$ because some of its members will fall
beyond the magnitude limit.  In other words, within the SDSS redshift
range of $0.005 < z < 0.2$, the typical distance to the 5th nearest
neighbour ranges from 0.34 $h^{-1}$ to 4.0 $h^{-1}$ comoving Mpc (for
$0.005 < z < 0.05$ and $0.15 < z < 0.2$ respectively).  \jwb{In contrast,
$\Dist$ does not behave differently in two modes that depend on the
number of fellow group members.  (Recall that $\Rvir$ depends on $\Mh$
which has been corrected for missing members).}

\jwb{Additionally, although both $\delta_N$ and $\Dist$ suffer from
projection effects, we expect these effects to be worse for $\delta_N$
than for $\Dist$.  The \cite{yang07} group finder includes or rejects
group members based on a $z$-window that is dependent on halo
properties (in an iterative way), but $\delta_N$ is computed for all
galaxies within a {\it fixed} $z$-window that is comparable only to
the velocity dispersions of the most massive clusters.  Since the
majority of galaxies are not members of such massive clusters, and
since the majority of satellites in the group catalog are not
interlopers \citep{yang07}, we expect $\delta_N$ to suffer from
projection more than $\Dist$.}

Given these difficulties in interpretating $\delta_N$ for satellites,
we prefer to use $\Dist$ as a measure of local ``environment''.  But
we have included the popular $\delta_N$ variable in our study of
quenching in \sec{quenching} in order to compare with previous work.

\begin{figure*}
\epsscale{2.2}
\plottwo{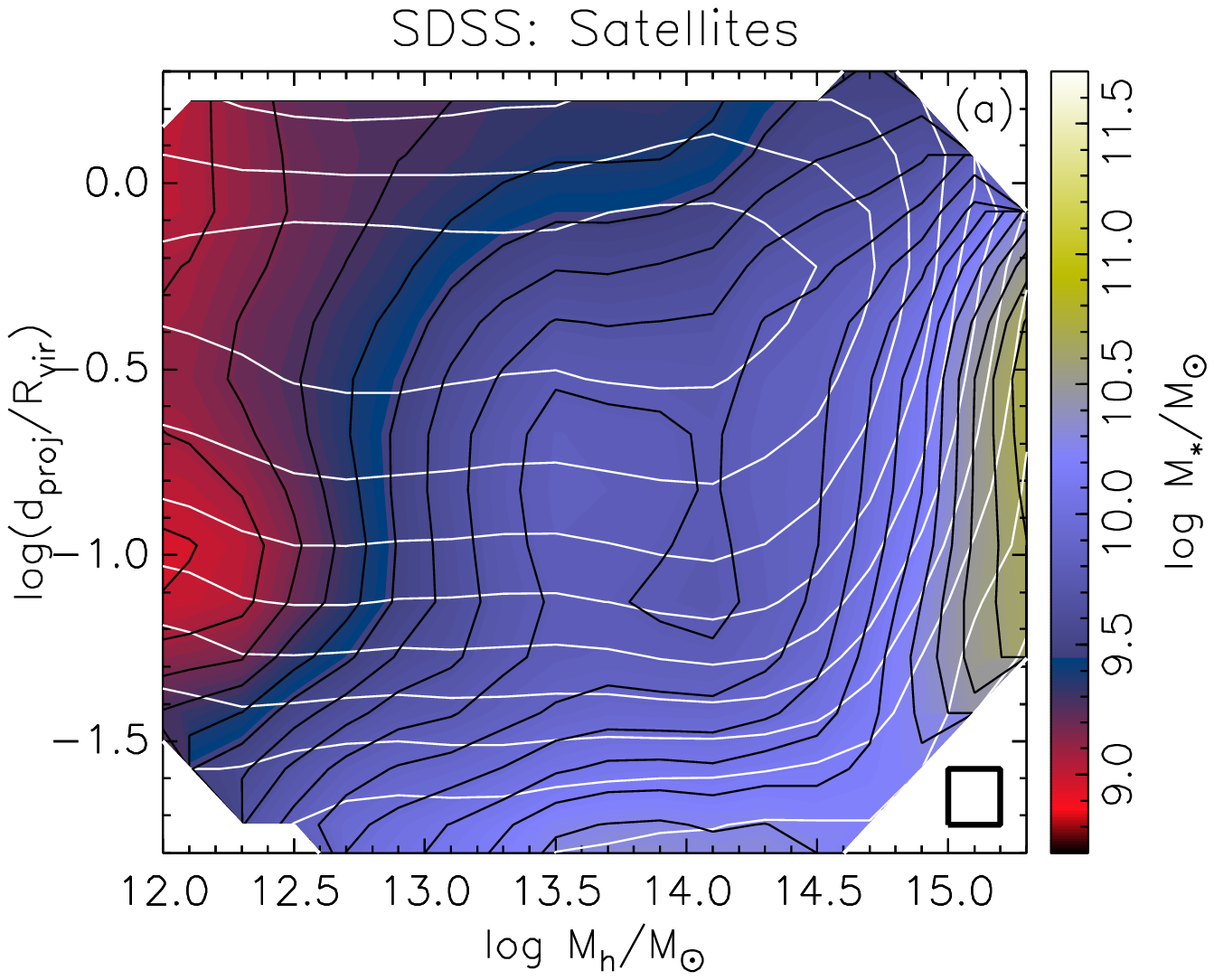}{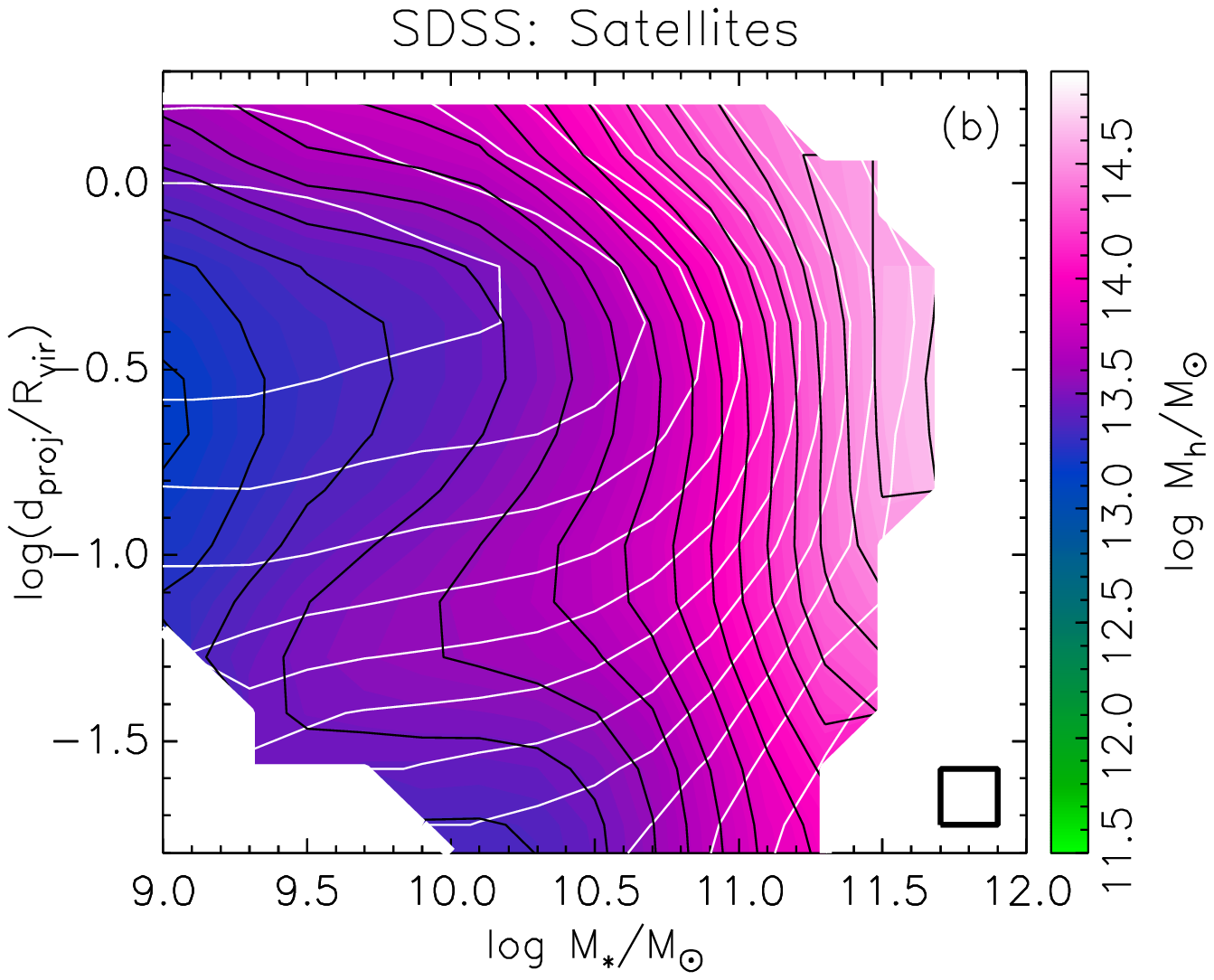}
\caption{\small \jwb{(a) Mean stellar mass $\Ms$ as a function of relative
  group distance $\Dist=\dproj/\Rvir$ and halo mass $\Mh$.  (b). Mean
  halo mass $\Mh$ as a function of relative group distance $\Dist$ and
  stellar mass $\Ms$.  The thin white contours in each panel mark the
  weighted number density of points and are separated by 0.25 dex in
  density.  The black contours in follow the coloured shading and are
  separated by 0.1 dex.  The means were computed in pixels of the
  displayed size and smoothed over adjacent pixels.  These panels show
  that: (1) the $\Dist$-$\Mh$ plane does not exhibit dual-mode
  behaviour as in \fig{tutsat}$b$; (2) the dynamical range of mean
  $\Mh$ in the $\Dist$-$\Ms$ plane is diminished compared to the
  $\delta_5$-$\Ms$ plane.}}
\label{tutsatd}
\end{figure*}

\jwb{For completeness, we replicate \fig{tutsat}$b$ and $c$ in
\fig{tutsatd}$a$ and $b$, except that we put $\Dist$ on the $y$-axis
instead of $\Dfive$.  In panel $a$, we see that the dual-modes of
$\delta$ that we saw in \fig{tutsat}$b$ disappear as expected.  For
example, the two regions of high $\Ms$ in \fig{tutsat}$b$ become one
region in \fig{tutsatd}$a$.  }

\jwb{In \fig{tutsatd}$b$, the dynamical range of $\Mh$ diminishes compared
to \fig{tutsat}$c$.  This is because low-mass haloes will never host
very many satellites (nor very massive satellites) and $\delta_5$ for
these satellites will almost always be in the cross-halo mode; 90\% of
satellites in haloes of masses $10^{12-12.7}\Msun$ live in haloes with
fewer than six members, so their $\delta_5$ is forced to be low.  This
is why the lower left corner of \fig{tutsat}$c$ is coloured deep green
(low $\Mh$).  But these same satellites have no such restriction on
$\Dist$.  Some of them could reside very near their group centres even
if their $\delta_5$ is low.  Furthermore, if their masses are low,
they could reside in haloes of all masses.  For these two reasons,
there are no regions in \fig{tutsatd}$b$ with very low average $\Mh$.}

\section{Main Results: Quenching and Halo Environment}
\label{quenching}

\begin{figure}
\epsscale{0.9}
\plotthreecol{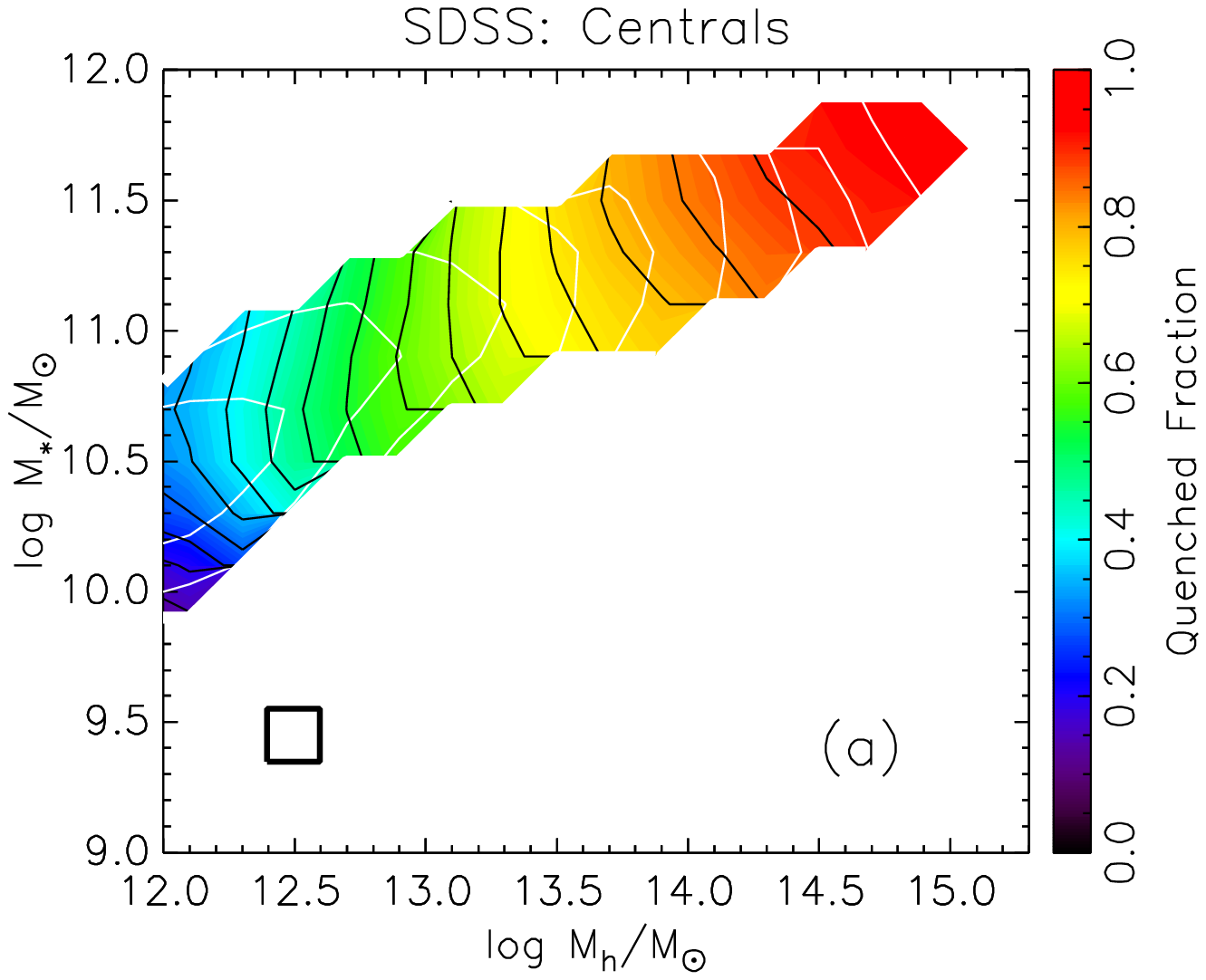}{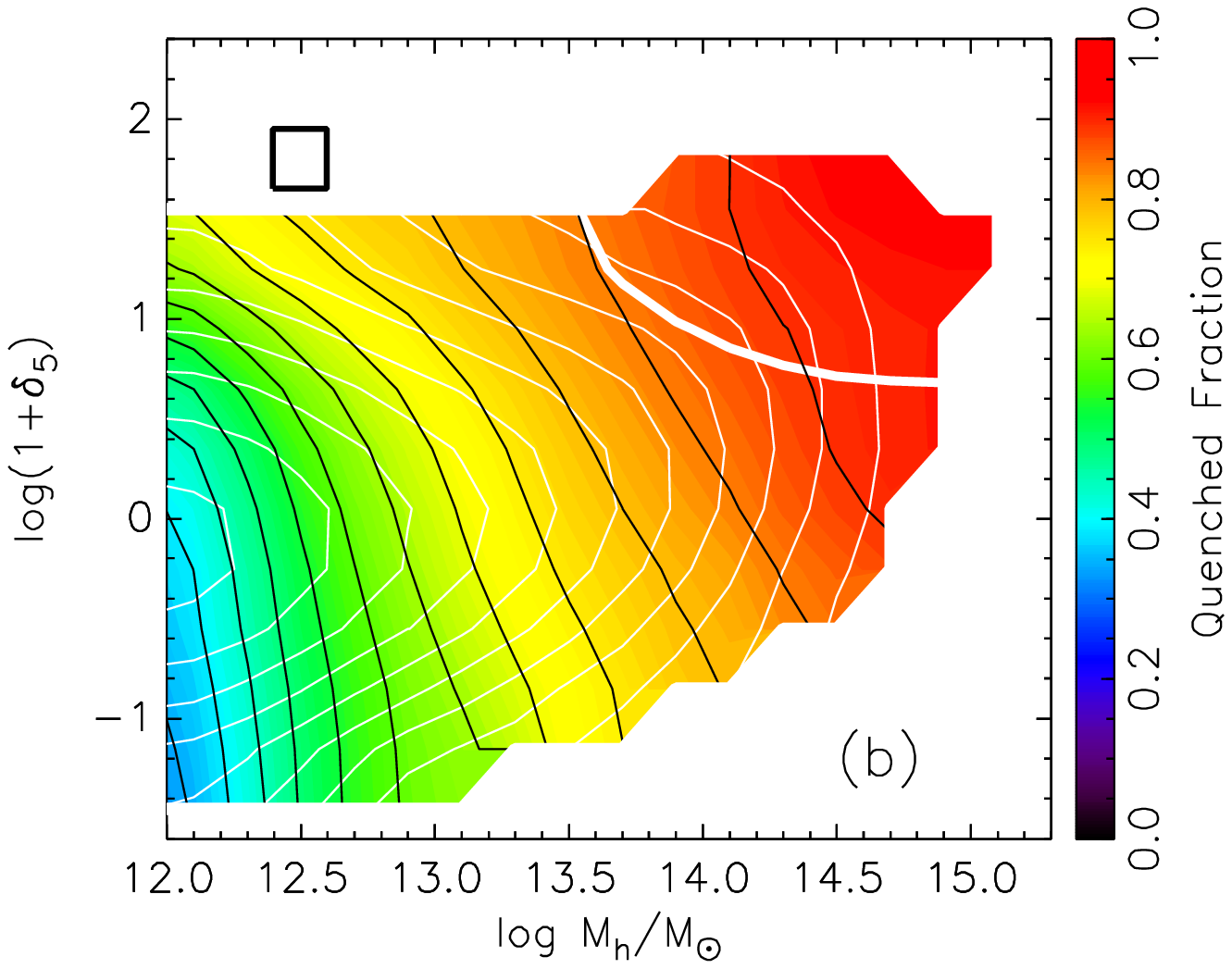}{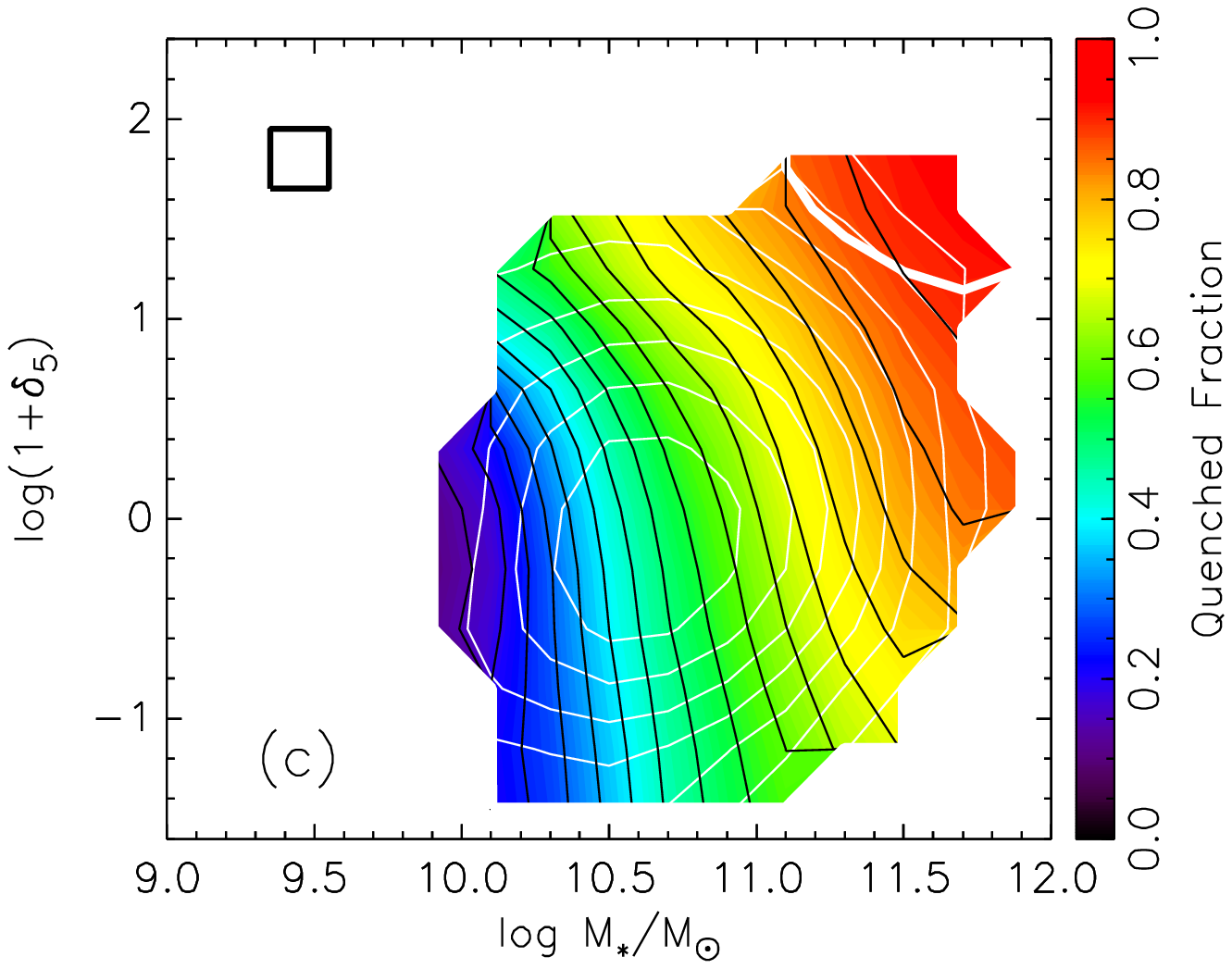}
\caption{\small The mean quenched fraction as a function of $\Ms$ and
  $\Mh$ (a), of $\delta_5$ and $\Mh$ (b), and of $\delta_5$ and $\Ms$
  (c) for central galaxies in the SDSS ($0 < z < 0.2$).  The thin
  white contours in each panel mark the weighted number density of
  points and are separated by 0.25 dex in density.  The thick white
  line marks where the median membership of a galaxy's group is six.
  Membership is greater above this line.  The black contours follow
  the coloured shading and are separated by 0.05.  The means were
  computed in pixels of the displayed size and smoothed over adjacent
  pixels.  To first approximation, the quenched fraction for centrals
  correlates with $\Mh$ more strongly than with $\Ms$ or $\delta_5$ as
  shown by the vertical black contours in panel $a$ and $b$.
  \jwb{Quenching in $c$ roughly follows mean $\Mh$ (see \fig{tutcen}$c$).}
}
\label{qfrac_cen}
\end{figure}

\begin{figure}
\epsscale{0.9}
\plotthreecol{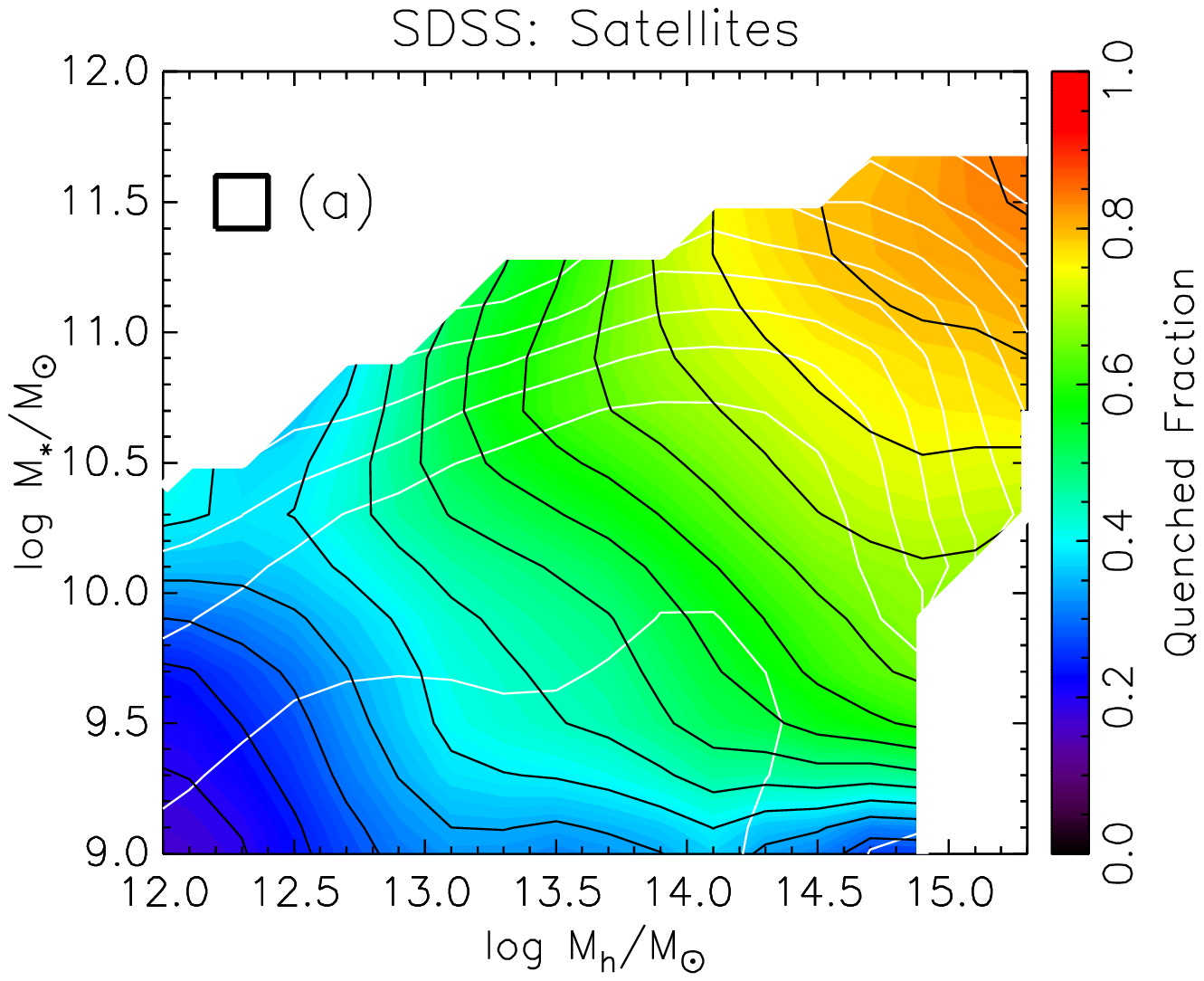}{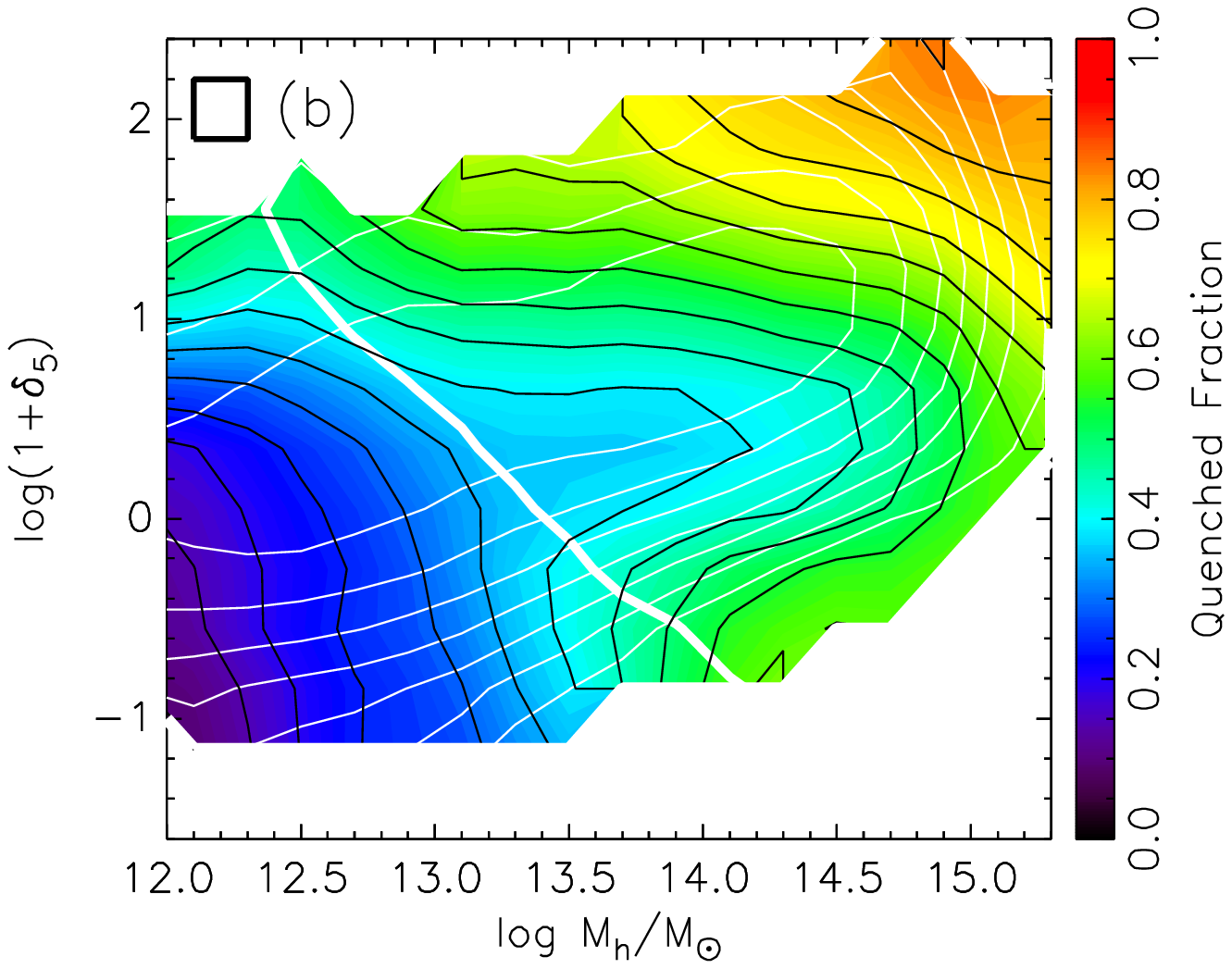}{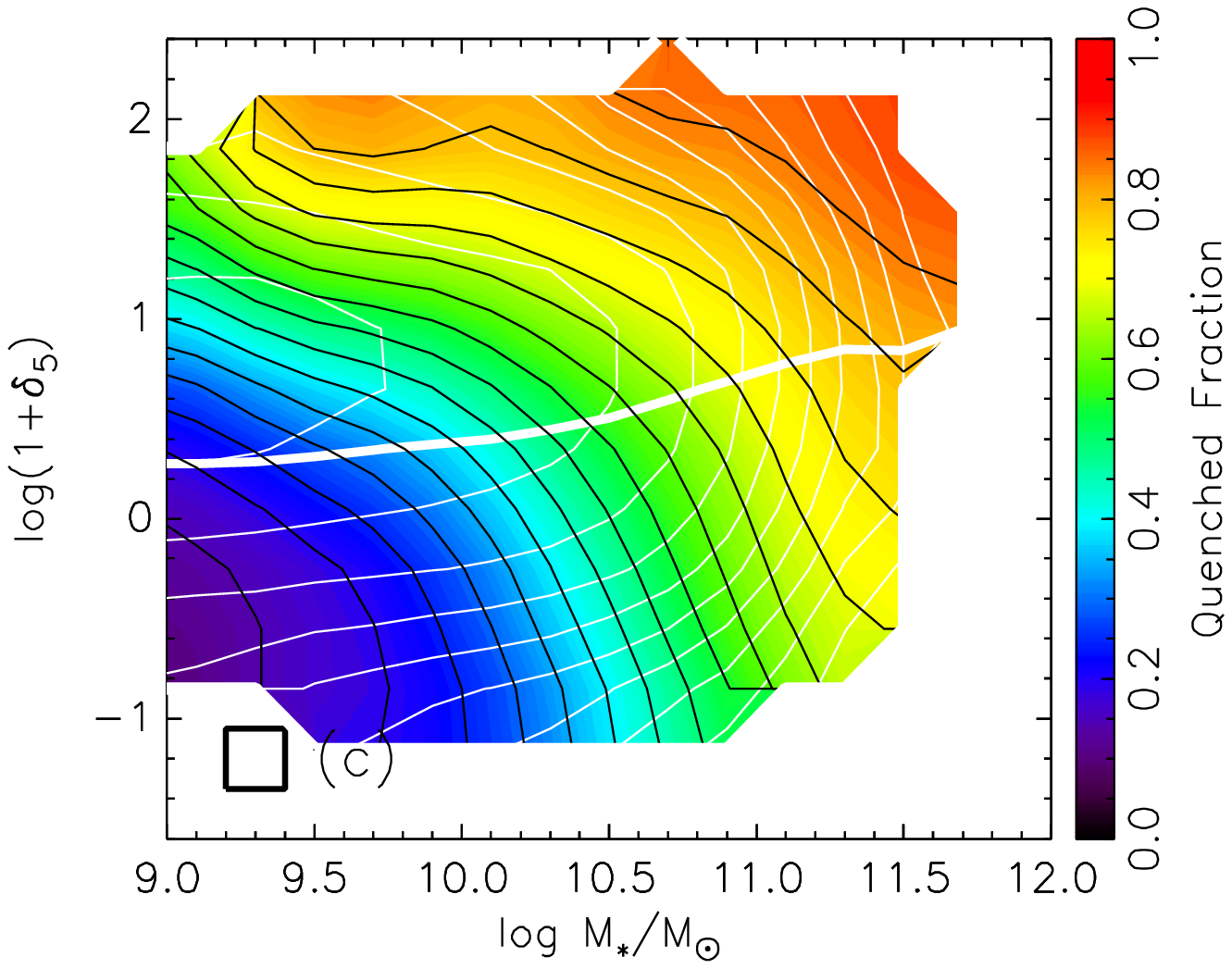}
\caption{\small The mean quenched fraction as a function of $\Ms$ and
  $\Mh$ (a), of $\delta_5$ and $\Mh$ (b), and of $\delta_5$ and $\Ms$ (c)
  for satellite galaxies in the SDSS ($0 < z < 0.2$).  The thin white
  contours mark the weighted number density of points and are
  separated by 0.25 dex in density.  The thick white line marks where
  the median membership of a galaxy's group is six.  Membership is
  greater above the thick line.  The black contours follow the
  coloured shading and are separated by 0.05.  The means were computed
  in pixels of the displayed size and smoothed over adjacent pixels.
  Quenching for satellites correlates with all three quantities $\Mh$,
  $\Ms$ and $\delta_5$.  None of these panels contains regions of very
  high mean quenched fraction, indicating that, in contrast to
  centrals, the combination of $\Ms$, $\Mh$ and $\delta_5$ is not
  optimum for expressing the quenched fraction of satellites.
}
\label{qfrac_sat}
\end{figure}

In the previous section we explored the relations between stellar
mass, halo mass and environment density for satellite and central
galaxies in the SDSS.  We showed that for central galaxies, $\Ms$
correlates with $\Mh$ for low masses, but massive centrals of a given
stellar mass occupy a large variety of halo masses.  We also showed
that the single-halo mode of $\delta_5$ for centrals is characterised
by increasing $\Mh$ with increasing $\delta_5$ while in the cross-halo
mode, there is little correlation between $\delta_5$ and $\Mh$.
\jwg{For satellite galaxies, there is a weak correlation between $\Ms$
  and $\Mh$ in the cross-halo mode only. } In the single-halo mode,
$\delta_5$ for satellites correlates with $\Mh$ as expected for
virialised haloes with a halo-occupation function that scales with
$\Mh$.  Since we now know how these quantities relate to each other,
we are in a position to explore quenching as a function of these
quantities.

\subsection{Quenching for centrals}
\label{quenching_cen}

\fig{qfrac_cen} repeats the panels of \fig{tutcen} ($\Ms$ and $\Mh$,
$\delta_5$ and $\Mh$, and $\delta_5$ and $\Ms$) but now shows the
fraction of quenched galaxies as the third parameter represented by
the colour gradient.  \fig{qfrac_cen}$a$ is the key result of this
paper.  \jwb{It shows that the quenched fraction appears to increase more
strongly with $\Mh$ at fixed $\Ms$ than with $\Ms$ at fixed $\Mh$ for
central galaxies.}
The contour lines of the quenching to first order follow $\Mh$, not
$\Ms$, \ie, they are roughly vertical above $\Ms \sim 10^{10.4}\Msun$.
The quenched fraction increases by less than $\sim 0.1$ over 0.8 dex,
or 40\% of the range in $\Ms$ (at constant $\Mh$), but increases by
almost 0.35 over 40\% of the $\Mh$ range (at constant $\Ms$).
\jwg{This result seems to prefer halo-quenching models over mechanisms
  that are more closely tied to $\Ms$ such as the QSO mode of AGN
  feedback.  Note that the halo environment heated by a virial shock
  is more likely to host the radio mode of AGN \citep{dekbir06}
  because the coupling of the AGN energy to the halo gas is more
  efficient when the gas is hot.}

\jwb{To second order, \fig{qfrac_cen}$a$ shows a small dependence of
  quenching on $\Ms$ at constant $\Mh$ for massive haloes.  This can
  be explained in the following way.  After a virial shock forms in a
  massive halo, residual cold gas may still form stars inside the
  galaxy after the shock has been triggered.  An $\Ms$ dependence
  could then naturally arise if at fixed $\Mh$, a central with higher
  stellar mass had converted more of any residual gas into stars,
  exhausting its gas supply, and making it more susceptible to
  quenching.}  However, this $\Ms$-dependence of the quenching seems
to be a secondary effect compared to the dominant trend of quenching
with $\Mh$ as seen in \fig{qfrac_cen}$a$.

\fig{qfrac_cen}$b$ shows quenching in the $\delta_5$-$\Mh$ plane.  It
reveals that the quenched fraction increases with $\Mh$ and only
weakly with $\delta_5$.  At constant $\Mh$, the quenched fraction
increases at the most by 0.2, and for most masses only by 0.1, over
two orders of magnitude in $\delta_5$.  However at constant
$\delta_5$, the quenched fraction increases by as much as 0.4 over two
orders of magnitude in $\Mh$.

When we examine quenching in the $\delta_5$-$\Ms$ plane
(\fig{qfrac_cen}$c$), especially at large $\delta_5$, we see that the
quenched fraction for centrals depends on both $\delta_5$ and $\Ms$.
The increase of the quenched fraction with $\Ms$ is consistent with
the increase of $\Mh$ with $\Ms$ below $\Ms \sim 10^{11}\Msun$ (see
\fig{tutcen}$c$).  Above this mass, the quenched fraction increases
with $\delta_5$ in a way similar to the increase of $\Mh$ with
$\delta_5$ in \fig{tutcen}$c$.  Thus, apart from the low $\Ms$ and
high $\delta_5$ region, the quenching pattern in \fig{qfrac_cen}$c$ is
explainable by a basic trend of quenched fraction with $\Mh$.  In
\sec{tutorial} we discussed that the $\Mh$ pattern in \fig{tutcen}$c$
is the result of the interplay between the two modes of $\delta_N$ and
the weakening of the $\Ms$-$\Mh$ relation for massive centrals.  Thus
the quenching pattern in the $\delta_5-\Ms$ plane is consistent with a
correlation of quenching with $\Mh$.  Note that \cite{peng11} did not
find the $\delta_5$ dependence of the quenched fraction at high $\Ms$
mostly because they defined quenching by colour and not by SFR - more
on this in \sec{comppeng}.

The dependence of quenching on $\delta_5$ at low to intermediate $\Mh$
and at high $\delta_5$ (see \fig{qfrac_cen}$b$, also reflected in $c$
at low $\Ms$) is not a neccessary trend in the simple theoretical
framework of halo quenching that we have been advancing.  This may
possibly be attributed to misclassified satellites (an explanation
suggested by \citealp{peng11}).  However, we can offer another
explanation that takes the classification at face value and appeals to
the halo quenching mechanism.  Halo masses on the order of $\Mcrit$
may be capable of sustaining a stable virial shock without actually
containing one.  It has been seen in cosmological simulations that,
once the halo mass is $\sim \Mcrit$, the trigger for the formation of
a shock that quickly expands to the virial radius is typically an
occasional minor merger.  Small centrals that have a high $\delta_N$
must have one or more external haloes nearby, and are therefore more
likely to capture loose satellites of these neighbouring haloes, which
may trigger the virial shock that causes quenching.  \jwg{This
  explanation predicts a higher quenched fraction at the highest
  values of $\delta_5$ in halo somewhat below $\Mcrit$, as is
  observed.}

An astute reader may notice that the quenched fraction seems to reach
lower levels when using $\Ms$ (\fig{qfrac_cen}$c$) than when using
$\Mh$ (\fig{qfrac_cen}$b$).  \jwb{These lower levels of quenching at
  low $\Ms$ compared to low $\Mh$ reflect the fact that the quenching
  contours in \fig{qfrac_cen}$a$ are close to horizontal for $\Ms
  \ltsima 10^{10.4}\Msun$ and $\Mh \ltsima 10^{12.5}\Msun$.  This is
  the regime where $\Mh$-dependent quenching is not expected to be
  important (since $\Mh$ is below or around $\Mcrit$).  It is here
  that residual $\Ms$-dependent quenching becomes important, and this
  regime will be explored further in \sec{quenchingaegis}.}

\begin{figure*}
\epsscale{2.0}
\plottwo{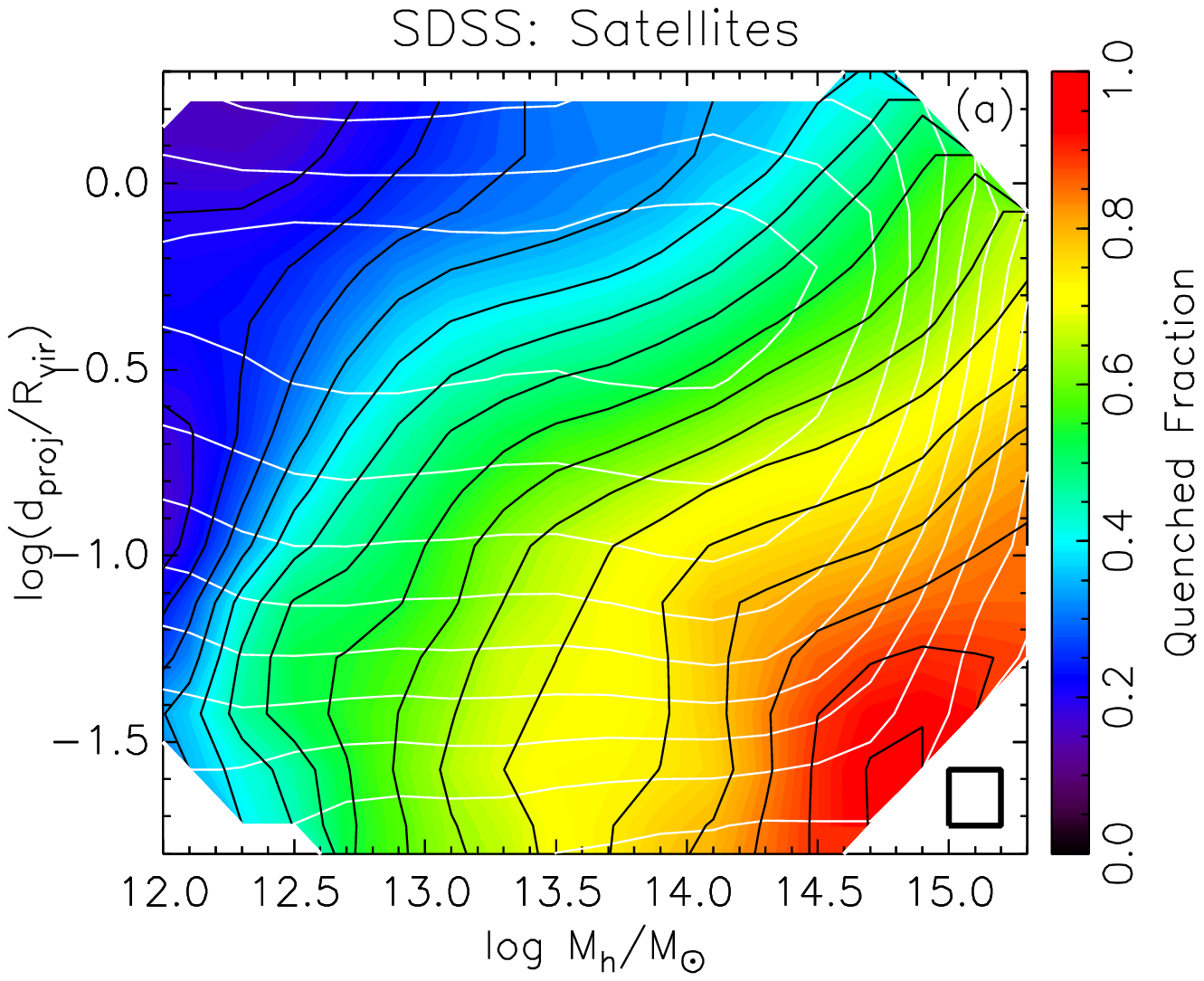}{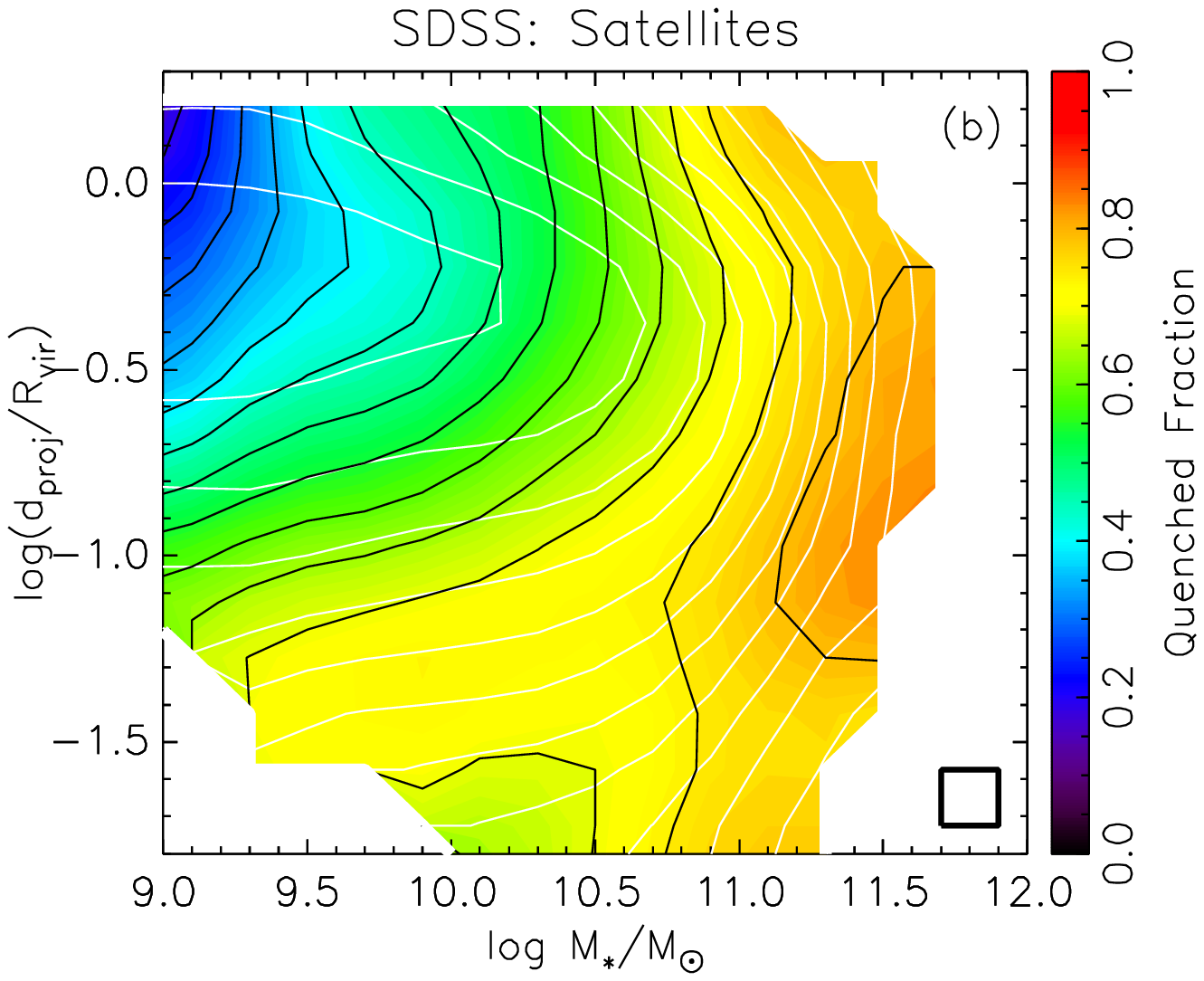}
\caption{\small The mean quenched fraction of SDSS satellites ($0 < z
  < 0.2$) as a function of the projected distance to the group centre
  relative to the group virial radius ($d_{\rm proj}/R_{vir}$) and the
  halo mass $\Mh$ (a) and stellar mass (b).  The thin white contours
  in each panel mark the weighted number density of points and are
  separated by 0.25 dex in density.  The black contours follow the
  coloured shading and are separated by 0.05.  The means were computed
  in pixels of the displayed size and smoothed over adjacent pixels.
  Quenching for satellites increases with $\Mh$ and decreases with
  $\Dist=d_{\rm proj}/\Rvir$.  Quenching also increases with $\Ms$ at
  large $\Dist$ where $\Mh$-quenching is weakest and where the
  satellites likely fell in recently, suggesting a quenching
  dependence on the sub-halo.  Satellites closer to the group centre
  than $\sim 0.2\Rvir$ likely fell in less recently, and their
  quenched fraction depends strongly on host $\Mh$ and almost not at
  all on satellite $\Ms$. }
\label{dproj}
\end{figure*}

\jwr{A correlation between the quenched fraction and halo mass at
  fixed stellar mass, for centrals, is expected on theoretical grounds
  based on the critical mass for virial shock heating, as in
  \cite{bir03}, \cite{ker05}, and \cite{dekbir06}.  We note,
  however, that a detected correlation does not necessarily imply
  causation, and that one should apply caution because the halo mass
  and the stellar mass are not independent of each other.  Indeed, at
  low masses, $M_*<10^{11}\msun$ and $M_h < 10^{12}\msun$, halo mass
  and stellar mass are expected to grow in concert, while at larger
  masses the stellar mass tends to a constant as the halo mass keeps
  growing (\eg, \citealp{cooray05,yang12}).  Can this introduce an apparent
  correlation of the quenched fraction with $M_h$ at a fixed $M_*$ in
  an alternative scenario where the "true" cause of quenching is
  associated with $M_*$ and not with $M_h$?  It can clearly introduce
  a general trend, where the average halo mass of quenched galaxies is
  larger than for un-quenched galaxies, but it is not expected to
  introduce a continuous gradient of the quenched fraction with $M_h$
  at a given $M_*$ all the way to $10^{15}\msun$, and no correlation
  of the quenched fraction with $M_*$ at a given $M_h$, as in
  \fig{qfrac_cen}$a$.  We conclude from \fig{qfrac_cen}$a$ that the
  direct contribution of halo mass to quenching seems to be more
  fundamental than the contribution of stellar mass.}


\subsection{Quenching for satellites}
\label{quenching_sat}

\Fig{qfrac_sat} shows the equivalent of \fig{qfrac_cen}, but for
satellites.  In all three panels of \fig{qfrac_sat}, the quenched
fraction for satellites increases with all three quantities $\Mh$,
$\Ms$ and $\delta_5$ suggesting that these three projections are not
the most helpful way to understanding the main drivers of quenching
for satellites.  As argued in \sec{tutorial}, $\delta_N$ for
satellites is a complicated quantity, whose properties change
according to the choice of $N$, on limiting magnitudes.  Furthermore,
$\Ms$ for satellites is expected to correlate with their position in
their host halo, \eg, because dynamical friction causes massive
satellites to spiral inward faster than less massive ones.  Moreover,
less massive satellites suffer from colour-dependent incompleteness.
A mass-limited cut on the satellite sample which would make it
complete for red galaxies (following the cut applied by
\citealp{dutton11}: $\Ms > 10.7 + \log(z/0.1)$) includes very few
galaxies smaller than $\Ms \sim 10^{10.3}\Msun$ and excludes more than
half the satellite sample.

In light of these difficulties in interpreting \fig{qfrac_sat}, we
complement Figs. \ref{qfrac_sat}$b$ and $c$ with Figs. \ref{dproj}$a$
and $b$, replacing $\delta_5$ on the $y$-axis with the simpler
projected radial distance from the centre ($\Dist \equiv
\dproj/\Rvir$).  (Recall from \fig{satdist} that $\delta_N$ correlates
with $\Dist$ in the single-halo mode of $\delta_N$.)

\fig{dproj}$a$ shows that the quenched fraction for satellites
increases with both halo mass and with proximity to the group centre.
The quenching trend with radial distance is strongest for distances of
greater than $\sim$0.1-0.2$\Rvir$.  Below this, the motions of
satellites and/or projection effects may smear out the quenching trend
with $\Dist$, causing quenching to appear to increase only with $\Mh$.
However we do not expect projection effects to be strong, especially
in massive haloes which contain many satellites at small radial
distance so the signal-to-noise for small $\Dist$ is higher.

\fig{dproj}$b$ shows that the quenched fraction also increases with
satellite stellar mass $\Ms$ for constant $\Dist$.  The $\Ms$
dependence is strongest for satellites at the farthest distances from
the centre, where the $\Mh$ dependence is weakest (but even here, the
$\Mh$-dependence of the quenched fraction is not that weak, increasing
by more than 0.3 over 2.5 dex).  If we assume that these galaxies at
large $\Dist$ have only just fallen into a larger halo, their
quenching dependence on $\Ms$ is consistent with a quenching
dependence on the mass of their {\it own} halo before infall, now sub-halo,
since $\Mh$-dependent quenching operated on them while they were still
centrals (\fig{qfrac_cen}$a$).  The $\Ms$ dependence in \fig{dproj}$b$
virtually disappears for satellites closer to the centre than $\sim
0.2 \Rvir$, where the $\Mh$ dependence is the strongest.  If these
satellites have resided in their host haloes for some time, the
quenching processes which operate in the host halo and which depend on
the host $\Mh$ will eventually affect the satellites so that their
quenched fraction will depend strongly on host $\Mh$ rather than on
the satellite $\Ms$ as we see in \fig{dproj}.

\jwb{\fig{dproj}$a$ succeeds in identifying the properties of
  satellites that produce high quenched fractions of $\gtsima 0.95$,
  {\it namely high $\Mh$ and close proximity to the halo centre}.
  None of the panels in \fig{qfrac_sat} or \fig{dproj}$b$ produce
  regions of the plot with such high quenched fraction.  This seems to
  suggest that it is a combination of halo mass and distance to the
  halo centre that governs quenching in satellites.}

\jwb{In terms of dynamical range of quenching, however, the
  combination of $\Dist$ and $\Mh$ produces a range as broad as the
  pair $\delta_5$ and $\Mh$ (\fig{qfrac_sat}$b$) or even the pair
  $\delta_5$ and $\Ms$ (\fig{qfrac_sat}$c$).  However, quenching
  trends with $\delta_5$ are difficult to interpret due to the dual
  nature of $\delta$.  Which mode of $\delta$ a satellite belongs to
  depends on redshift as discussed earlier, in that a galaxy in the
  single-halo mode at low $z$ will appear to be in the cross-halo mode
  at high $z$ because some members of its group will drop out of the
  magnitude limit.  Furthermore, these satellites with spuriously low
  $\delta_5$ will likely be bluer on average than the true population,
  especially at lower mass, since low mass red galaxies are selected
  against at higher $z$.  Indeed, when we checked the sub-sample with
  $z<0.05$, satellites with $-1 < \delta_5 < 0$ and $9 < \log
  \Ms/\Msun < 10.5$ had quenched fractions 0.1 higher than the whole
  redshift sample in these $\delta_5$ and $\Ms$ ranges, whereas the
  quenched fraction in the $\Dist$-$\Mh$ plane did not change.  For
  these reasons, quenching trends with $\delta_5$ may be somewhat
  exaggerated, and $\Dist$ is the preferred predictor of quenching.}

\jwb{As for whether $\Mh$ or $\Ms$ is better at predicting quenching for
satellites, \fig{qfrac_sat}$a$ seems to indicate that they are equally
good.  However, quenching seems to follow {\it average} $\Mh$ better
than {\it average} $\Ms$.  Specifically, the quenching contours in
\fig{qfrac_sat}$c$ roughly follow the $\Mh$ contours in
\fig{tutsat}$c$, while the quenching contours in \fig{qfrac_sat}$b$ do
not follow the $\Ms$ contours of \fig{tutsat}$b$.  Similarly, the
quenching contours in \fig{dproj}$a$ do not follow the $\Ms$ contours
in \fig{tutsatd}$a$ (except at high $\Dist$) while the quenching
contours in \fig{dproj}$b$ do somewhat follow the $\Mh$ contours in
\fig{tutsatd}$b$ (except at high $\Dist$).  (Quenching at the high
$\Dist$ as we argued above may be following sub-halo mass more than
host halo mass.)}

\jwb{In summary, we have found that the combination of halo mass and
distance to the halo centre is an excellent predictor of quenching for
satellites, producing high quenched fractions of $\gtsima 0.95$.
Combinations involving $\delta_5$ and $\Ms$ also produce large dynamic
range of quenching, but this may be due to a lucky combination of 
redshift effects and sub-halo effects. }

The trend of the quenched fraction with radial distance from the group
centre is consistent with external mechanisms of quenching such as the
halo quenching model, where virial shock heating is most efficient for
large halo masses and in the centres of haloes.  In such a scenario,
quenching is expected to be a function of the density of the hot gas
which increases with decreasing radial distance to the centre as well
as with $\Mh$.  This trend of quenching with proximity to the centre
is also consistent with other processes such as strangulation, tidal
stripping and ram pressure stripping.

\subsection{Comparisons with previous work}
\label{comppeng}

Our finding that quenching increases primarily with $\Mh$ rather than
$\Ms$, and is also correlated with $\Dist$ for satellites, is
consistent with the findings of \cite{weinmann06} who find that halo
mass is more important than luminosity in predicting galaxy properties
(and improves on their result as $\Ms$ is a more intrinsic
internal property than $L$).  

Our study improves on that of \cite{kimm09} \jwr{who} show
similar plots as \fig{qfrac_cen}$a$ and \fig{qfrac_sat}$a$ and could
not decide which of $\Ms$ and $\Mh$ was more important for quenching
in centrals.  \jwr{For} satellites they argued that $\Ms$ and $\Mh$
were equally important for quenching.  We have improved on their
\jwr{analysis} by combining $\Ms$- and
\jwb{$\delta_5$-dependent quenching into a dependence on $\Dist$ and $\Mh$.
For centrals, we have improved on their result by smoothing the
quenched fraction over the $\Ms$-$\Mh$ plane in order to discern the
overall trend of quenching with $\Mh$.  The slight $\Ms$-dependence of
the quenched fraction for centrals with high $\Mh$ is similar to that
found in \cite{kimm09}, but we have shown that the the overall
quenching trend with $\Mh$ is stronger.}

\jwb{Note also that \cite{kimm09} use stellar masses that were
  computed using the models of \cite{bel03} which can underestimate
  $\Ms$ for dusty, star-forming galaxies (for a typical dust model
  with attenuations of 1.6 and 1.3 in the $g$ and $r$ bands, $\Ms$
  will be underestimated by 0.2 dex: see \citealp{bel03}).
  Underestimating $\Ms$ for star forming galaxies but not for
  quiescent galaxies has the effect of exaggerating quenching
  trends with $\Ms$ at fixed $\Mh$.  In contrast, the values of $\Ms$
  that we use are estimated by SED fitting taking dust into account
  (see the references in \sec{mssdss}).  }

\jwr{One may worry that the group stellar masses of \cite{yang07},
  computed using \cite{bel03}, also result in exaggerated quenching
  trends with $\Mh$.}  However, we do not expect this effect to be
  large especially above $\Mh \sim 10^{13}\Msun$ where the centrals
  are more than 70\% quenched and their masses contribute less than
  75\% of their group mass.  To test the effect of dust on the $\Mh$
  estimates, we performed a rough rescaling of the Yang et al. $\Mh$
  values to reflect group masses based on Brichmann et al. $\Ms$
  estimates, and find that \jwr{the quenched fraction increases more
  than twice as much with $\Mh$ at fixed $\Ms$ than with $\Ms$ at
  fixed $\Mh$ over the same range of mass.  This is despite the
  uncertainties in $\Mh$ being much larger ($> 0.3$ dex) than the
  typical uncertainties in $\Ms$ ($< 0.05$ dex).  So, after
  considering dust, our result that $\Mh$ predicts quenching more than
  $\Ms$ appears to be robust.  (More on dust below.)}

\jwb{Our result that quenching of satellites correlates with $\Mh$ and
anticorrelates with $\Dist$ is consistent with the results of
\cite{blantonberlind07} and \cite{hansen09} (\jwr{who use
group luminosity and group richness} rather than
$\Mh$).  The \jwr{decreased quenching with $\Dist$ also
agrees} with several other studies of SFR or the blue/red fractions
within groups and clusters
\citep{gomez03,bal04,tanaka04,rines05,haines07,wolf09,vonderLinden10}.
(See also \citealp{diaferio00,spr01,weinmann10} for semi-analytic
model recipes that reproduce these trends).}

Our results also agree with those of \cite{wetzel12} who find that
quenching for satellites correlates with $\Mh$ and $\Dist$ at fixed
$\Ms$.  \jwr{\cite{wetzel12b} also demonstrated that the distribution
of specific SFR for satellites can be explained by considering the
satellites' time since first infall.}  \jwr{\cite{delucia12} also
  showed} using a semi-analytic cosmological model that satellite
distance from the group/cluster centre is correlated with the time of
accretion.  This is consistent with our finding that the satellite
quenched fraction increases with $\Ms$ at higher $\Dist$ values.
\jwr{Since these satellites recently fell in,} their quenched fraction
reflects their sub-halo mass rather than their host halo mass.  But
some time after the first infall (2-4 Gyr: \citealp{wetzel12}; 5-7
Gyr: \citealp{delucia12}), halo quenching takes over so that quenching
for satellites closer than $\sim 0.2 \Rvir$ depends strongly on their
host halo mass.

\jwb{While \cite{bamford09} found that the red fraction of galaxies
  increases with $\Ms$ and environment (as measured by $\delta_N$ and
  the distance to the nearest galaxy group), they also found that the
  red fraction does not vary strongly with group mass (see their
  Fig. 13).  \jwr{This at first sight seems to contradict our result that
  quenching depends on $\Mh$ (derived from group mass).}  However,
  these authors use the C4 Cluster Catalog of \cite{miller05} which
  does not include isolated galaxies.  The membership of each group
  ranges from 10 to more than 200 galaxies so that Fig. 13 of
  \cite{bamford09} is dominated by satellites.  As we showed in
  \fig{dproj}$a$, both $\Mh$ and $\Dist$ are needed to predict
  quenching for satellites.  Most satellites lie at 0.3-1$\Rvir$ from
  their group centres \jwr{(white contours of \fig{dproj}$a$)} which is where the
  $\Mh$-dependence of quenching is weakest.  Therefore, measuring the
  quenched fraction of satellites as a function of $\Mh$, regardless
  of $\Dist$, will result in a weak $\Mh$ dependence of quenching.
  This also seems to be the main reason why \cite{vandenBosch08} find
  only weak correlation between satellite colour and $\Mh$ and between
  satellite colour and $\Dist$.}

Our analysis improves on the interpretation of \cite{peng10,peng11}
who find that mass determines quenching for both centrals and
satellites, while $\delta_5$ also governs quenching for satellites.
They do not distinguish between stellar mass and halo mass for
centrals (for satellites, see below).  While quenching for
centrals apparently depends on $\Ms$ (\fig{qfrac_cen}$c$), we have
found that \jwr{quenching correlates more strongly with $\Mh$ at fixed
  $\Ms$ than with $\Ms$ at fixed $\Mh$} (\fig{qfrac_cen}$a$).  We also
related the $\delta_5$ and $\Ms$ dependence of quenching for
satellites to \jwr{a quenching relation with $\Dist$ and $\Mh$ and
  argued that these are the preferred predictors of quenched
  fraction.}

\begin{figure*}
\epsscale{2}
\plottwo{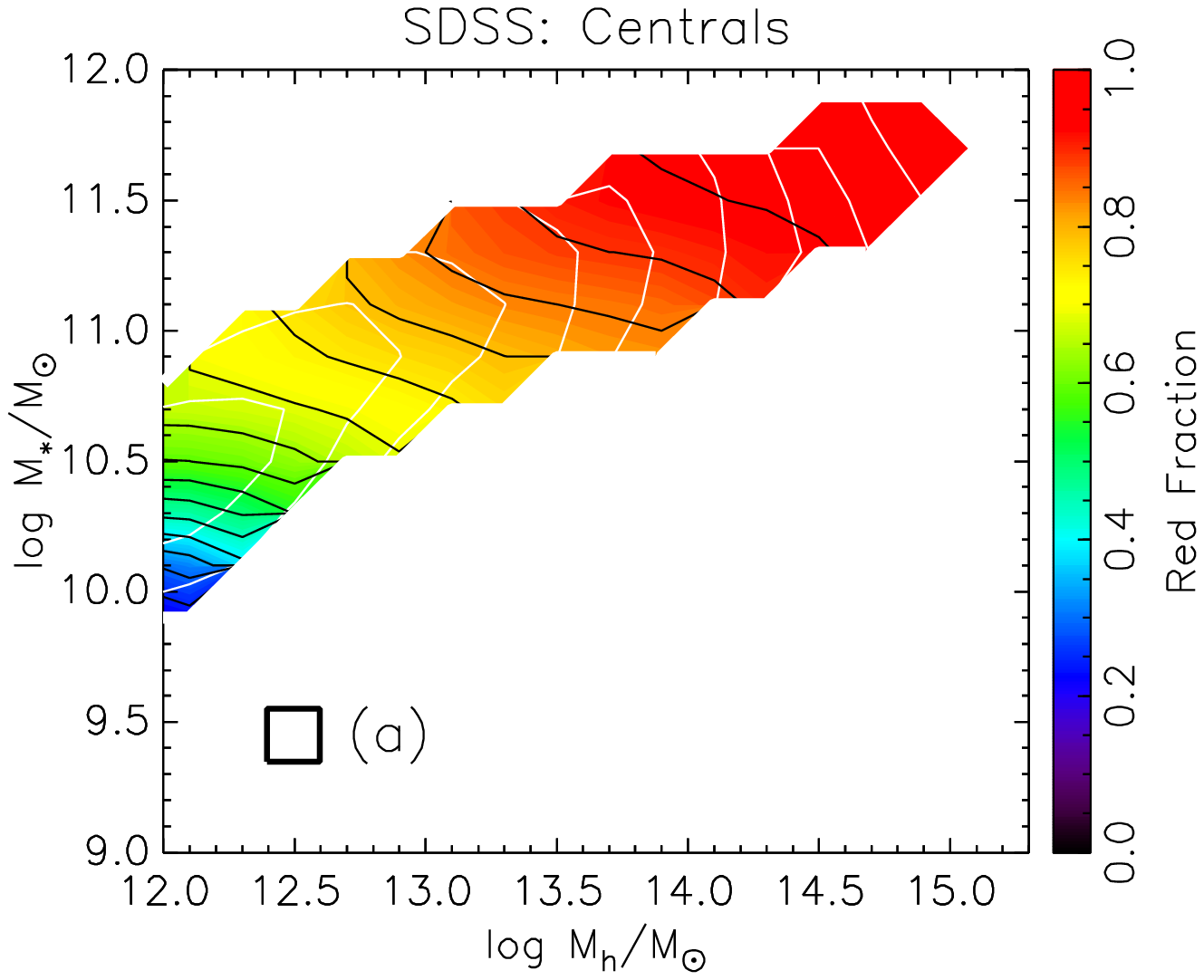}{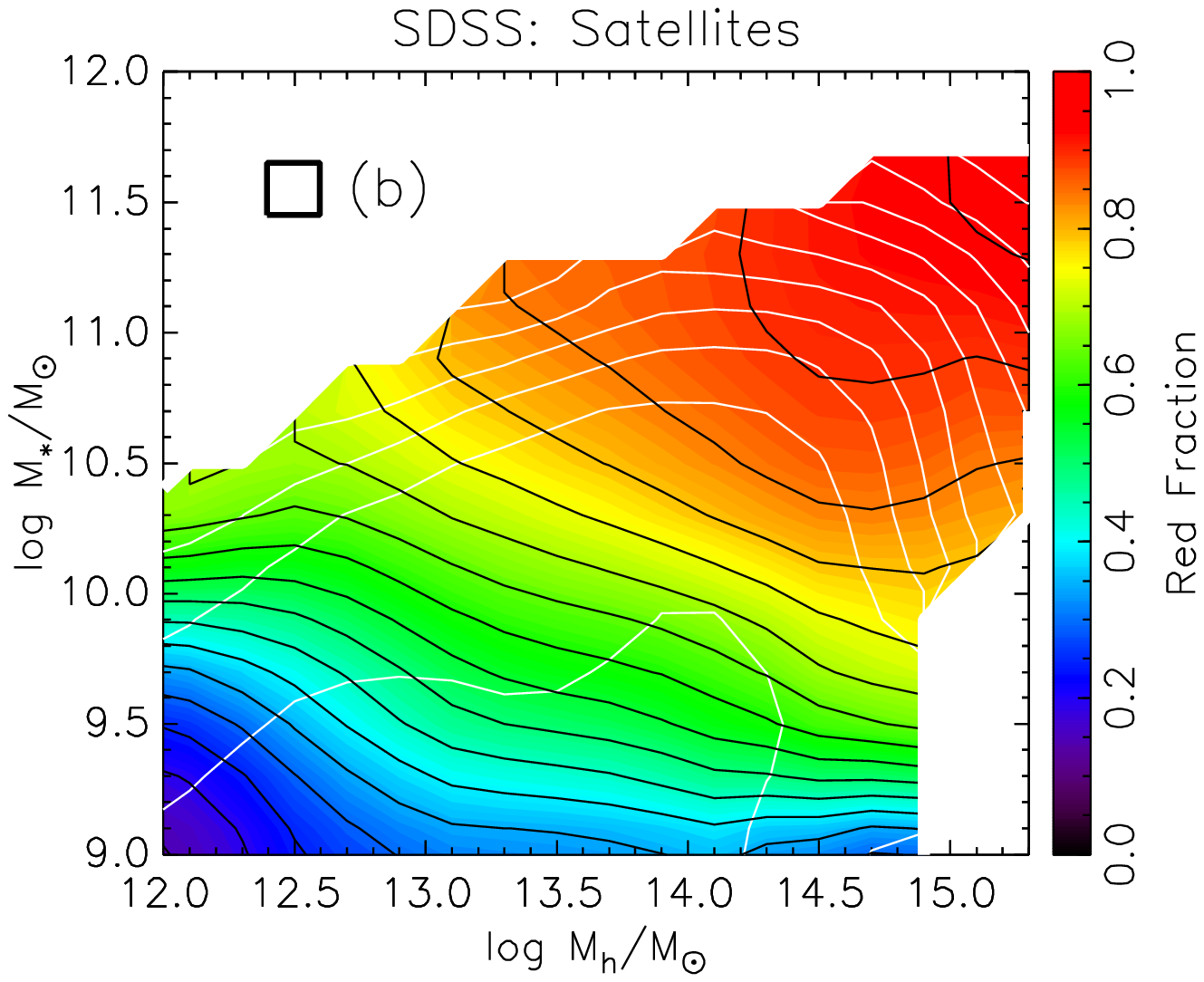}
\caption{\small The mean red fraction as a function of $\Ms$ and $\Mh$
  for centrals (a) and satellites (b) in the SDSS ($0 < z < 0.2$).
  The thin white contours in each panel mark the weighted number
  density of points and are separated by 0.25 dex in density.  The
  black contours follow the coloured shading and are separated by
  0.05.  The means were computed in pixels of the displayed size and
  smoothed over adjacent pixels.  To first approximation, the red
  fraction for centrals correlates with $\Ms$ more strongly than with
  $\Mh$ for both centrals and satellites.  The red fraction behaves
  differently from the quenched fraction (\fig{qfrac_cen}$a$,
  \ref{qfrac_sat}$a$) because red galaxies which are still forming
  stars are on average more massive than the star-forming population.
}
\label{rfrac}
\end{figure*}

However, \cite{peng11} find that for satellites, $\Ms$-quenching is
more important than $\Mh$-quenching at a fixed $\delta_5$, namely for
$1.0 < \Dfive < 1.3$.  \jwr{But this is not} the whole quenching picture.
Using \fig{satdist}, this narrow range of $\delta_5$ corresponds to a
$\log \Dist$ of about -0.4 which is in the region of \fig{dproj} where
\jwr{satellites only recently fell in.  Here, quenching relates to the
sub-halo and the host halo mass will not matter much.}  But we
demonstrated that quenching of satellites strongly increases with
$\Mh$ (and almost not at all with $\Ms$) for satellites closer to the
group centre than $\sim 0.2\Rvir$ where we expect the time since
infall to be long.

\subsection{\jwg{SFR vs. Colour}}
\label{sfrvscolour}

Lastly, we caution that measuring quenching by colour rather than by
SFR causes $\Ms$-dependent quenching to appear stronger and
$\Mh$-dependent quenching to appear weaker than they actually are.
The reason is that colour as a quenching proxy is severely
contaminated by dust.  As described in \sec{SFvscolour}, one third of
the red sequence galaxies are in fact not quenched but are dusty
star-formers.  These predominantly lie on the massive end of the SF
sequence (but share a similar mass distribution to that of the red
sequence).  Thus, dividing the galaxies by colour removes many of the
massive galaxies from the ``star forming'' sample, leaving an overall
less massive ``blue'' sample (by about 0.4 dex).  (This same exercise
does not render the ``red'' sample with a significantly different mass
distribution as the ``passive'' sample).  Therefore, the red fraction
will appear to correlate with $\Ms$ more than the quenched fraction.
Furthermore, since $\Ms$ and $\Mh$ do not correlate for satellites,
and only very weakly for centrals at the massive end, the red fraction
will appear to correlate more with $\Ms$ than with $\Mh$.  And this is
the reason why \cite{peng11} missed the significant
$\delta_5$-dependent quenching for centrals at the massive end
(\fig{qfrac_cen}$c$) that we showed follows the increase of $\Mh$ with
$\delta_5$ at these masses (\fig{tutcen}$c$), reflecting the stronger
trend of quenching with $\Mh$ than with $\Ms$.  This is also the
reason why they are led to find that $\Mh$ is unimportant to the red
fraction for satellites.

To illustrate this effect of dust, we show the red fraction for
centrals as a function of $\Ms$ and $\Mh$ in \fig{rfrac}$a$,
reproducing \fig{qfrac_cen}$a$, but using the red fraction as the
third parameter denoted by the coloured shading.  The red fraction
increases almost entirely in the $\Ms$ direction in contrast to the
quenched fraction (\fig{qfrac_cen}$a$).  \fig{rfrac}$b$ is the
analogous figure for satellites.  Comparing \fig{rfrac}$b$ with
\fig{qfrac_sat}$a$, the red fraction increases more strongly with
$\Ms$ than with $\Mh$, but the quenched fraction seems to increase
equally strongly for both masses (which agrees with \citealp{kimm09}).
In both panels of \fig{rfrac}, for constant $\Mh$, the galaxies with
higher $\Ms$ will have a higher fraction of dusty star-formers than
those with lower $\Ms$, so that correcting for dust will make the
quenching lines become more vertical.  The striking difference between
the red fraction and the quenched fraction demonstrates that dust can
mislead one's interpretation of the primary drivers of quenching.

Our result that $\Mh$ is more important than $\Ms$ for quenching
represents a significant physical difference from the interpretation
of \cite{peng10,peng11}, but their analytic represention of quenching
might still be a useful mathematical tool.  These authors proposed
that the dual dependence of the quenched fraction on $\delta_5$ and
$\Ms$ is separable, with $\delta_5$ governing the quenching of
satellites and $\Ms$ governing the quenching of centrals.  Our results
improve their model by substituting $\Ms$ with $\Mh$ with no
dependence of quenching on $\delta_5$ for centrals (the
$\delta_5$-dependence at constant $\Ms$ or $\Mh$ is very weak except
at the highest $\delta_5$ - see \fig{qfrac_cen}$b$ and $c$ and the
discussion in \sec{quenching_cen}).  For satellites, one may
substitute $\delta_5$ and $\Ms$ with $\Dist$ and $\Mh$ assuming that
$\Dist$-quenching and $\Mh$-quenching are separable.

\section{The evolution of the relations between stellar mass, halo
  mass and density}
\label{tutorialaegis}

The relations between stellar mass, halo mass and environment density
as described in \sec{tutorial} are expected to change with redshift.
The most important physical reason for such a change is that the halo
mass function is expected to grow between $z\sim 1$ to $z\sim 0$.  The
\cite{pressschechter74} scale mass $M^{*}_{PS}$ of the halo mass
function is about $10^{12}\Msun$ at $z=0.75$ and about $10^{13}\Msun$
at $z=0$ for the concordance cosmology.  \jwb{The number density of haloes
with $\Mh = 10^{13.6}$, $10^{14.5}$, and $10^{15} \Msun$ grows by a
factor of 2, 6, and 30 from $z=1$ to $z=0$ (using the halo mass
function of \citealp{tinker08} and the mass transfer function of
\cite{eisenstein98}).  Therefore many more haloes at $z\sim 1$ than at
$z\sim 0$ will be below $\Mcrit$.}  In the AEGIS field, 85$\%$ of the
galaxies at $0.75 < z < 1$ live in haloes less than $\Mh =
10^{12.5}\Msun$, in contrast to only 39$\%$ of the SDSS galaxies below
this mass.  \jwb{Thus at $z\sim 1$ we expect to see if any
$\Mh$-independent quenching becomes dominant in the absence of
$\Mh$-dependent quenching.}

A second difference between $z\sim 1$ and $z\sim 0$ is due to the
brighter magnitude limit at the higher redshift resulting in fewer
satellites observed at $z\sim 1$ than at $z\sim 0$.  The weighted
satellite fraction for the AEGIS sample in the redshift bin $0.75 < z
< 1$ is 10$\%$ compared to 30\% in the SDSS above $\Ms=10^{10}\Msun$.
Since the number of satellites in the AEGIS sample is small (162) any
conclusions concerning quenching for these satellites would be dubious
and so we choose not to address satellites for this redshift.

In order to see how the relations between $\Mh$, $\Ms$ and $\delta_N$
($N=3$ for AEGIS; see \sec{data}) might change at $z\sim 1$ due to
reduced halo masses, we performed the same analysis on the AEGIS
centrals as we performed on the SDSS centrals.  (Note the following
figures depicting a third quantity as a function of the $x$- and
$y$-axes are computed using the same smoothing as in the SDSS with
$m=2$; see \sec{tutorial}.)

\fig{tutcen_aegis} is the high-$z$ ($0.75 < z < 1$) analogue to
\fig{tutcen} (\ie, the $\Ms$-$\Mh$, $\delta_3$-$\Mh$ and
$\delta_3$-$\Ms$ planes).  \fig{tutcen_aegis}$a$ shows that the
stellar mass scales with the halo mass for most of the $\Mh$ range,
and starts to flatten around $\Mh \sim 10^{12.5}\Msun$.  We show the
best fit of Eq. 7 of \cite{yang08} for central galaxies in
\fig{tutcen_aegis}$a$ as the green dashed curve.  The high-end slope
of $0.21$ is similar to that of SDSS for centrals (blue dotted curve
of \fig{tutcen_aegis}), but the knee point is 0.5 dex higher, placing
the upper part of the $\Ms$-$\Mh$ relation for AEGIS higher than that
of the SDSS.  This is consistent with halo growth since $z\sim 1$
around centrals that do not grow much in $\Ms$.  

\jwb{Our $\Ms$-$\Mh$ relation at $z\sim 1$, and its normalization relative
to the $\Ms$-$\Mh$ relation at $z\sim 0$, are in good agreement with
the findings of \cite{leauthaud12} who measure this relation for the
COSMOS survey at $z\sim 0.9$ in comparison with the SDSS.  Our
relation at $z\sim 1$ is also consistent with that of \cite{moster10}
quantitatively, however its relative normalisation compared to the
$\Ms$-$\Mh$ relation at $z\sim 0$ differs from their findings.
\cite{moster10} find that the upper portion of the $z\sim 1$
$\Ms$-$\Mh$ relation roughly coincides with the relation at $z\sim 0$
while the lower portion shifts to lower $\Ms$ at $z\sim 1$ at fixed
$\Mh$ relative to the relation at $z\sim 0$.  \cite{leauthaud12}
speculate that this discrepancy is due to underestimated errors on the
part of \cite{moster10}.  Further discussion on the origin of the
discrepancy is beyond the scope of this paper.}

The number of observed group members is almost always below $N=3$ for
centrals (fewer than 1\% of centrals reside in groups of more than
three observed members, and these live in the most massive haloes).
Therefore we expect to see in AEGIS only a muted form of the
two-component distribution in $\delta_N$ vs. $\Mh$ that we saw for
centrals in the SDSS data (\fig{tutcen}$b$).  Most of the galaxies
being in the cross-halo mode, $\delta_3$ should not correlate with
$\Mh$ except at the very highest masses.  As expected,
\fig{tutcen_aegis}$b$ shows that the distribution of AEGIS centrals
shows no correlation between $\delta_3$ and $\Mh$, except that the
centrals in the most massive haloes never reside in low densities.
For the same reason, the coloured shading of $\Mh$ in
\fig{tutcen_aegis}$c$ runs with $\Ms$ and not with $\delta_3$, except
perhaps weakly at the very massive end.

\subsection{Quenching at high-$z$}
\label{quenchingaegis}

\begin{figure*}
\epsscale{2.2}
\plotthree{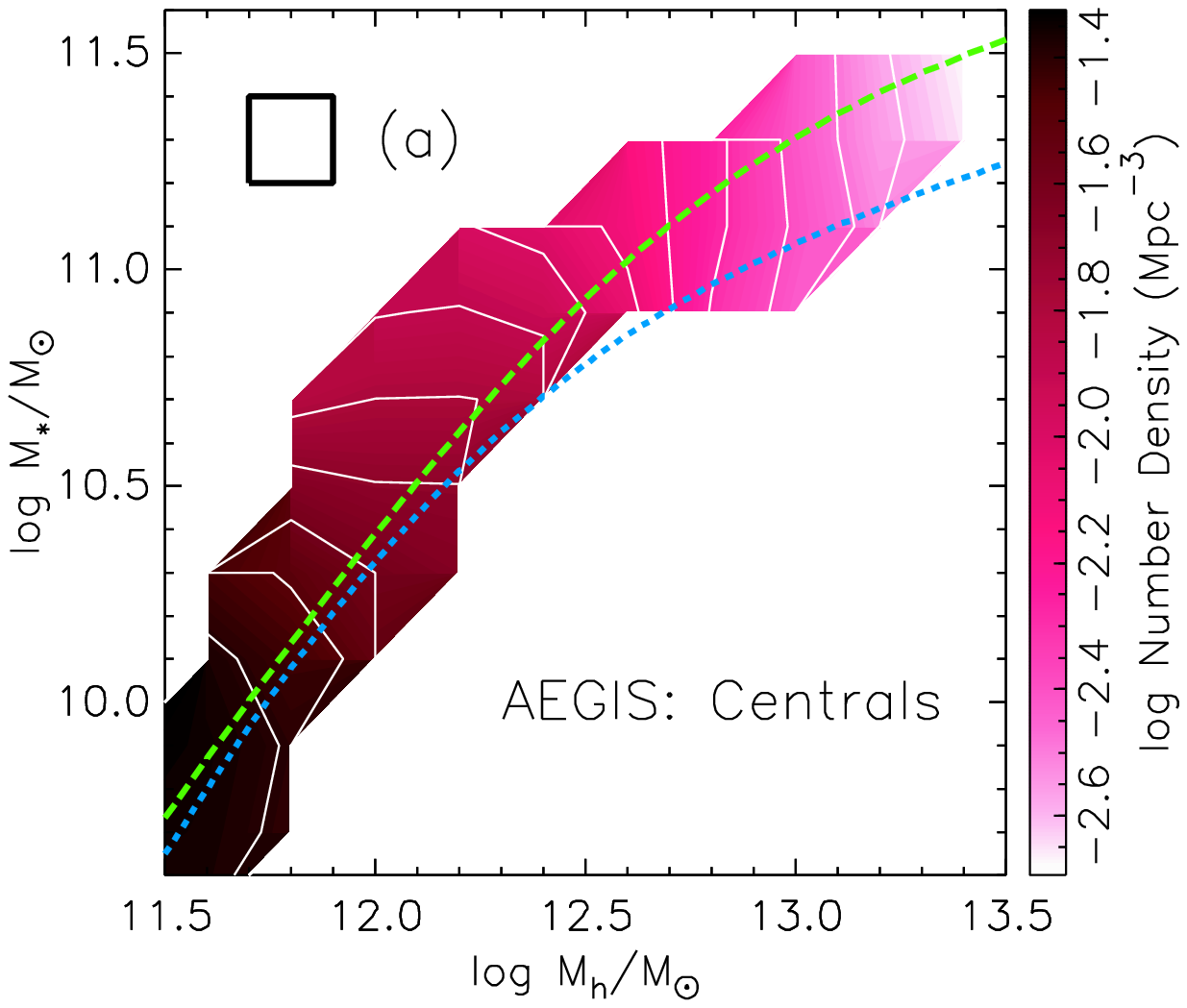}{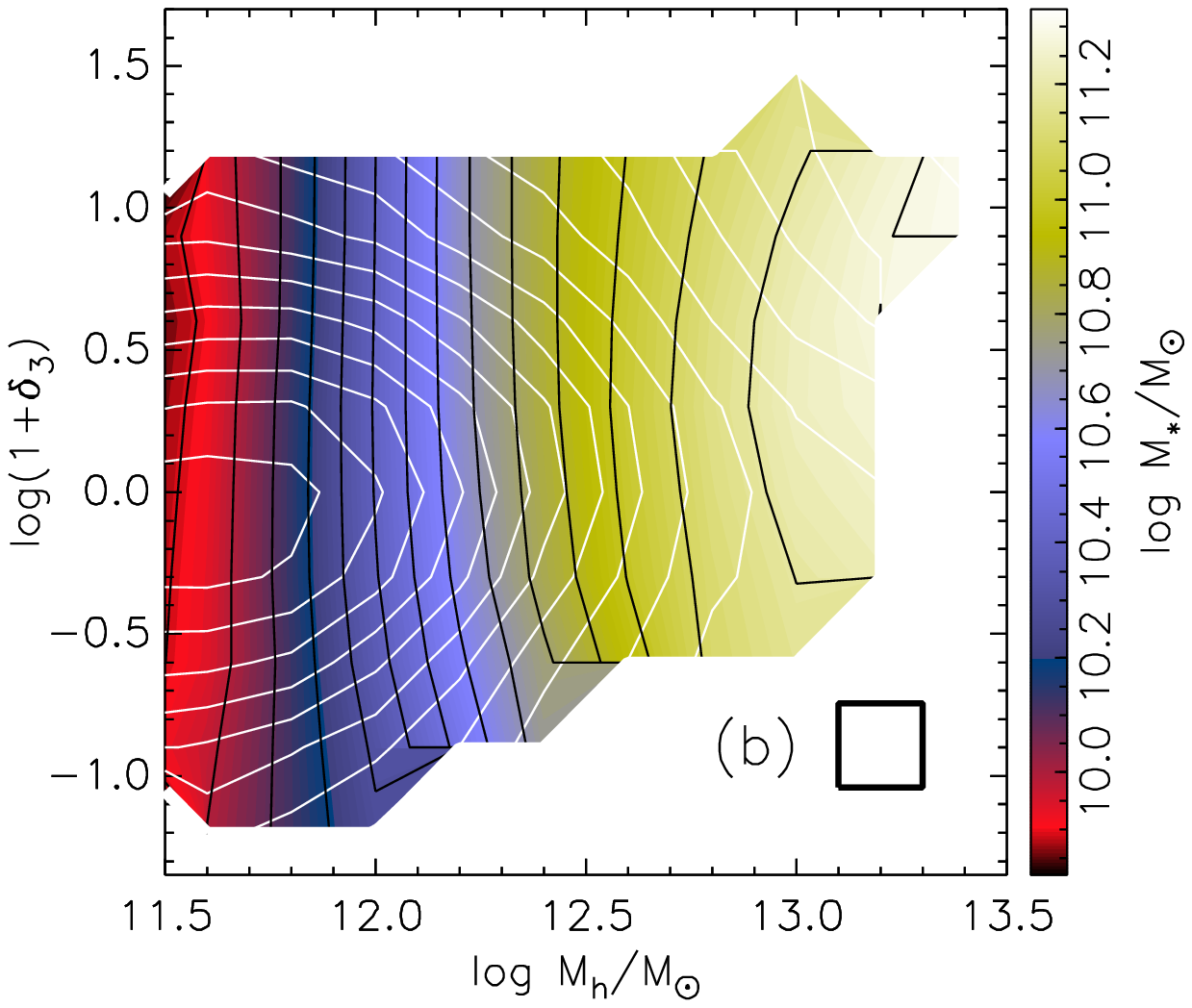}{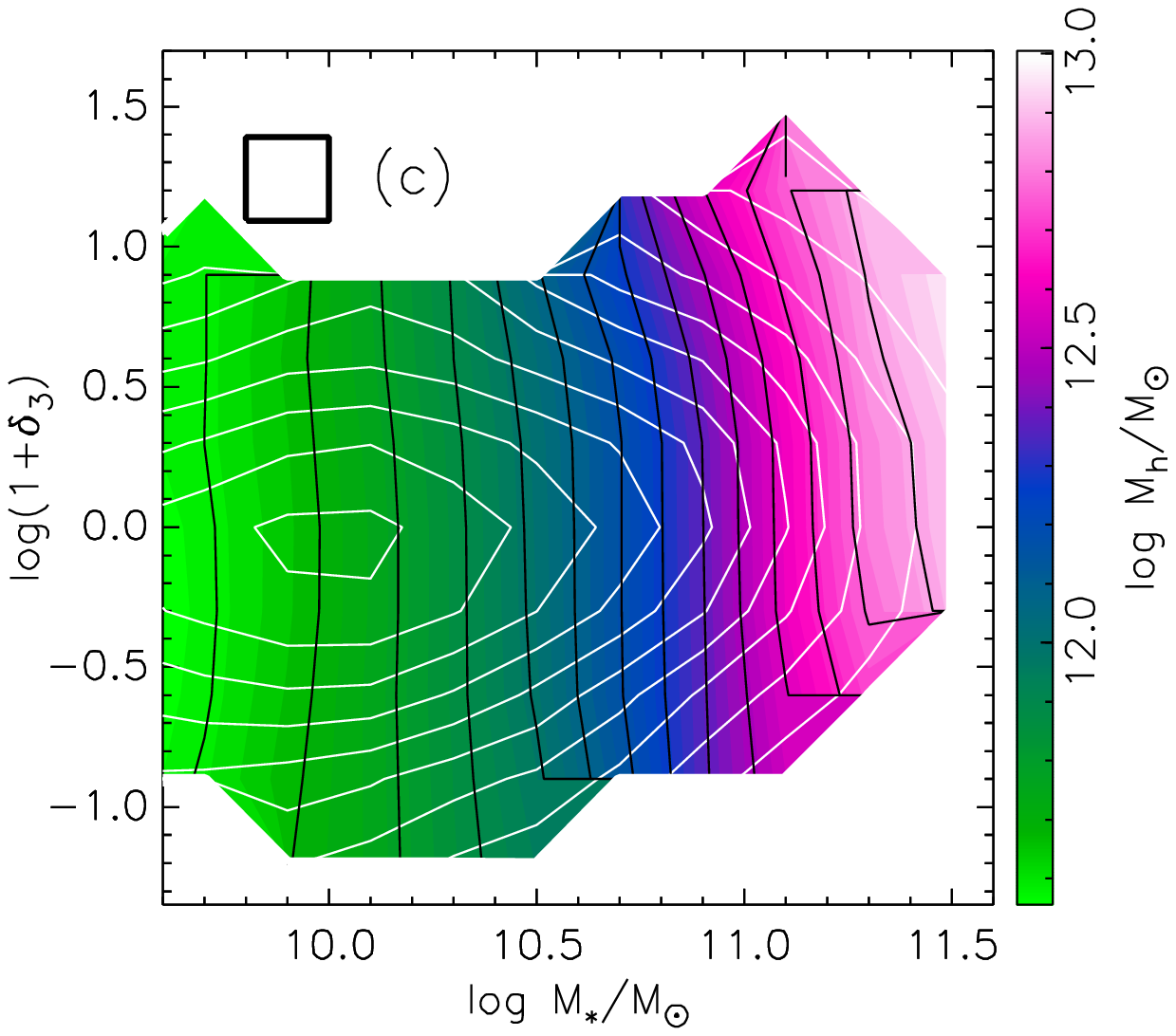}
\caption{\small (a). Stellar mass $\Ms$ versus halo mass $\Mh$ for the
  AEGIS ($0.75 < z < 1$).  The green dashed curve is a fit to Eq. 7 of
  Yang et al. (2008).  The blue dotted line is the analogous fit for
  the SDSS.  (b) Mean stellar mass $\Ms$ as a function of density
  $\delta_3$ and halo mass $\Mh$.  (c). Mean halo mass $\Mh$ as a
  function of density $\Dthree$ and stellar mass $\Ms$.  All panels
  show only central galaxies in the AEGIS sample.  The thin white
  contours in each panel mark the weighted number density of points
  and are separated by 0.1 dex in density.  The black contours in (b)
  and (c) follow the coloured shading and are separated by 0.1 dex.
  The number densities in (a) and the means in (b) and (c) were
  computed in pixels of the displayed size and smoothed over adjacent
  pixels.  The $\Ms$-$\Mh$ relation for massive centrals at $z\sim 1$
  lands higher than that of the centrals at $z\sim 0$, consistent with
  halo growth since $z\sim 1$.  $\delta_3$ and $\Mh$ do not correlate
  except that centrals of massive haloes do not live in sparce
  environments.}
\label{tutcen_aegis}
\end{figure*}

\begin{figure*}
\epsscale{2.2}
\plotthree{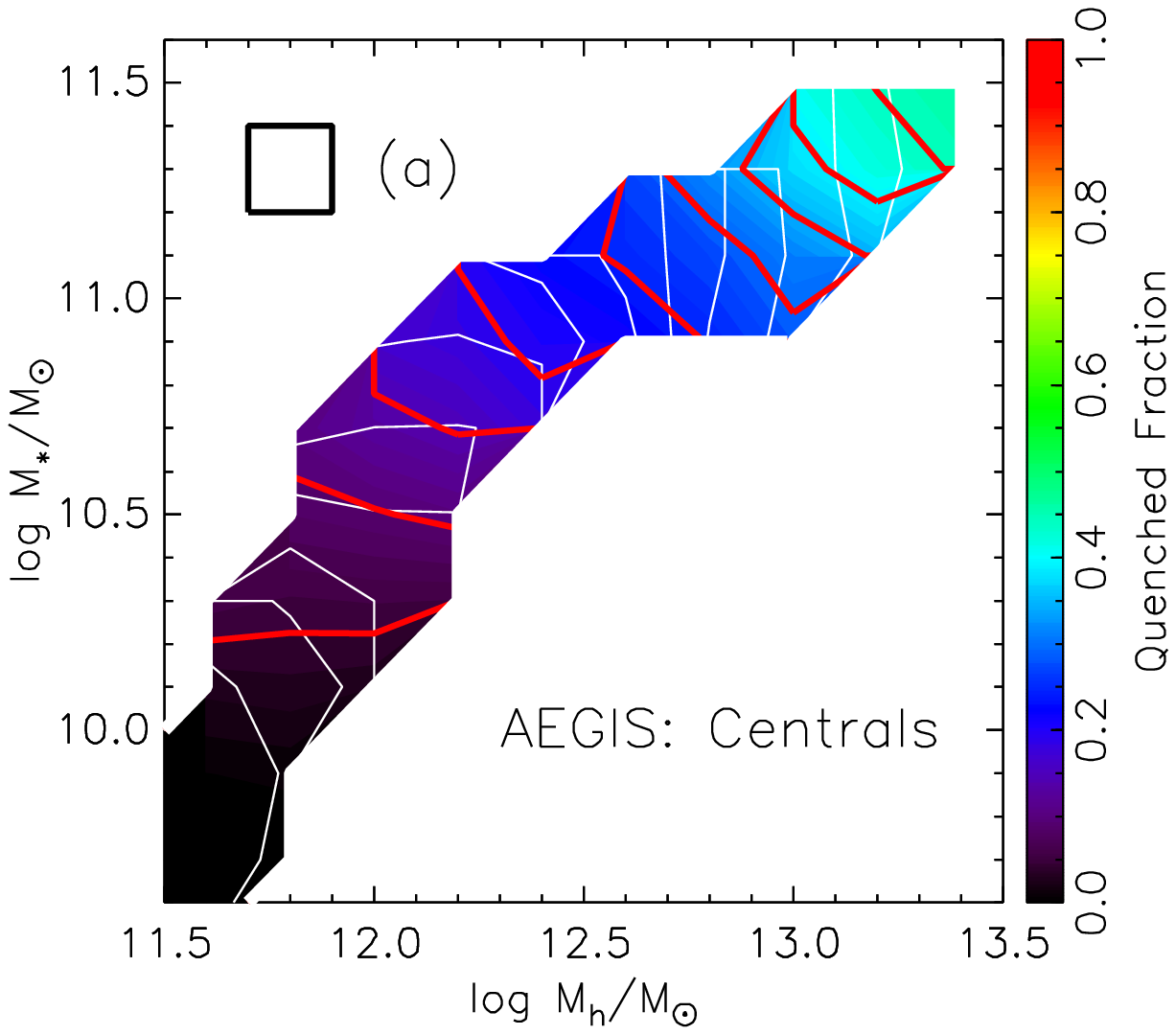}{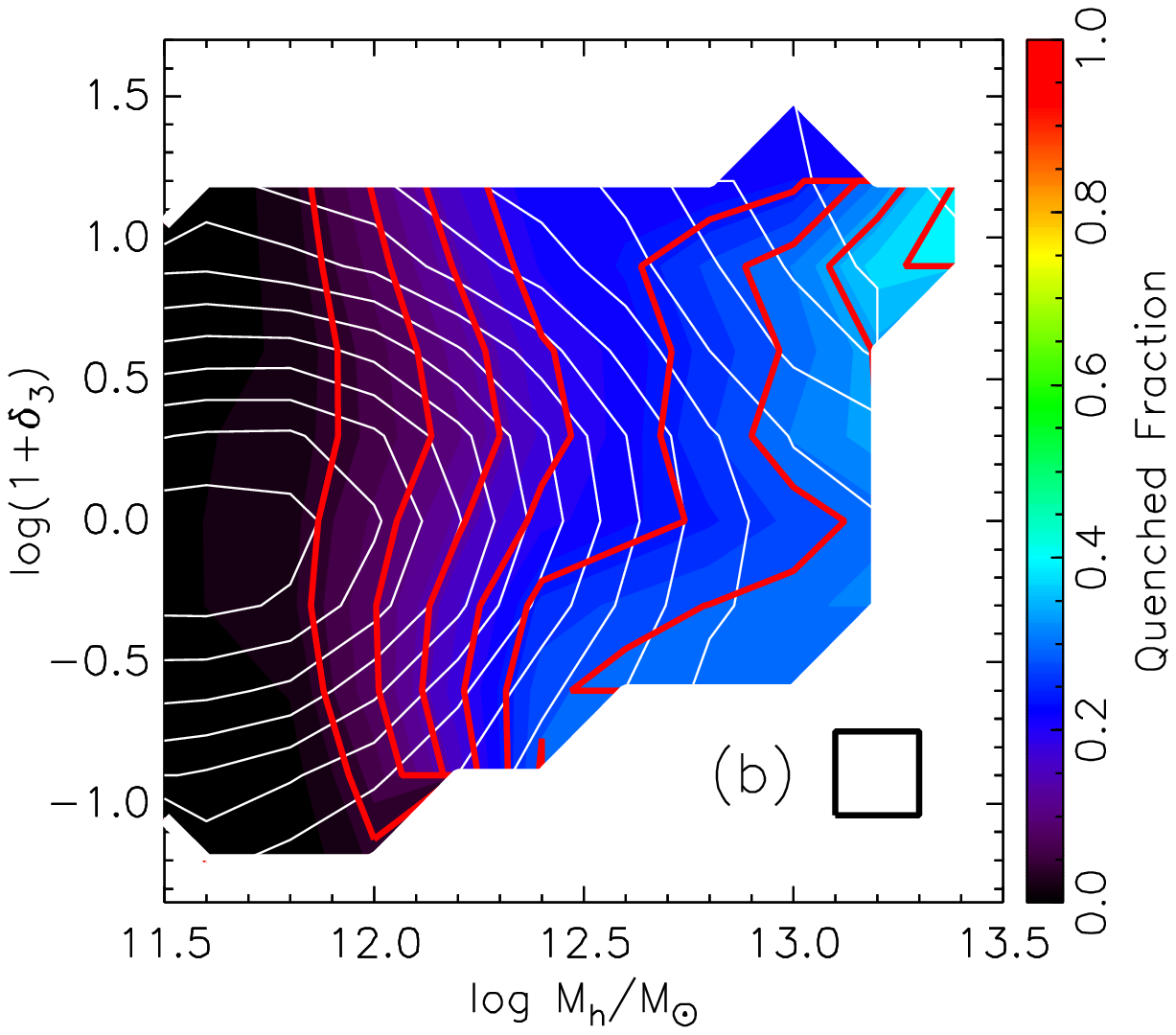}{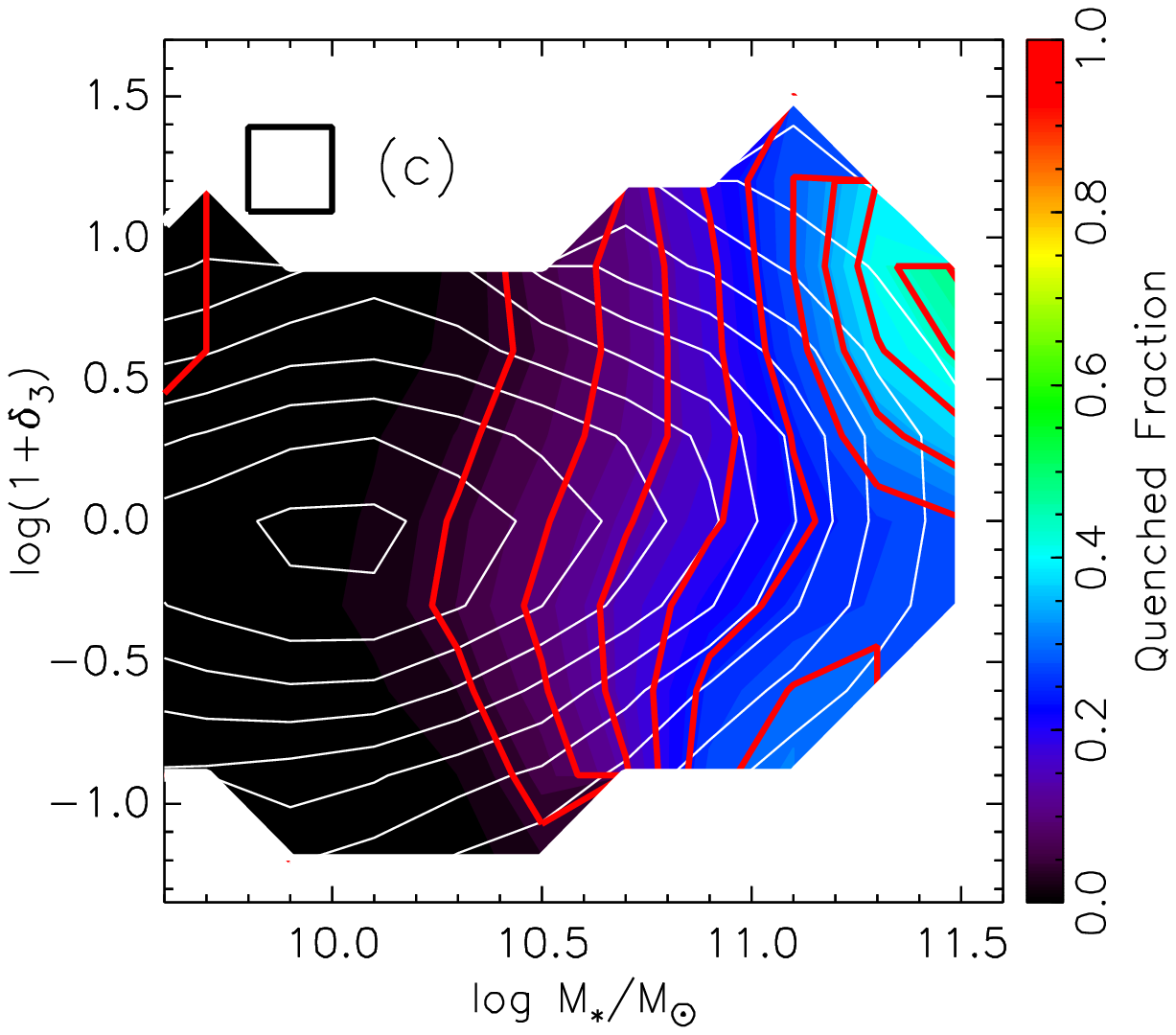}
\caption{\small The mean quenched fraction as a function of $\Ms$ and
  $\Mh$ (a), of $\delta_3$ and $\Mh$ (b), and of $\delta_3$ and $\Ms$
  (c) for central galaxies in the AEGIS ($0.75 < z < 1$).  The thin
  white contours mark the weighted number density of points and are
  separated by 0.1 dex in density.  The red (dark) contours follow the
  coloured shading and are separated by 0.05.  The means were computed
  in pixels of the displayed size and smoothed over adjacent pixels.
  Below $\Mh \sim 10^{12.5}\Msun$, quenching for centrals at $z\sim 1$
  increases primarily with $\Ms$.  Above this mass, quenching
  increases with $\delta_3$, corresponding with the increase of $\Mh$
  with $\delta_3$ in \fig{tutcen_aegis}$c$ just as for the SDSS
  centrals.  But the overall quenched fraction remains low.}
\label{qfrac_cen_aegis}
\end{figure*}

As we have argued, halo quenching via virial shock seems to dominate
the quenching of galaxies in haloes above $\Mh \sim 10^{12.5}\Msun$,
\ie, those that are large enough to support a virial shock.  However,
haloes smaller than this are less likely to support a stable virial
shock and such haloes would be ideal for testing whether
non-halo-dependent quenching mechanisms operate when virial shock
heating is relatively unimportant.  As described above, the majority
(85$\%$) of galaxies at $z\sim 1$ reside in haloes smaller than $\Mh
\sim 10^{12.5}\Msun$ (corresponding to $\Ms \sim 10^{11}\Msun$ for
centrals - see \fig{tutcen_aegis}$a$) while the stellar mass function
of galaxies has remained relatively constant over time since $z\sim 1$
\citep[\eg, ][]{bundy06}.  While the halo quenching model predicts
very low quenched fractions at all masses below $\Mh \sim
10^{12.5}\Msun$, the behaviour of the quenched fraction in this regime
will indicate the strength of other quenching mechanisms when halo
quenching is expected to be less important.

\fig{qfrac_cen_aegis} shows the same panels as \fig{tutcen_aegis}
(\ie, central galaxies at $z\sim 1$) with the quenched fraction as the
third parameter denoted by the coloured shading.  Panel $a$ shows that
the quenched fraction depends only on $\Ms$ (horizontal red lines) for
low-mass centrals but begins to depend also on $\Mh$ at higher
masses (tilted red lines).  The increase of the quenched fraction with
$\Ms$ right below $\Mh \sim 10^{12.5}\Msun$ reflects a similar
behaviour at $z\sim 0$ (see \fig{qfrac_cen}$a$ at the lowest masses)
and points to an $\Ms$-dependent quenching mechanism that operates
where halo quenching is not expected to dominate.

\jwb{Note that our AEGIS sample is complete only down to $\Ms \sim
10^{10.7}\Msun$ \citep{noe07}, and the missing galaxies below this
mass are likely to be red or quiescent and of lower mass.  Therefore,
for fixed $\Mh$ in \fig{qfrac_cen_aegis}$a$, regions of lower $\Ms$ in
the incomplete region may actually be more quenched than observed.  In
other words, the quenching contours in \fig{qfrac_cen_aegis}$a$ may in
reality be more tilted than horizontal.  However we do not expect the
effect to be very large since there is evidence \citep[\eg,
][]{bel04,brown07,faber07} that there are fewer red galaxies at $z\sim
1$ to be missed at all.}

In panel $c$ of \fig{qfrac_cen_aegis}, the quenched fraction increases
with $\Ms$ for centrals less massive than $\Ms \sim 10^{11}\Msun$,
but begins to increase also with $\Dthree$ above this mass.  For these
galaxies the increase of the quenched fraction with $\delta_3$
coincides with an increase in $\Mh$ (\fig{tutcen_aegis}$c$) similar to
what we saw for the SDSS centrals (\sec{quenching_cen}).

\jwb{In order to ensure that the $\Ms$-dependence of quenching we observe
here is not due some artefact of the group finding process, \ie,
contamination by satellites, we tested the AEGIS mock catalogues in
the following simple manner.  For each mock catalogue, we assumed a
mass-independent quenching such that 20\% of all centrals and 80\% of
all satellites are quenched regardless of $\Ms$ or $\Mh$.  We then
computed the quenched fraction in 10 bins of $\Ms$ and $\Mh$ for
centrals as determined by the group finder.  The Pearson correlation
coefficients between the quenched fraction and mass were on average
$0.02\pm 0.08$ for the quenched fraction vs. $\Ms$ and $0.14\pm 0.08$
for the quenched fraction vs. $\Mh$.  For the real AEGIS data, the
Pearson correlation coefficients are about 0.9 between the quenched
fraction and both $\Ms$ and $\Mh$.  In other words, satellite
contamination in the quenching trends is too low to explain the
quenching correlations with mass seen in \fig{qfrac_cen_aegis}.}

Regardless of these trends of the quenched fraction, the total
quenched fraction remains low at all masses.  The weighted quenched
fraction for all the centrals at ($0.75 < z < 1$) is 0.11.  Since the
DEEP2 $R$-band magnitude limit corresponds to a lower $\Ms$ for blue
galaxies than red galaxies, we compare these weighted quenched
fractions with those of the SDSS after applying a completeness cut
that samples the same complete region of colour-mass space out to
$z=1$ (refer to \citealp{gerke07} for details).  The
colour-mass-complete weighted quenched fraction is 0.41 for centrals,
almost four times those of the AEGIS galaxies.  This dearth of
quenched galaxies at $z\sim 1$ compared to $z\sim 0$ is consistent
with the growth of the red sequence over the last $\sim 7$ Gyr
\citep[\eg, ][]{bel04,brown07,faber07}.  Since the halo mass function
is predicted to have grown an order of magnitude in mass since $z\sim
1$, the quadrupling of the colour-mass-complete quenched fraction from
$z\sim 1$ to $z\sim 0$ can naturally be explained by halo mass
dependent quenching.  

\jwb{Our result that $\Ms$-dependent quenching dominates over $\Mh$- and
$\delta_3$-dependent quenching for centrals at $z\sim 1$ is consistent
with the predictions of \cite{mcgee09}.  These authors used a
semi-analytic model to predict that the fraction of ``environmentally
affected'' galaxies should be very low at $z\sim 1.5$.}

Although $\Ms$-dependent quenching seems to operate in regimes where
virial shock heating is expected to be unimportant, it seems to be a
weak effect compared to the halo quenching which seems to have
eventually overtaken $\Ms$-dependent quenching by $z\sim 0$.

\section{Summary and Conclusions}
\label{discuss}

Using the large sample size of the SDSS at $z\sim 0$ and the
multi-wavelength nature of the AEGIS dataset at $z\sim 1$ we have
examined the relations between stellar mass, halo mass, environment
density and distance of satellites from the group centre over the
redshift range $0 \ltsima z \ltsima 1$.  We showed that there are two
keys to understanding these relations.  First, the density estimator
to the $N$th-nearest neighbour behaves in two modes, one in which the
number of observed group members is less than $N$ so that the density
measures the distance to the nearest haloes, and the other in which
the number is greater than $N$ so that the density measures distances
within a given halo.  These are the cross-halo and single-halo modes
respectively, \jwg{and they are typically $\delta_N \ll 1$ and $\delta_N
\gg 1$ for a given $\Ms$}.  Second, the relations between $\Ms$, $\Mh$
and $\delta_N$ behave differently for satellites than for central
galaxies.  We argued that for satellites, the projected distance from
the halo centre $\Dist$, which anti-correlates with $\delta_N$ in the
single-halo mode, is a better measure of ``environment'' than
$\delta_N$ because the former does not behave in two modes that depend
on both $N$ and the limiting magnitude of the survey.  \jwg{$\Dist$
measures density, but also time spent in the cluster.}

Armed with the understanding of these relations, we explored
quenching as a function of the quantities $\Mh$, $\Ms$, $\delta_N$,
and $\Dist$.  We used the wealth of SFR data from the SDSS and
the AEGIS surveys to measure quenching as the fraction of low-SFR
galaxies, rather than the fraction of red galaxies which can be
severely affected by dust.  We separated the analysis into samples of
central and satellite galaxies since the predictions for the behaviour
of the quenched fraction differ for these populations.

We also discussed the effects of the halo mass function at $z\sim 1$
having fewer massive haloes compared to $z\sim 0$ on the relations
between $\Ms$, $\Mh$ and $\delta_N$, and on the behaviour of the
quenched fraction as a function of these quantities.

Our main results are listed in detail below:

\begin{enumerate}

\item {\bf Colour: an imperfect surrogate for quenching}

~

 About one third (30\% - volume weighted) of the red sequence of the
 colour-magnitude diagram for the SDSS lies on the SF sequence.  These
 dusty red star forming galaxies tend to be the most massive star
 forming galaxies and are on average 0.4 dex more massive than the
 whole population of SF sequence galaxies (see \fig{colourvssfrresid}
 and the discussion in \sec{SFvscolour}).  These results are
 consistent with those of \cite{maller09} who find that a third of red
 sequence galaxies move to the blue cloud after an
 inclination-dependent dust correction, and that this dust correction
 is larger for galaxies with brighter $M_K$.  The presence of
 dust-reddened galaxies on the red sequence means that colour does not
 correlate perfectly with star-formation quenching, which we showed
 has important implications for quenching correlations vs. stellar and
 halo mass.  For both centrals and satellites, the red fraction will
 appear to correlate more with $\Ms$ (due to the massiveness of dusty
 star formers) and not much with $\Mh$ (due to the weak correlation
 between $\Ms$ and $\Mh$ at the high end for centrals and no
 correlation at all for satellites), which is the opposite of the
 behaviour of the quenched fraction (see \fig{rfrac} and the
 discussion in \sec{comppeng}).

~

\item {\bf Relations between $\Mh$, $\Ms$, $\delta_N$ and $\Dist$}

\begin{enumerate}
  
\item Environment density as measured by $\delta_N$, \ie, the density
  based on the distance to the $N$th nearest neighbour, is
  characterised by two modes of behaviour, a cross-halo mode in which
  the number of observed group members is fewer than $N$ and a
  single-halo mode where the membership is greater than $N$.  

\item For central galaxies at $z\sim 0$, $\Ms$ correlates with $\Mh$
  below about $\Mh \sim {\rm few} \times 10^{12}\Msun$ above which
  $\Ms$ only weakly depends on $\Mh$, consistent with the findings of
  \cite{yang09} (\fig{tutcen}$a$).  At $z\sim 1$, the relation between
  $\Ms$ and $\Mh$ coincides with the relation at $z\sim 0$ at $\Mh\sim
  10^{12}\Msun$, but rises above the $z\sim 0$ relation for haloes
  more massive than that (\fig{tutcen_aegis}$a$).  This is consistent
  with findings of \cite{leauthaud12}, and with halo growth around
  central galaxies which grow very little in $\Ms$.  Most (85$\%$) of
  all galaxies at $z\sim 1$ reside in haloes smaller than $\Mh =
  10^{12.5}\Msun$.

\item Central galaxies in the single-halo mode at $z\sim 0$ are
  characterised by high $\delta_5$ and high $\Mh$, while those in the
  cross-halo mode exhibit little relation between $\delta_5$ and $\Mh$
  (\fig{tutcen}$b$).  Central galaxies at $z\sim 1$ behave similarly,
  but the number of observed group members rarely exceeds $N=3$
  (\fig{tutcen_aegis}$b$).

\item For satellite galaxies at $z\sim 0$ in the single-halo mode
  (which is 58$\%$ of these satellites), $\Ms$ does not strongly
  correlate with $\Mh$.  In the cross-halo mode
  for satellites, $\Ms$ correlates with $\Mh$ as expected from
  statistical and dynamical friction constraints (\fig{tutsat}$b$,$c$).

\item For satellite galaxies in the single-halo mode at $z\sim 0$,
  $\delta_5$ correlates with $\Mh$ (\fig{tutsat}$b$), as expected from
  considerations of the universal halo profile and from the
  predictions of the spherical collapse model of a virialised halo
  combined with the known halo occupation function.

\item For satellites at $z\sim 0$, $\delta_5$ anti-correlates with
  $\Dist = d_{\rm proj}/R_{\rm vir}$ in the single-halo mode but shows
  no dependence in the cross-halo mode (\fig{satdist}).  This is
  consistent with the expected statistical behaviour of $\delta_5$.

\end{enumerate}

\item {\bf Quenching, Mass, Density and Distance}

\begin{enumerate}

\item \jwb{For central galaxies at $z\sim 0$, the quenched fraction appears
  more strongly correlated with $\Mh$ (at fixed $\Ms$) than with $\Ms$
  (at fixed $\Mh$) for centrals more massive than $\Ms \sim
  10^{10.4}\Msun$ (\fig{qfrac_cen}$a$).}

\item At $z\sim 1$, where most of the galaxies reside in haloes less
  massive than the halo-quenching scale, quenching correlates with
  $\Ms$ rather than $\Mh$ for haloes below $\sim 10^{12.5}\Msun$
  (which host 85\% of these centrals), but the quenched fraction
  remains low ($\ltsima 0.2$) (\fig{qfrac_cen_aegis}$a$).  This
  behaviour may signal a $\Ms$-dependent quenching mode at high
  redshift that is overtaken by $\Mh$-dependent quenching as the
  haloes grow by $z\sim 0$.

\item For central galaxies at $z\sim 0$, in the $\delta_5$-$\Ms$
  plane, the quenched fraction increases with both $\Ms$ and
  $\delta_5$ (especially at high $\Ms$) in such a way that is
  consistent with the increase in mean $\Mh$ with both of these
  quantities (compare \fig{qfrac_cen}$c$ with \fig{tutcen}$c$).  In
  other words, the increase of quenching with both $\Ms$ and
  $\delta_5$ for centrals largely reflects the increase of quenching
  with $\Mh$ alone.  At $z\sim 1$ the central galaxies behave in a
  similar way (compare \fig{qfrac_cen_aegis}$c$ with
  \fig{tutcen_aegis}$c$).

\item For satellite galaxies at $z\sim 0$, the quenched fraction
  increases with $\Mh$ and decreases with the distance to the group
  centre $\Dist$ (\fig{dproj}$a$).  \jwb{The combination of $\Mh$ and
  $\Dist$ is an excellent predictor of quenching for satellites,
  producing quenched fractions as high as 95\%.  $\Mh$ and $\Dist$ are
  preferred over $\Ms$ and $\delta_5$ as predictors of quenching
  because 1) $\delta_5$ behaves differently in two modes and suffers
  from redshift effects making quenching trends with $\delta_5$
  difficult to interpret, and 2) quenching follows average $\Mh$
  better than average $\Ms$ (compare \fig{qfrac_sat}$c$ with
  \fig{tutsat}$c$ and \fig{qfrac_sat}$b$ with \fig{tutsat}$c$).
  $\Ms$-dependent quenching seems to matter more than $\Mh$ only at
  large distances from the group centre (where satellites are likely
  to have just recently fallen in) and may reflect a quenching
  dependence on the mass of the sub-halo (compare \fig{dproj}$a$ with
  $b$).  But nearer to the group centre than $\sim 0.2 \Rvir$, the
  quenching depends strongly on host $\Mh$ and almost not at all on
  satellite $\Ms$.}

\end{enumerate}

\end{enumerate}

Our main conclusion is that the quenched fraction of galaxies at
$z=0-1$ seems to be primarily driven by the dark-matter halo mass.
This is consistent with the theoretical prediction for an accretion
shutdown due to virial shock heating once the halo grows above a
threshold mass, which possibly triggers other quenching mechanisms,
such as stellar and AGN feedback \citep{bir03,dekbir06}.  For
satellite galaxies, the quenching also depends on the proximity to the
halo center, which reflects the local density and the history of the
satellite within the halo. Stellar mass is only a secondary factor,
and $\delta_N$ is a poor measure of environment.  As discussed in
\sec{comppeng}, our results are consistent with those of previous work
\citep[eg.,][]{weinmann06,blanton07,hansen09,kimm09,wetzel12}.  Our
results can be brought into agreement with the conclusion of
\cite{peng10,peng11} once stellar mass is substituted by halo mass.
The difference is driven by our proper use of SFR rather than color to
define quenching, and it has the advantage of unifying all phenomena
under a common halo quenching model.  However, marginal evidence is
also found for an $\Ms$-related quenching mode at earlier epochs when
$\Mh$ is low.  It will be valuable to study the quenching mechanism
with comparisons to galaxy properties beyond $\Ms$, such as
bulge-to-disk ratio and central density (Cheung et al., Fang et
al. and Barro et al., in preparation).  Combined with the accretion
shutdown associated with halo mass, internal mechanisms related to the
bulge and the central black hole may serve as the direct cause for shutdown
in star formation \citep{dekel09,martig09}.

\section*{Acknowledgements}
\jwb{We thank the anonymous referee for comments that greatly improved
  this paper.}  We thank Marcello Cacciato for helpful discussions and
for kindly sharing his halo mass function code.  We also thank Frank
van den Bosch and Xiaohu Yang for kindly providing the group catalog
for the SDSS DR7 and for helpful discussions.  We acknowledge the
helpful and stimulating discussions with Andrew Wetzel, Surhud More,
Loren Hoffman, Simon Lilly and Shannon Patel.  All plots in this paper
were produced using the open source GNU Data Language
(http://gnudatalanguage.sourceforge.net/).  This research has been
supported by the ISF grant 6/08, by GIF through grant
G-1052-104.7/2009, by a DIP grant, and by an NSF grant AST-0808133 at
UCSC.  MCC acknowledges support from NASA through Hubble Fellowship
grant \#HF-51269.01-A, and support from the Southern California Center
for Galaxy Evolution.

\bibliographystyle{mn2e}
\bibliography{jobib}

\label{lastpage}
\end{document}